\documentclass[twocolumn,twocolappendix,usenames,dvipsnames]{aastex631}

\usepackage{comment}

\begin{document}

\title{Complex Velocity Structure of Nebular Gas in Active Galaxies Centred in Cooling X-ray Atmospheres}

\author{Marie-Jo\"{e}lle Gingras}
\affiliation{Department of Physics and Astronomy, University of Waterloo, Waterloo, ON N2L 3G1, Canada}
\affiliation{Waterloo Centre for Astrophysics, Waterloo, ON N2L 3G1, Canada}

\author{Alison L. Coil}
\affiliation{Department of Astronomy and Astrophysics, University of California, San Diego, La Jolla, CA 92093, USA}

\author{B.R. McNamara}
\affiliation{Department of Physics and Astronomy, University of Waterloo, Waterloo, ON N2L 3G1, Canada}
\affiliation{Waterloo Centre for Astrophysics, Waterloo, ON N2L 3G1, Canada}

\author{Serena Perrotta}
\affiliation{Department of Astronomy and Astrophysics, University of California, San Diego, La Jolla, CA 92093, USA}

\author{Fabrizio Brighenti}
\affiliation{Dipartimento di Fisica e Astronomia, Universitá di Bologna, Via Gobetti 93/2, 40122, Bologna, Italy}

\author{H.R. Russell}
\affiliation{School of Physics \& Astronomy, University of Nottingham, University Park, Nottingham NG7 2RD, UK}

\author{Muzi Li}
\affiliation{Department of Physics and Astronomy, University of Waterloo, Waterloo, ON N2L 3G1, Canada}
\affiliation{Waterloo Centre for Astrophysics, Waterloo, ON N2L 3G1, Canada}

\author{S. Peng Oh}
\affiliation{Department of Physics, University of California, Santa Barbara, Santa Barbara, CA 93106, USA}

\author{Wenmeng Ning} 
\affiliation{Department of Astronomy and Astrophysics, University of California, San Diego, La Jolla, CA 92093, USA}
\affiliation{Department of Physics and Astronomy, University of California, Los Angeles, Los Angeles, CA 90095, USA}



\begin{abstract}

[OII] emission maps obtained with the Keck Cosmic Web Imager (KCWI) are presented for four galaxies centered in cooling X-ray cluster atmospheres.  Nebular emission extending tens of kpc is found in systems covering a broad range of atmospheric cooling rates, cluster masses, and dynamical states. Abell 262's central galaxy hosts a kpc-scale disk. The nebular gas in RXJ0820.9+0752 is offset and redshifted with respect to the central galaxy by $10-20$ kpc and 150 km s$^{-1}$, respectively.  The nebular gases in PKS 0745-191 and Abell 1835 are being churned to higher velocity dispersion by X-ray bubbles and jets. 
The churned gas is enveloped by larger scale, lower velocity dispersion (quiescent) nebular emission.  The mean line-of-sight speeds of the churned gas, quiescent gas, and the central galaxy each differ by up to $\sim 150$ km s$^{-1}$; nebular speeds upward of $800$ km s$^{-1}$ are found. Gases with outwardly-rising speeds upward of several hundred km s$^{-1}$ are consistent with being advected behind and being lifted by the rising bubbles.  The peculiar motion between the galaxy, nebular gas, and perhaps the hot atmosphere from which it presumably condensed is affecting the bubble dynamics, and may strongly affect thermally unstable cooling, the dispersal of jet energy, and the angular momentum of gas accreting onto the galaxies and their nuclear black holes.

\end{abstract}



%\section*{\textbf{********Changes*********}}
%\color{Green}
%\textbf{Things to do}
%\begin{itemize}
%    \item 
%\end{itemize}

%\color{Red}
%\textbf{About comments from referee}
%\begin{itemize}
%    \item I added errorbars to Fig. 14-17. I mention the errorbars in the caption of Fig. 14 and in the discussion.
%    \item I fixed Fig. 12 and fixed the discussion part about it.
%    \item I fixed Fig. 18 (position of central galaxy).
%    \item Made comment about the metallicity of the ELs being free.
%    \item Made Appendix 2 on the uncertainties. Fig. 11 and uncertainties comment taken care of.
%    \item Plot for v$_{02}$ and v$_{98}$.
%    \item Added VLA beam size to A262.
%    \item Made it clear that both components are real.
%    \item Clearly explained v02 and v98.
%\end{itemize}

%\color{black}
%\textbf{************************}

\section{Introduction} \label{sec:intro}

Clusters of galaxies host the largest and most luminous elliptical galaxies at their centres.  These galaxies are often referred to as Brightest Cluster Galaxies (BCGs) or central dominant galaxies. 
They also lie at the bases of hot X-ray atmospheres with luminosities of 10$^{43}-10^{45}$ erg s$^{-1}$ and temperatures between 10$^7$ to 10$^8$ K.   
Gas densities surrounding central galaxies are often upwards of $10^{-2}$ cm$^{-3}$.  At these densities atmospheres are expected to cool and condense into molecular clouds and form stars \citep{Fabian1994,Cavagnolo2008,Rafferty2008}. 
Star formation is observed in central galaxies, as in the systems studied here; however, the star formation rates are more than an order of magnitude lower than those expected by unimpeded cooling \citep{Fabian1994,Peterson2006}. This discrepancy between cooling models and the products of cooling suggests that one or more heating mechanisms are suppressing cooling.

Energetic feedback from active galactic nuclei (AGN) is a likely heating mechanism due to the prevalence of radio jets with sufficient power to compensate for radiative cooling losses \citep{McNamara2007,McNamara2012}. 
AGN-driven jets and radio lobes also displace $10^7$ to $10^{10}$ M$_{\odot}$ of hot gas surrounding them, carving X-ray cavities or bubbles that rise buoyantly through the atmosphere \citep{Birzan2004,Rafferty2006,Pope2010}. Several processes, including enthalpy released by rising bubbles \citep{McNamara2007}, turbulence \citep{Scannapieco2008,Gaspari2012}, shocks \citep{Heinz1998,David2001}, sound waves \citep{Fabian2003a,Ruszkowski2004a,Ruszkowski2004b,Voit2005,Nulsen2013} and cosmic rays \citep{Ruszkowski2008,Guo2008,Mathews2008}, contribute to the propagation and thermalization of energy from AGN feedback throughout hot atmospheres
\citep{Donahue2022}.

A spectrum of gas phases is observed surrounding central galaxies, including cold molecular gas with temperatures of a few tens of Kelvins \citep{Edge2001,SC2003}, warm molecular gas with temperatures less than 400 K \citep{Johnstone2007}, warm ionized nebular gas with $\sim10^4$ K temperatures, coronal gas at $10^6$ K \citep{Canning2011}, and hot $10^7-10^8$ K X-ray emitting gas. 
The discovery of bright nebular line emission surrounding galaxies and often extending tens of kpc into the bright X-ray atmospheres was the earliest evidence of atmospheric cooling \citep{Heckman1981,Hu1985,Heckman1989,Crawford1999,Hatch2007,McDonald2010,Fabian1977}. The brightest systems are associated with $10^9-10^{11}~\rm M_\odot$ of molecular gas, often associated with star formation at rates of tens of solar masses per year \citep{Rafferty2008,Fogarty2019,Russell2019,Olivares2019}.

Observations of gas cooling through all phases place strong constraints on thermally unstable cooling models.  Bright nebular emission and star formation in  central cluster galaxies and giant ellipticals are generally associated with systems with atmospheric cooling times lying below 1 Gyr.  Other cooling indicators include central entropy lying below 30 keV cm$^2$ \citep{Cavagnolo2008,Rafferty2008} and ratios of cooling time to free-fall time (t$_{\textrm{cool}}$/t$_{\textrm{ff}}$) below $\sim 35$. These thresholds are believed to be closely related to the onset of thermal instabilities \citep{Nulsen1986,Pizzolato2005,McCourt2012,Sharma2012,Gaspari2012,Voit2015,McNamara2016,Hogan2017,Pulido2018,Donahue2022}.
Thermally unstable cooling from hot atmospheres fuels star formation and could provide a continual fuel source to maintain the feedback cycle.  

In addition to heating the atmosphere, buoyantly rising radio bubbles and jets may enhance thermally unstable cooling by lifting cooling parcels of gas to higher altitudes \citep{Revaz2008,McNamara2016,Donahue2022}.
In this scenario, as X-ray cavities rise through the hot atmosphere, they entrain low entropy gas from the center of the galaxy to higher altitudes, mixing the gas and transporting high metallicity material outside the cluster core \citep{Churazov2001,Simionescu2009,Pope2010,Kirkpatrick2011,Gitti2011}. Uplift may accelerate thermally unstable cooling by increasing the infall time of the gas, lowering t$_{\textrm{cool}}$/t$_{\textrm{ff}}$ below the threshold required for thermal instabilities to occur. 

Observations of nebular emission are central to understanding the conditions under which thermally unstable cooling proceeds, the forms of ionization and heating, and the relationships between cooling inflows and AGN-driven outflows that regulate the growth of galaxies.  While nebular emission is sensitive to a negligible fraction of the cooling mass, its high surface brightness probes the full extent of the unstable cooling region as well as the gas dynamics with respect to the central galaxy and its radio jets and lobes.  
To investigate the behaviour of the warm nebular gas in and around the central galaxies and its connection to both multiphase gas and AGN feedback, we undertook new observations of four cooling cluster cores using the Keck Cosmic Web Imager \citep[KCWI]{Morrissey2018}. The integral field spectroscopy (IFS) capabilities of KCWI provide high sensitivity and high spectral and spatial resolution over a large field of view (FOV). From our observations, we obtain detailed maps of the fluxes and the kinematics of various emission lines, such as [OII]3726,9 \AA, [OIII]4959 \AA, [OIII]5007 \AA, H$_{\beta}$, H$_{\gamma}$, [NeIII]3869 \AA, and of the central galaxy's stellar population. Combined with previous studies and archival observations from the Atacama Large Millimeter/submillimeter Array (ALMA), Very Large Array (VLA) and Chandra X-ray Observatory, we study the dynamics of different phases of the intracluster medium (ICM) and their interactions with the central galaxy, radio jets and X-ray cavities.

In this paper we analyze IFS observations of the central regions of Abell 262, RXJ0820.9+0752, Abell 1835 and PKS 0745-191 focusing on the morphology and kinematics of the [OII]3726,9 \AA \ emission line doublet. The [OII]3726,9 \AA \ emission doublet is the brightest emission line in the KCWI data for these targets. Its high contrast against the declining U-band stellar continuum below the 4000~\AA~ break adds sensitivity to the imagery. The high sensitivity images presented here trace the nebular emission to larger radii than earlier studies. Its relatively high spectral and spatial resolutions permit detailed studies of nebular gas velocities with respect to stellar velocities in the central galaxy. Our targets are part of a pilot program on Keck chosen for their range of properties and their complex structures.

In Section~\ref{sec:Observations and data reduction} we provide a comprehensive account of our observations, data reduction techniques, and fitting methodology. We also discuss the complementary archival observations that are used throughout the paper and describe the analysis of the complementary data. In Section~\ref{sec:Results} we present flux and kinematic maps for each of our targets (Abell 262: Section~\ref{sec:Results:A262}, RXJ0820.9+0752: Section~\ref{sec:Results:RXJ0820}, Abell 1835: Section~\ref{sec:Results:A1835}, PKS 0745-191: Section~\ref{sec:Results:PKS0745}). In Section~\ref{sec:Discussion}, we discuss the implications of our results and compare them to AGN feedback models. Finally, in Section~\ref{sec:Summary}, we summarize the key takeaways. Throughout this paper we use a flat $\Lambda$CDM cosmology with H$_0 = 70$ km s$^{-1}$ Mpc$^{-1}$ and $\Omega_{\textrm{m}} = 0.3$.

\section{Observations and data reduction} \label{sec:Observations and data reduction}

\begin{deluxetable*}{ccccccccc}
\tabletypesize{\scriptsize}
\tablewidth{0pt} 
\tablecaption{KCWI Observations
\label{tab:Observation_parameters}}
\tablehead{ \colhead{Target} & \colhead{Date} & \colhead{Central wavelength} & \colhead{Blue filter} & \colhead{\# of pointings} & \colhead{Total observing time} & \colhead{Seeing} & \colhead{Total FOV} & \colhead{Total extent} \\ \colhead{} & \colhead{(DD/MM/YYYY)} &\colhead{(\AA)} &\colhead{} &\colhead{} &\colhead{(hrs)} &\colhead{(\arcsec)} &\colhead{($\arcsec \times \arcsec$)} &\colhead{($\textrm{kpc} \times \textrm{kpc}$)}}
\colnumbers
\startdata 
Abell 1835 & 19/03/2021 & 5300 & None & 6 & 2 & 0.8 & $35 \times 33$ & $141 \times 134$ \\
PKS 0745-191 & 12/02/2021 & 4650 & KBlue & 11 & 3.67 & 1.2 & $26 \times 31$ & $52 \times 61$ \\
Abell 262 & 14/01/2022 & 4300 & None & 10 & 2.5 & 0.9 & $30 \times 32$ & $10 \times 11$ \\
RXJ0820.9+0752 & 14/01/2022 & 4650 & None & 9 & 2.25 & 0.9 & $21 \times 24$ & $44 \times 50$
\enddata
\tablecomments{Columns: (1) Target name. (2) Date of KCWI observation. (3) Central wavelength. (4) Blue filter used. (5) Number of pointings. (6) Total observing time. (7) Seeing. (8) Total FOV obtained from the different pointings in arcseconds. (9) Same as previous column in units of kpc.
}
\end{deluxetable*}

\begin{splitdeluxetable*}{ccccccccccBcccccccc}
        \tablecaption{Targets properties \label{tab:target_properties}}
    \tabletypesize{\scriptsize}
    \tablewidth{0pt} 
    \tablehead{ \colhead{Target} & \colhead{R.A.} & \colhead{Dec.} & \colhead{z}& \colhead{R$_{2500}$} & \colhead{M$_{2500}$} & \colhead{t$_{\textrm{cool}}$}& \colhead{L$_{\text{x}}$} & \colhead{r(L$_{\text{x}}$)} & \colhead{L$_{\text{\textrm{H}}_{\alpha}}$}  & \colhead{L$_{\text{radio}}$} & \colhead{M$_{\text{mol}}$} & \colhead{M$_{*}$} &  \colhead{SFR$_{\text{IR}}$} &  \colhead{Cooling rate} &  \colhead{\# of X-ray cavities} & \colhead{$E_{cav}$}& \colhead{$P_{cav}$}
    \\ \colhead{} & \colhead{(J2000)} & \colhead{(J2000)} & \colhead{}& \colhead{(kpc)} & \colhead{($10^{13} M_{\odot}$)} & \colhead{(Gyr)}& \colhead{(erg s$^{-1}$)} & \colhead{(kpc)} & \colhead{(erg s$^{-1}$)} & \colhead{(erg s$^{-1}$)} &\colhead{(M$_{\odot}$)} &\colhead{(M$_{\odot}$)} &\colhead{(M$_{\odot}$ yr$^{-1}$)} &\colhead{(M$_{\odot}$ yr$^{-1}$)}& \colhead{} & \colhead{($10^{57}$erg)}& \colhead{($10^{43}$erg s$^{-1}$)}
    } 
\colnumbers
\startdata 
Abell 1835 & 14:01:02.1 & +02:52:43 & 0.2514 & 711 & $61.0_{-2.5}^{+2.5}$ & $0.39_{-0.03}^{+0.04}$ & $1.6\times10^{44}$ & 20 & $1.7\times10^{42}$ & $3.71\times10^{41}$ & $5\times10^{10}$ & $6\times10^{11}$ & $125\pm25$ & 1175 & 2 & $156^{+62}_{-43}$ & $165^{+102}_{-55}$\\
PKS 0745-191 & 07:47:31.3 & $-$19:17:40 & 0.1024 & 614 & $36.5_{-2.8}^{+2.9}$ & $0.35_{-0.02}^{+0.02}$ & $4\times10^{43}$ & 14 & $1.75\times10^{42}$ & $3.87\times10^{42}$ & $(4.6\pm0.3)\times10^{9}$ & $5\times10^{11}$ & 17.2 & $776\pm18$ & 2 & $69^{+29}_{-21}$ & $51_{-16}^{+25}$\\
Abell 262 & 01:52:46.5 & +36:09:07 & 0.0160 & 286 & $3.40_{-0.10}^{+0.10}$ & $0.37^{+0.02}_{-0.02}$ & $1.69\times10^{41}$ & 3.5  & $3.16\times10^{39}$ & $2.77\times10^{39}$ & $3.4\times10^{8}$ & $3.3\times10^{11}$ & 0.54 & $\sim 3$ & 2 & $2.5^{+1.0}_{-0.7}$ & $1.6^{+1.2}_{-0.6}$\\
RXJ0820.9+0752 & 08:21:02.3 & +07:51:47 & 0.1103 & 357 & $6.9_{-1.3}^{+1.4}$ & $0.42^{+0.04}_{-0.04}$ & $6.5\times10^{42}$ & 15 & $4.00\times10^{41}$ & $7.28\times10^{39}$ & $3.9\times10^{10}$ & $1.4\times10^{11}$ & 37 & $34\pm10$ & 1 & $1.1^{+0.6}_{-0.4}$ & $1.1^{+0.9}_{-0.5}$\\
\enddata
\tablecomments{ Columns: (1) Galaxy name. (2) Right ascension (RA). (3) Declination (Dec). (4) Redshift obtained from stellar absorption lines. Heliocentric correction has been applied. (5) $R_{2500}$, radius within which the mean density is 2500 times the critical density ($\rho_c$). (6) $M_{2500}$, where M$_{2500}=4\pi \bar{\rho}\textrm{R}_{2500}^3/3$ and $\bar{\rho}=2500\rho_c$. (7) Cooling time within $0.02R_{2500}$. (8) X-ray luminosity within the radius listed in column 9. (9) Radius chosen for the X-ray luminosity listed in the previous column such that it covers a similar extent as the detected [OII] emission. (10) H$_{\alpha}$ luminosity. (11) Radio luminosity between 10 MHz and 5 GHz. (12) Molecular gas mass. (13) Stellar mass. (14) Star formation rate from infrared luminosity. (15) Cooling rate obtained from X-ray (Chandra) observations. (16) Number of known X-ray cavities. (17) Total energy required to inflate known X-ray cavities. (18) Total AGN power required to inflate the cavities.\\
References: (5)-(9), (17)-(18): \citep{Hogan2017,Pulido2018}. (10): \citep[Abell 1835, Abell 262]{Crawford1999}, \citep[PKS 0745, RXJ0820.9]{Hamer2016}. (11): \citep{Pulido2018}. (12): \citep[Abell 1835]{McNamara2014}, \citep[PKS 0745]{Russell2016}, \citep[Abell 262]{Olivares2019}, \citep[RXJ0820.9]{Edge2001}. (13): \citep[Abell 1835, RXJ0820.9]{Olivares2019}, \citep[PKS 0745]{Donahue2011}, \citep[Abell 262]{Main2017}. (14) \citep[Abell 1835]{Egami2006}, \citep[PKS 0745, Abell 262, RXJ0820.9]{ODea2008}. (15) \citep[Abell 1835, PKS 0745, Abell 262]{McDonald2018}, \citep[RXJ0820.9]{Vantyghem2018}. (16): \citep[Abell 1835]{McNamara2006}, \citep[PKS 0745]{Sanders2014,Russell2019}, \citep[Abell 262]{Clarke2009}, \citep[RXJ0820.9]{Vantyghem2019}.
}
\end{splitdeluxetable*}

\subsection{Observations}

Our four cluster targets were observed with KCWI \citep{Morrissey2018}, which is installed on the Keck II telescope. For all targets we used the blue low-dispersion (BL) grating with the medium slicer, which results in a FOV of $16.5 \arcsec \times 20.4 \arcsec$ for each pointing. It has a spaxel scale of $0.29 \arcsec \times 0.69 \arcsec$, and a spectral resolution of $\rm R=1800$. We used a detector binning of $2 \times 2$ for all targets. Table~\ref{tab:Observation_parameters} lists additional observational parameters for each target.
The exposure time for each pointing is between $15-20$ minutes. The number of pointings and the total area covered for each object were chosen based on published observations of nebular gas extent and morphology, when available, to focus on regions of interest.  Abell 262 did not have a published nebular gas map, so we used an archival cold molecular gas distribution map instead. Previous observations of similar central galaxies have shown that detected nebular gas and molecular gas distributions often spatially overlap, with nebular emission being more extended than for the molecular gas \citep{Olivares2019,Russell2019}. 

\subsection{Data reduction}

We reduced the data using the KCWI Data Extraction and Reduction Pipeline (KDERP) in IDL. This pipeline applies various corrections, including bias subtraction, gain corrections and cosmic-ray rejections, and executes sky subtraction. We manually masked any source emission from our images to create blank sky images which we subtracted from our observations. In cases where the source covered the entire pointing such that obtaining a blank sky image was not possible, we used the sky image of a nearby pointing. The pipeline also corrects for differential refraction. To resample our data on a $0.29\arcsec \times 0.29 \arcsec$ spaxel grid and create a mosaic for each target, we used the IFSR\_KCWIRESAMPLE and IFSR\_MOSAIC routines from the IFSRED library \citep{Rupke2014a}. Table~\ref{tab:Observation_parameters} lists the resulting dimensions of each mosaic. As Abell 262 is at a much lower redshift than the other targets, a smaller central region is observed ($\sim$100 kpc$^2$), whereas for the other targets we observed areas at least 25 times larger. On the other hand, the low redshift of Abell 262 provides exquisite spatial resolution with spaxel sizes of less than $0.1~\rm kpc \times 0.1~kpc$.

\subsection{Stellar continuum and emission line fitting}

\label{sec:Observations and data reduction: Data analysis: Fitting}

From the reduced mosaics, we used the IFSFIT IDL library \citep{Rupke2014b} to fit the spectrum in each spaxel. This library uses the Penalized Pixel-Fitting method (PPXF) \citep{PPXF} and MPFIT \citep{MPFIT} to fit the stellar continuum and the emission lines, respectively. The fitting procedure followed five steps:

\begin{enumerate}
    \item \textbf{Creating a spatially-integrated spectrum:}\\
    To obtain a high quality spectrum of the stellar component of the BCG, we first summed the spectra in the spaxels with strong stellar continuum emission.
    \item \textbf{Spatially-integrated Stellar Population Synthesis (SPS) model:}\\
    Using IFSFIT on the spatially-integrated spectrum, we obtain a SPS best fit model for the stellar continuum at the very center of the galaxy.
    First the emission lines are masked from the spectrum, which is then fitted using the PPXF method of \cite{PPXF}. The fit is a linear combination of SPS models from \cite{SSP2005} and Legendre polynomials. 
    The SPS models are allowed to vary in age, but their metallicity is fixed to that of the sun. The SPS models are then summed over all stellar templates to obtain the stellar continuum model fit.
    A linear combination of Legendre polynomials are used to account for issues in data reduction such as scattered light, atmospheric effects, and/or residual features in the stellar continuum. We used Legendre polynomials of order 15.
    
    From this fit, we obtain the stellar velocity in the centermost region of the galaxy, which is then used as the systemic velocity of the system.

    \item \textbf{Fitting each spaxel:}\\
        For each spaxel, the emission lines are masked and the starlight spectrum is fitted, using the same method as in the previous step. 
        MPFIT \citep{MPFIT} is then used to fit the emission lines in the continuum-substracted spectrum. Before fitting, the model line profiles are convolved with the spectral resolution. Each emission line is fit with 1 or 2 Gaussian profiles, as needed. All emission lines are fit simultaneously with the same number of velocity components. The gaussian components of the different emission lines are all tied in redshift and linewidth to those of the [OIII]5007 emission line. The ratios of the different emission lines are free to vary, except for the ratio of [OII]3729/3726 which is fixed to 1.2, assuming an electron density of 400 cm$^{-3}$ \citep{Pradhan2006}.

  \item \textbf{Determining the number of emission line velocity components:}\\
  For each target, IFSFIT is run twice, allowing one or two Gaussian components for the emission lines. If the addition of a second Gaussian component does not improve the fit, a one component fit is returned. The two velocity component fits are used when: a) the flux for each component is positive and has a $\text{S/N} \geq 2$, b) both components have velocity dispersions of $\sigma < 500$ km s$^{-1}$, c) the flux ratio of the components is less than 10, and d) the addition of a second component significantly improves the fit. This last criterion is evaluated using the Bayesian Information Criterion (BIC):
        \begin{equation}           
        \text{BIC} = \chi^2+k\text{ln}(n)
        \end{equation}
  where $k$ is the number of parameters ($k = 3$ for 1 component and $k = 6$ for 2 components) and $n$ is the number of data points. A second component is used when the BIC of the two components fit is smaller than the BIC of the one component fit. Finally, we assume that spaxels requiring two Gaussian components should be spatially clustered together and not distributed randomly across the mosaic. Two component fits are used when neighbouring spaxels also use two component fits. Of the four galaxies observed, Abell 1835 and PKS 0745 have regions where the fits require a second component. Abell 262 and RXJ0820 can be modelled using only one kinematic component.

\end{enumerate}

\subsection{Archival observations}

Complimentary archival observations are used to compare the properties of the ionized gas in the ICM with the other gas phases, as well as to study the interactions between AGN feedback and gas dynamics. Table~\ref{tab:target_properties} lists various properties of our targets obtained from multi-wavelength observations.

Cold molecular gas maps showing CO emission were obtained from ALMA archival observations for all four targets. Table~\ref{tab:ALMA:info} gives the details of the ALMA observations. Maps of the CO fluxes were measured by integrating over the targeted CO emission lines. The resulting CO contours are consistent with results from previous papers \citep{McNamara2014,Russell2016,Vantyghem2017,Russell2019,Olivares2019,Vantyghem2019}.

Detected X-ray bubbles are also shown in the maps presented in this paper. As discussed in Section~\ref{sec:intro}, X-ray cavities are believed to play an important role in determining ICM dynamics. Combining previous cavity detections \citep{McNamara2006,Rafferty2006,Sanders2014,Vantyghem2019} with nebular gas flux and kinematics maps, we can study how cavities impact nebular gas dynamics as well as investigate whether different AGN feedback models are consistent with our observations. 

\begin{deluxetable*}{ccccccc}
        \tablecaption{ALMA observations}
    \tabletypesize{\scriptsize}
    \tablewidth{0pt} 
    \tablehead{ \colhead{Target}&\colhead{Date of observation}&\colhead{Observing time}&\colhead{CO line}&\colhead{$\nu_{\textrm{obs}}$}&\colhead{Beam size}&\colhead{Reference}\\ \colhead{} & \colhead{} & \colhead{(min)} & \colhead{} & \colhead{(GHz)} & \colhead{$\arcsec \times \arcsec$} & \colhead{}}
\label{tab:ALMA:info}
\startdata 
Abell 1835 & 2012-03-28 & 59.5 & J$ = 3-2$ & 276.190 & $0.60 \times 0.48$ & \cite{McNamara2014}\\
PKS 0745-191 & 2014-08-19 & 56.4 & J$ = 3-2$ & 313.562 & $0.27 \times 0.19$ & \cite{Russell2016}\\
Abell 262 & 2016-06-27 & 11.1 & J$ = 2-1$ & 226.863 & $0.95 \times 0.61$ & \cite{Olivares2019}\\
RXJ0820.9+0752 & 2016-10-01 & 22.7 & J$ = 3-2$ & 311.248 & $0.21 \times 0.165$ & \cite{Vantyghem2017}\\
\enddata
\end{deluxetable*}

\section{Results}\label{sec:Results}

Using KCWI integral field spectroscopy, we study the morphology and kinematics of the warm ionized gas in four cooling cluster cores.
While various emission lines are detected, including [OII], [NeIII], [OIII], and numerous hydrogen lines, the [OII]3726,9 \AA \ doublet is the most prominent line in our data and is therefore the focus of this paper. The analysis of other emission lines will be the focus of a future paper. For all flux and kinematics maps of the [OII]3726,9 \AA \ doublet, a $\textrm{S/N} \geq 5$ flux is applied for every spaxel. 

\begin{deluxetable*}{ccccc}\label{tab:OII_lum}
        \tablecaption{[OII]3726+[OII]3729 integrated luminosity}
    \tabletypesize{\scriptsize}
    \tablewidth{0pt} 
    \tablehead{ \colhead{Target}&\colhead{L$_{[OII]}$} & \colhead{$\textnormal{M}_{\textnormal{ion}}$} & \colhead{Area [OII]} & \colhead{Area CO}\\ \colhead{} & \colhead{erg s$^{-1}$} & \colhead{M$_{\odot}$} & \colhead{kpc$^2$} & \colhead{kpc$^2$}
    }
\colnumbers
\startdata 
Abell 1835&$4.0 \times 10^{42}$ & $1.1\times10^8$ & 1467 & 84 \\ %fixed
PKS 0745-191&$4.6 \times 10^{41}$ & $1.3\times10^7$ & 547 & 3.2 \\ %fixed
Abell 262&$1.4 \times 10^{40}$ & $4.2\times10^5$ & 24 & 1\\ %fixed
RXJ0820.9+0752&$1.8 \times 10^{41}$ & $5.2\times10^6$ & 266 & 27 \\ %fixed
\enddata
\tablecomments{Columns: (1) Target name. (2) Total [OII] luminosity observed with $\text{S/N} \geq 5$.
(3) Ionized gas mass inferred from total [OII] luminosity.
(4) Total area with [OII] emission where $\text{S/N} \geq 5$ for the [OII] flux.
(5) Total area with detected CO emission, using a flux cut of 10\% of the peak CO flux.
}
\end{deluxetable*}

\subsection{Abell 262}\label{sec:Results:A262}

The central galaxy in Abell 262 has the lowest redshift and the lowest mass in our sample (see Table~\ref{tab:target_properties}). Abell 262 is different than our other targets as it hosts a molecular circumnuclear disk with a radius of 1.5 kpc \citep{SC2003,Russell2019} embedded in a more extended disk of ionized gas \citep{Hatch2007}. Circumnuclear disks in the core of central galaxies are rare, but not unheard of \citep[HydraA]{Rose2019}.

The central galaxy in Abell 262 has well-known AGN jets oriented to the east and west, both with linear sizes of $\sim 11~\rm kpc$ by $\sim 4$ kpc \citep{Clarke2009}. Extended radio lobes and jet inflated X-ray cavities \citep{Blanton2004,Birzan2008,Clarke2009} are found between 8 to 29 kpc from the core of the central galaxy. Due to the low redshift of the source, our FOV is too small to include the X-ray cavities, the closest of which are $8-10$ kpc east and west of the nucleus.

\subsubsection{Abell 262: Morphology}

\begin{figure*}
    \centering
    \includegraphics[width=\textwidth]{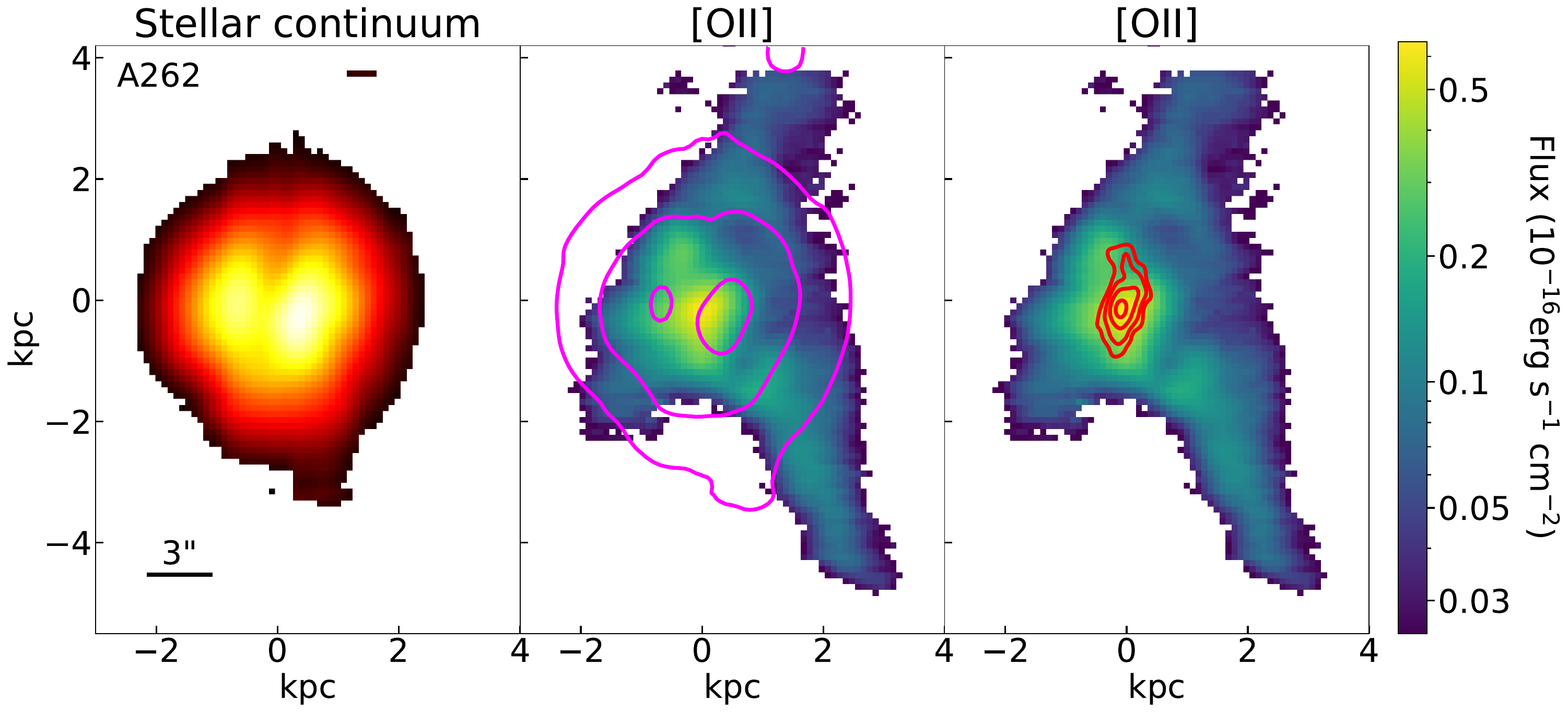}
    \caption{Left panel: Smooth stellar continuum map for Abell 262. The continuum flux colorbar ranges between $(15-75) \times 10^{-16}$ erg s$^{-1}$cm$^{-2}$. Middle and right panels: Continuum-subtracted flux maps of [OII]3726,9$\lambda\lambda$ for Abell 262. Magenta contours show the stellar flux at 20\%, 40\%, and 80\% of the maximum stellar flux. Red contours show the flux of CO(2-1) emission, where the contours represent 20\%, 40\%, and 80\% of the maximum CO(2-1) \citep{ALMA_A262}. The two closest X-ray cavities are located $8-10$ kpc east and west of the nucleus \citep{Clarke2009}, which lie beyond the region with detected nebular emission. The positions of these cavities are shown in Fig.~\ref{fig:sig_radio:A262}.
    }
    \label{fig:flux:A262}
\end{figure*}

Abell 262 has the lowest redshift in our sample such that the KCWI FOV only covers the inner part of the central galaxy. The KCWI spectra cover rest-frame wavelengths between $\rm 3445~\AA$ and $\rm 5118~\AA$. The left panel of Fig.~\ref{fig:flux:A262} shows the stellar flux map for Abell 262.

For our analysis the ``nucleus" of the central galaxy, which is the origin in the various flux and velocity maps, is taken to be the spaxel with the brightest stellar emission. However, the central galaxy in Abell 262 has a well-known dust lane, which can be seen in the left panel of Fig.~\ref{fig:flux:A262} cutting vertically through the center of the galaxy. Absorption from the central dust lane causes a decrease of the flux in the central region. Therefore, contrary to the other targets in our sample, the brightest stellar continuum spaxel does not provide an accurate position of the nucleus in this case. Instead, the location of the nucleus is defined by the spaxel with the largest stellar velocity dispersion. 

The stellar map shown in Fig.~\ref{fig:flux:A262} uses a threshold for the stellar flux per spaxel of $1.5 \times 10^{-15}$ erg s$^{-1}$ cm$^{-2}$. Overall, we detect a total stellar luminosity of $3.6 \times 10^{42}$ erg s$^{-1}$ in a roughly circular region with a radius of 2.5 kpc. Half of the total stellar flux is emitted within a radius of 1.43 kpc around the nucleus.

The middle and right panels of Fig.~\ref{fig:flux:A262} show the continuum-subtracted [OII] flux map, with S/N $ \geq 5$. Stellar emission and CO(2-1) emission are shown by magenta and red contours, respectively. While X-ray cavities have been detected in this system, they are not shown in Fig.~\ref{fig:flux:A262} as they lie beyond the FOV of our maps. The two closest X-ray cavities are roughly $8-10$ kpc from the nucleus along the E-W axis. 

Within our FOV we measure a total [OII] luminosity of $1.4 \times 10^{40}$ erg s$^{-1}$. From the [OII] luminosity, we can estimate the mass of ionized gas ($\textnormal{M}_{\textnormal{ion}}$) using:
\begin{equation}\label{eq:Mion}
    \textnormal{M}_{\textnormal{ion}} = \frac{\mu \textrm{m}_{\textrm{H}} \textrm{L}_{\textrm{[OII]}}}{\gamma_{\textrm{[OII]}} \textrm{n}_\textrm{e}}
\end{equation}
where $\mu$ is the mean mass per hydrogen atom which we set to $\mu=1.4$, m$_{\textrm{H}}$ is the mass of the hydrogen atom, n$_{\textrm{e}}$ is the electron density number and $\gamma_{\textrm{[OII]}}$ is the effective line emissivity \citep{Osterbrock2006}. Assuming an electron temperature of $\sim 10 000$ K and an electron density of $\sim 400$ cm$^{-3}$, we obtain an effective line emissivity for the [OII] doublet of $\sim 10^{-25}$ cm$^3$ erg s$^{-1}$ \citep{PyNeb}.
It corresponds to an ionized gas mass of $\sim 4 \times 10^5$ M$_{\odot}$ (see Table~\ref{tab:OII_lum}).

Half of the [OII] emission comes from a central region with a radius of 1.32 kpc. The detected [OII] emission spans 5.45 kpc in RA and 8.58 kpc in Dec. The [OII] emission is highly spatially asymmetric. Fig.~\ref{fig:flux:A262} shows prominent filaments extending to the SW and NW of the nucleus, with another small filament to the SE. The nebular gas is more extended along the north-south axis, which is perpendicular to the axis of the cavities. 

The middle panel of Fig.~\ref{fig:flux:A262} shows that most (64\%) of the area with detected [OII] emission also has detected stellar emission. The stellar continuum and the [OII] emission have different morphologies. The asymmetry of the nebular gas is not present in the stellar continuum map. The upper half of the NW filament and the bottom half of the SW filament are the main regions where nebular gas is detected without a stellar counterpart. Since the regions with the brightest stellar continuum and brightest [OII] emission overlap, only 20\% of the [OII] flux is emitted from regions without detectable starlight.

The right panel of Fig.~\ref{fig:flux:A262} superimposes CO(2-1) contours, in red, onto the [OII] flux map \citep{ALMA_A262}. Most of the detected molecular gas ($> 70\%$) lies very close to the nucleus, in a roughly elliptical region with major and minor axes of $\sim 2$ kpc and $\sim 1$ kpc, respectively. Previous studies of the dynamics of the molecular gas in this system concluded that it is consistent with a rotating circumnuclear disk \citep{Russell2019,Olivares2019}. 

The spatial distributions of molecular gas and nebular gas are consistent with each other. CO(2-1) emission is detected where the nebular gas is the brightest. As detailed in Table~\ref{tab:OII_lum}, [OII] emission is much more extended than CO(2-1) emission, spanning a projected area 24 times larger. The morphologies of the nebular gas and the cold gas in this system are consistent with models where molecular gas cools out of its warmer surroundings.

\subsubsection{Abell 262: Kinematics}

\begin{figure*}
    \centering
    \includegraphics[width=\textwidth]{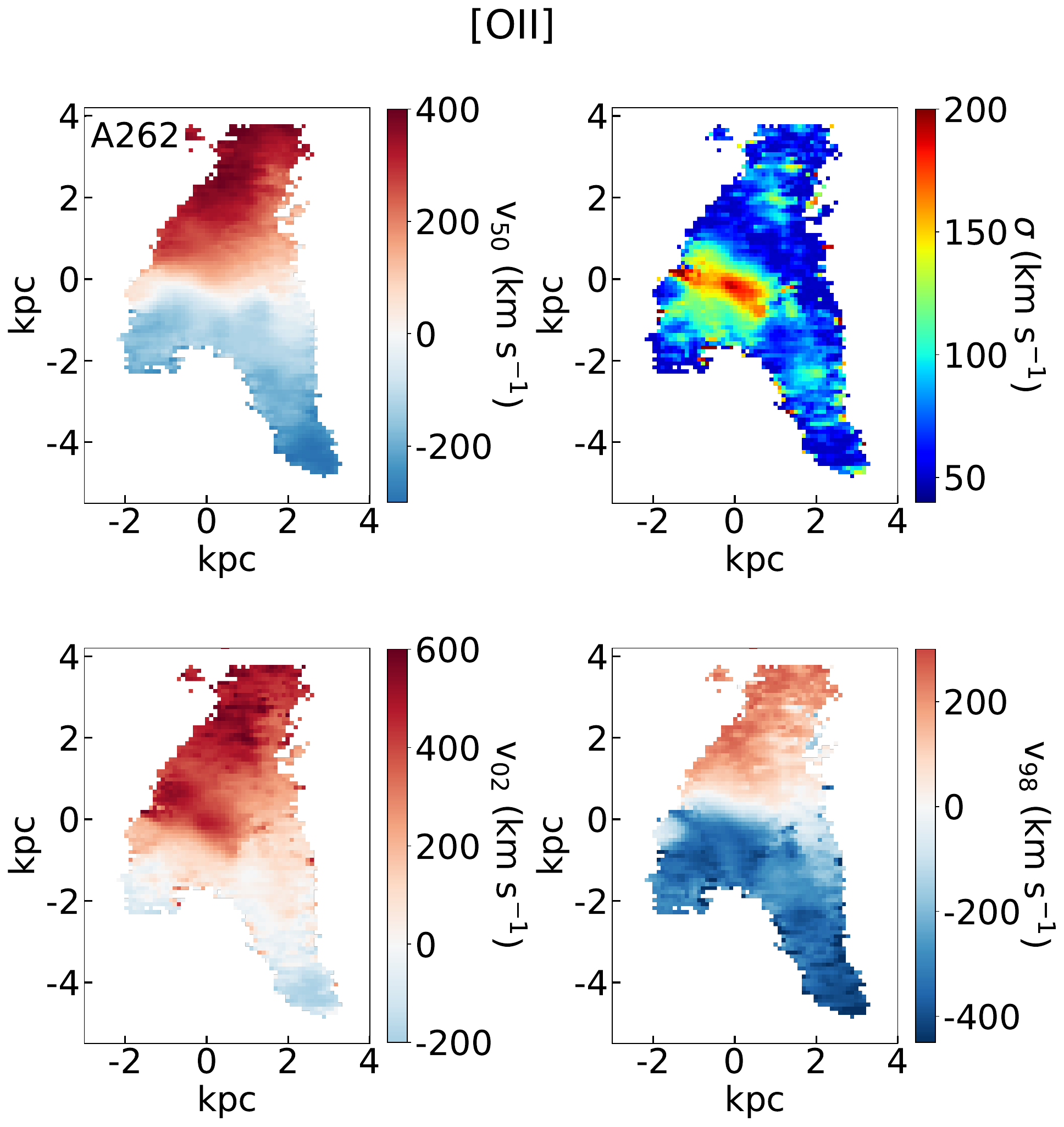}
    \caption{
    Line-of-sight kinematics of the [OII]3726,9$\lambda\lambda$ emission line doublet in Abell 262. Upper left panel: Median velocity ($v_{50}$). Upper right panel: Velocity dispersion ($\sigma$). Lower left panel: Maximum redshifted velocity ($v_{02}$), defined as the velocity which includes 2\% of the cumulative velocity distribution. Lower right panel: Maximum blueshifted velocity ($v_{98}$), defined as the velocity which includes 98\% of the cumulative velocity distribution. 
    }
    \label{fig:kinematics:A262}
\end{figure*}

\begin{figure}
    \centering
    \includegraphics[width=\columnwidth]{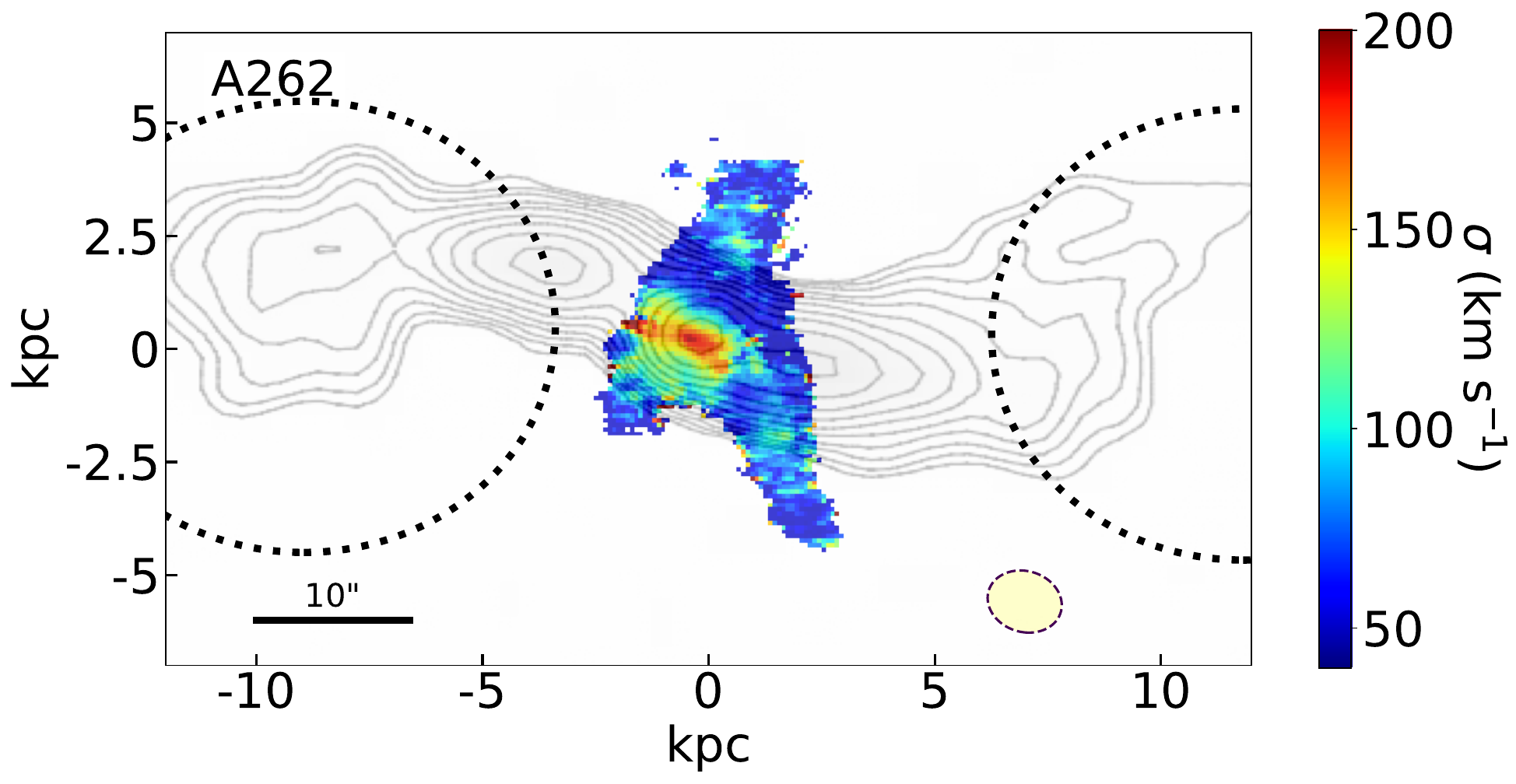}
    \caption{Velocity dispersion map of the [OII] nebular gas as in the top right panel of Fig.~\ref{fig:kinematics:A262}. Contours of the flux density at 1.4 GHz are overlaid as solid grey lines on the $\sigma$ map (taken from Figure 2 of \cite{Clarke2009}). The lowest flux density contour is at 3$\sigma$, where $\sigma=0.11$ mJy/beam, increasing by $\sqrt{2}$ for other contours. The beam size of the 1.4 GHz VLA observations is shown in the lower right by a black dashed ellipse}. The black dotted ellipses show the position of the X-ray cavities \citep{Blanton2004,Clarke2009}.
    \label{fig:sig_radio:A262}
\end{figure}

Fig.~\ref{fig:kinematics:A262} presents the line-of-sight (LOS) kinematics of the [OII] emission at each spaxel in the innermost region of the central galaxy of Abell 262. The panels shown are the median velocity (upper left), the velocity dispersion (upper right), the maximum redshifted velocity, $v_{02}$, (lower left), and the maximum blueshifted velocity, $v_{98}$, (lower right). Positive velocities are redshifted with respect to the systemic redshift, and negative velocities are blueshifted. The maximum redshifted (blueshifted) velocities are obtained by determining the velocity at 2\% (98\%) of the cumulative velocity distribution; these values probe the most extreme velocities of the ionized gas. The definitions of v$_{02}$ and v$_{98}$ are explained more in depth in Appendix~\ref{appendix:v02_v98}.
\color{black}
As discussed in Section~\ref{sec:Observations and data reduction: Data analysis: Fitting}, we attempted to fit the emission lines using both the one kinematic component model and the two kinematic component model. For this system, the emission lines are fitted using only one kinematic component. While the spectral fits hint to the possibility of a second kinematic component, especially for the more disturbed gas, we cannot robustly detect it using our current data. Additional observations are needed to further study the possibility of multiple kinematic components in this nebula.

To assess overall velocities over multiple spaxels, we define [OII] flux-weighted mean velocities as:
\begin{equation}\label{eq:flux-weighted_mean}
    \bar{\textrm{v}}=\frac{\sum_{\textrm{i}} \textrm{v}_{\textrm{i}} \times \textrm{f}^{\textrm{[OII]}}_{\textrm{i}}}{\sum_{\textrm{i}} \textrm{f}^{\textrm{[OII]}}_{\textrm{i}}}
\end{equation}
where v is velocity (e.g. v$_{50}$, $\sigma$, v$_{02}$ or v$_{98}$), and $\textrm{f}^{\textrm{[OII]}}_{\textrm{i}}$ is the [OII] flux within the i{\it th} spaxel of any given region. We use the flux-weighted mean velocity to obtain an average velocity within a spatial region as if analyzing it as a single spaxel.

The high spectral resolution of our data allows us to measure both nebular emission lines and absorption lines from the starlight of the central galaxy. The systemic redshift of the stars surrounding the nucleus is used to determine the zero point of the velocity plots, as discussed in Section~\ref{sec:Observations and data reduction: Data analysis: Fitting}. We observe a discrepancy between the systemic velocity and the central velocity of the nebular gas. Within a radius of $1 \arcsec$ around the nucleus, the nebular gas has a mean [OII] flux-weighted v$_{50}$ of $+105$ km s$^{-1}$, with median velocities in each spaxel ranging between $+10$ km s$^{-1}$ and $+172$ km s$^{-1}$. However, the [OII] flux-weighted v$_{50}$ across our entire FOV is closer to the systemic velocity, with a relative velocity of only $+39$ km s$^{-1}$.

The v$_{50}$, v$_{02}$ and v$_{98}$ maps (Fig.~\ref{fig:kinematics:A262}) show a clear gradient in velocities along the north-south axis. The nebular gas is redshifted in the north and blueshifted in the south, indicating ordered motion in a nascent disk.
This is consistent with the molecular gas dynamics previously observed by \cite{Russell2019}. Median velocity values range from $-314$ km s$^{-1}$ to $+422$ km s$^{-1}$. In addition to the main velocity gradient along the N-S axis, the northern ionized gas filament has an east-west gradient of $\sim 150$ km s$^{-1}$.

The top right panel of Fig.~\ref{fig:kinematics:A262} shows the LOS velocity dispersion of the ionized gas. The majority of the gas is fairly quiescent, with more than 50\% of the [OII] flux being emitted from gas with $\sigma < 100$ km s$^{-1}$. There is a high $\sigma$ region roughly centered on the nucleus and extending $\sim 25^{\circ}$ north of east. The gas is highly disturbed, with velocity dispersions up to 200 km s$^{-1}$. This region, where $\sigma \geq 140$ km s$^{-1}$, has a length of $\sim 2.5$ kpc and a width of $\sim 0.64$ kpc, emitting roughly one fifth of the total [OII] flux. The velocity dispersion gradually decreases away from the high $\sigma$ region. Most of the remaining gas is quiescent, with few slightly more disturbed knots of gas with $\sigma$ between 80 km s$^{-1}$ and 125 km s$^{-1}$. 

Fig.~\ref{fig:sig_radio:A262} shows VLA 1.4 GHz radio emission contours \citep{Clarke2009} superimposed onto the [OII] LOS velocity dispersion map. The dotted ellipses indicate the position of two detected X-ray cavities. Fig.~\ref{fig:sig_radio:A262} shows that the central high $\sigma$ region roughly extends along the axis of the radio jets, with the highest $\sigma$ nebular gas spatially coinciding with the brightest 1.4 GHz emission. This is consistent with the high $\sigma$ gas being disturbed by powerful AGN-driven jets. \color{black} Fig.~\ref{fig:sig_radio:A262} shows that the maximum extent of the 1.4 GHz radio emission coincides with the position of the X-ray cavities. This is also consistent with mechanical AGN feedback models where AGN-driven jets inflate X-ray cavities which rise buoyantly in the surrounding atmosphere.

\color{black}

\subsection{RXJ0820.9+0752}\label{sec:Results:RXJ0820}

This system is known for the significant offset between its central galaxy and its molecular and nebular gas reservoirs, which lie nearly 20 kpc NW of the galaxy's core \citep{Bayer-Kim2002, Salome&Combes2004, Vantyghem2017, Olivares2019}. \cite{Vantyghem2019} detected a possible X-ray cavity, which has yet to be confirmed from radio observations, located 6.4 kpc to the west of the nucleus (see Fig.~\ref{fig:flux:RXJ0820}, middle and right panels). With an estimated power of $1.14 \times 10^{43}$ erg s$^{-1}$, its cavity power is insufficient to produce the galaxy's massive cold gas reservoir solely through stimulated cooling of uplifted gas \citep{Vantyghem2019}. The $10^{10}$ M$_{\odot}$ molecular gas reservoir (see Table~\ref{tab:target_properties}) likely results from sloshing caused by galaxy interactions. Based on the presence of a secondary galaxy $\sim 8$ kpc away from the main galaxy and the significant offset of the molecular and nebular gas reservoirs, galaxy interactions are believed to play a crucial role in the dynamics of multiphase gas in the core of RXJ0820.9+0752.

\subsubsection{RXJ0820.9+0752: Morphology}\label{sec:RXJ0820:Results:Morphology}

\begin{figure*}
    \centering
    \includegraphics[width=\textwidth]{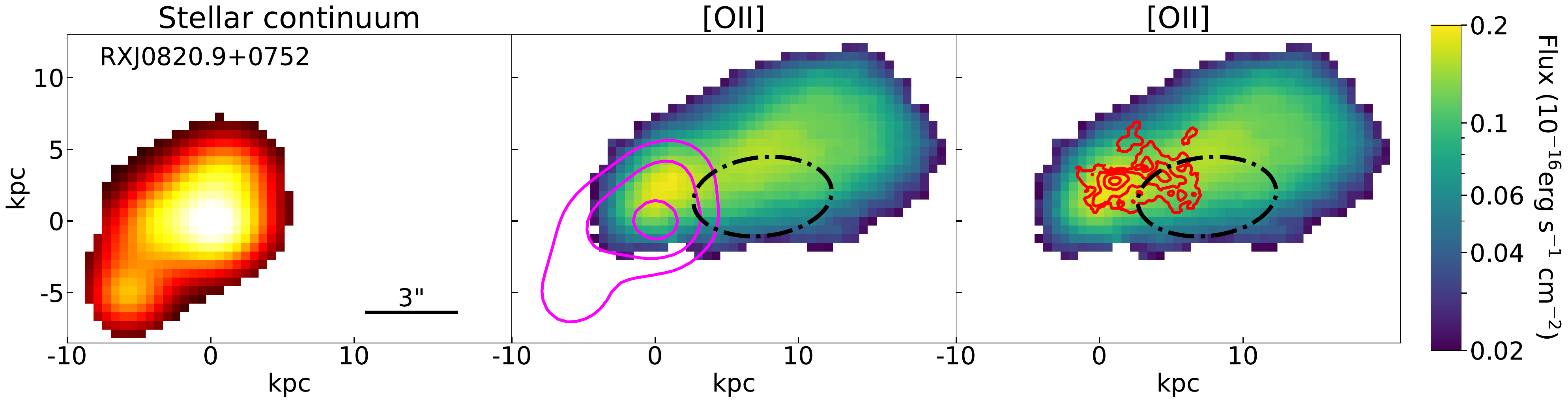}
    \caption{Left panel: Smooth stellar continuum map for RXJ0820.9+0752. The flux colorbar ranges between $(5-50) \times 10^{-16}$ erg s$^{-1}$cm$^{-2}$. Middle and right panels: Continuum-subtracted flux maps of [OII]3726,9$\lambda\lambda$. The magenta lines show stellar flux contours for 20\%, 40\%, and 80\% of the peak stellar flux. The red contours show the CO(3-2) flux, where the contours represent 10\%, 20\%, 40\%, and 80\% of the maximum CO(3-2) flux \citep{Vantyghem2017}. The black dash-dotted ellipse shows the position of the putative X-ray cavity \citep{Vantyghem2019}.}
    \label{fig:flux:RXJ0820}
\end{figure*}

The left panel of Fig.~\ref{fig:flux:RXJ0820} shows the stellar continuum map between $3300-5100$~\AA~. This map only includes spaxels with observed continuum flux of at least $6 \times 10^{-16}$ erg s$^{-1}$ cm$^{-2}$. The continuum emission spans $\sim 14$ kpc in RA and Dec, roughly in the shape of an ellipse with major and minor axes of 17 kpc and 11 kpc, respectively. The total stellar luminosity is $2.3 \times 10^{43}$ erg s$^{-1}$. Half of the continuum flux is emitted within a central circular region with a radius of 3.6 kpc.

This system has a secondary galaxy projected $\sim$7.8 kpc to the SE of the central galaxy. It is seen in Fig.~\ref{fig:flux:RXJ0820} as a secondary peak in the stellar continuum map. The stellar flux of the central galaxy is $\sim 3$ times greater than that of the secondary galaxy, with stellar luminosities within 5 kpc of their respective peaks of $1.7 \times 10^{43}$ erg s$^{-1}$ and $6.0 \times 10^{42}$ erg s$^{-1}$. The two galaxies have a relative velocity of almost 100 km s$^{-1}$, with the secondary galaxy being more redshifted. The proximity of the galaxies and their relative velocity indicate that they are likely interacting.

The middle and right panels of Fig.~\ref{fig:flux:RXJ0820} show the [OII] continuum-subtracted flux maps. Stellar continuum contours are overlaid onto the [OII] map in magenta for the middle panel, highlighting the spatial offset of the nebula. Similarly, the CO(3-2) emission, seen as red contours in the rightmost panel of Fig.~\ref{fig:flux:RXJ0820}, is significantly offset from the stellar peak.

The central galaxy has a total [OII] luminosity of $1.82 \times 10^{41}$ erg s$^{-1}$, half of which is emitted within a radius of 7.3 kpc from the [OII] peak. As listed in Table~\ref{tab:OII_lum}, an ionized gas mass of $\sim 5 \times 10^6$ M$_{\odot}$ can be inferred from the total [OII] luminosity, assuming T$_{\textrm{e}} = 10 000$ K and n$_{\textrm{e}} = 400$ cm$^{-3}$.

The [OII] flux map shows extended nebular gas spanning roughly 24 kpc in RA and 15 kpc in Dec. The nebular gas distribution is asymmetrical about the peak in [OII] emission, extending mostly to the NW of the peak. Contrary to the other systems presented here, the nebular gas extends beyond the X-ray cavity. 
The nebular gas is extended roughly along the major axis of the cavity.

The middle panel of Fig.~\ref{fig:flux:RXJ0820} demonstrates the significant morphological differences between the stellar continuum and the nebular emission. Less than 30\% of the area with detected [OII] emission also has detected stellar emission.
Although most of the nebular gas is offset from the stellar continuum, the [OII] flux peak is fairly close to the stellar peak, located $\sim 3.3$ kpc NNW of the nucleus.

For the molecular gas, most of the CO(3-2) emission lies in a region spanning $\sim 8.7$ kpc in RA and $\sim 6.3$ kpc in Dec. The molecular gas reservoir, which is not centered on the nucleus, mostly consists of a single filament, with two denser gas clumps. The brighter clump is located $\sim 3$ kpc NNW of the nucleus, while the fainter clump is $\sim 3.5$ kpc east of the brighter clump. The CO(3-2) emission is slightly east of the X-ray cavity, partially overlapping with it. 

As observed in our previous targets, the morphology of the CO(3-2) emission is quite similar to that of the nebular gas, although spanning a much smaller area. As detailed in Table~\ref{tab:OII_lum}, the projected area of the [OII] nebula is ten times greater than for the CO(3-2) emission. This central galaxy, which is known for its extensive molecular gas reservoir, has the smallest [OII] detection area to CO detection area ratio of the four objects studied in this paper.

\subsubsection{RXJ0820.9+0752: Kinematics}\label{sec:RXJ0820:Results:kinematics}

\begin{figure*}
    \centering
    \includegraphics[width=\textwidth]{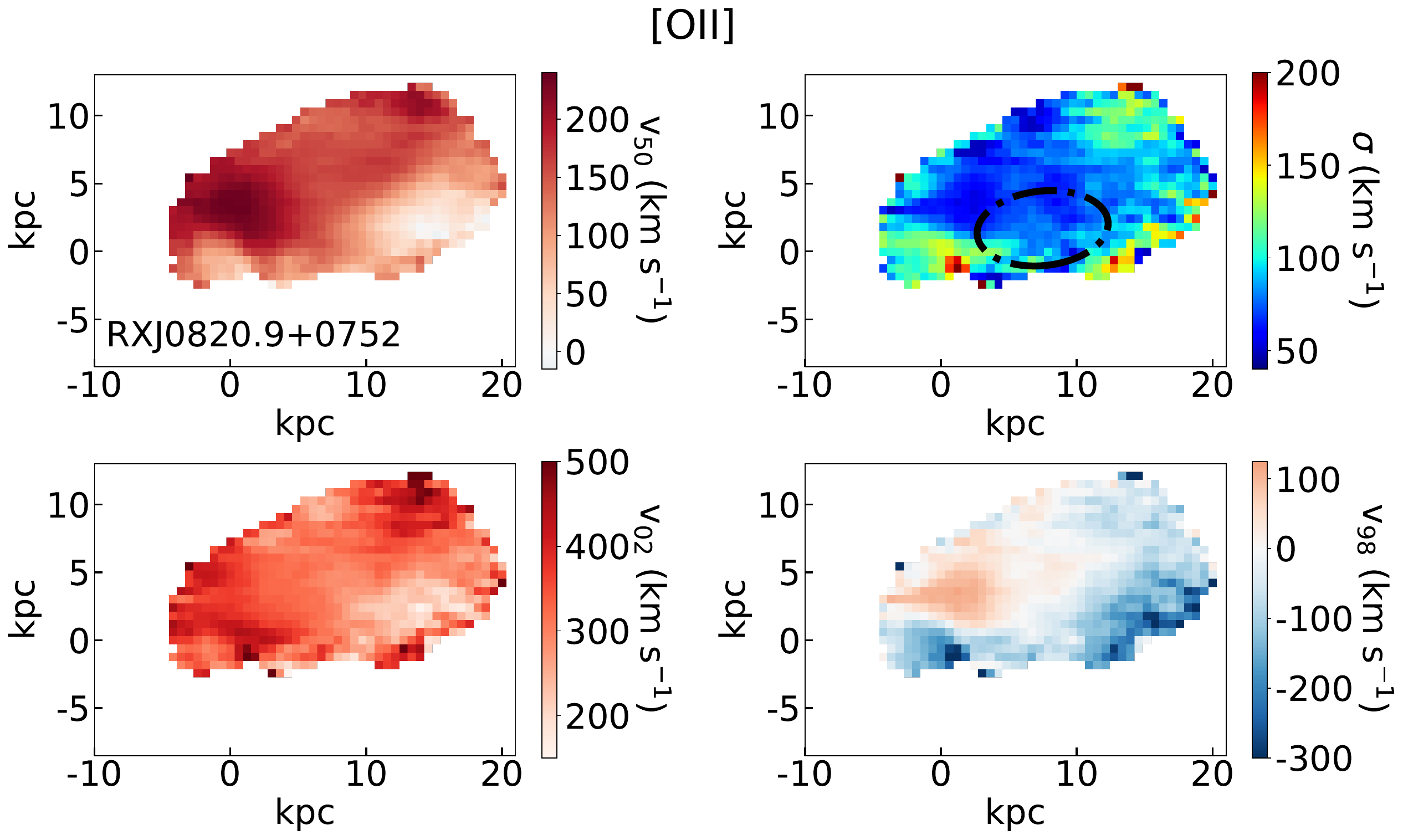}
    \caption{Kinematic maps of the [OII]3726,9$\lambda\lambda$ line for RXJ0820.9+0752. This figure is analogous to Fig.~\ref{fig:kinematics:A262}. In the upper right panel the dash-dotted ellipse shows the position of the tentative X-ray cavity.}
    \label{fig:kinematics:RXJ0820}
\end{figure*}

Fig.~\ref{fig:kinematics:RXJ0820} displays the LOS kinematic maps of the nebular gas in the central galaxy of RXJ0820.9+0752, which has a redshift of $z=$0.1103 (see Table~\ref{tab:target_properties}). As shown in the top left panel of Fig.~\ref{fig:kinematics:RXJ0820}, the nebular gas is substantially redshifted with respect to the systemic velocity. 
The [OII] median velocities range between $-12$ km s$^{-1}$ and $+286$ km s$^{-1}$, with a flux weighted mean v$_{50}$ of $\sim 150$ km s$^{-1}$. 

The v$_{50}$ map shows interesting dynamics for the gas with large v$_{50}$ ($\geq +150$ km s$^{-1}$). The two regions with largest v$_{50}$, located $\sim3.7$ kpc north and $\sim 16.9$ kpc north-west of the nucleus, are connected by a slightly lower velocity filament ($+146$ km s$^{-1} \leq$ v$_{50} \leq +174$ km s$^{-1}$). The gas with lowest v$_{50}$ is located $\sim 14$ kpc west of the nucleus.

The top right panel of Fig.~\ref{fig:kinematics:RXJ0820} shows that the gas is generally undisturbed. Almost 80\% of the nebular gas has $\sigma$ less than 100 km s$^{-1}$. The most disturbed gas lies in a small region located $\sim 1.6$ kpc south-west of the nucleus. The few regions where $\sigma \geq 100$ km s$^{-1}$ lie on the edge of the detected [OII] nebula. Appendix~\ref{appendix:RXJ0820} describes more in depth the properties of five regions of interest identified from the kinematic maps.

The bottom panels of Fig.~\ref{fig:kinematics:RXJ0820}, which show the most-redshifted and most-blueshifted velocities, do not reveal extreme velocities in this nebula. 
These velocity maps, as well as the $\sigma$ map,
support a quiescent scenario for the nebular gas, even though it is substantially offset from the center of the main galaxy.

\subsection{Abell 1835}\label{sec:Results:A1835}

Abell 1835 is an archetype cool core cluster whose central galaxy is experiencing powerful radio-mechanical (radio-mode) feedback. Its powerful mechanical AGN feedback, with power greater than $1.5\times10^{45}$ erg s$^{-1}$, is capable of offsetting the system's impressive 1175 M$_{\odot}$/yr of cooling \citep{McDonald2018} (see Table~\ref{tab:target_properties}). Despite the formidable jet power released over the past several tens of Myr, the current radio source is relatively faint and amorphous, with a flux density of 2.3 mJy at 4.76 GHz \citep{ODea2010}. \cite{ODea2010} detected a 1.5 arcsecond jet in the core of the central galaxy. Starting from the nucleus, the jet initially points to the west and then gradually bends in the NW direction.

\subsubsection{Abell 1835: Morphology}\label{sec:A1835:Morphology}

\begin{figure*}
    \centering
    \includegraphics[width=\textwidth]{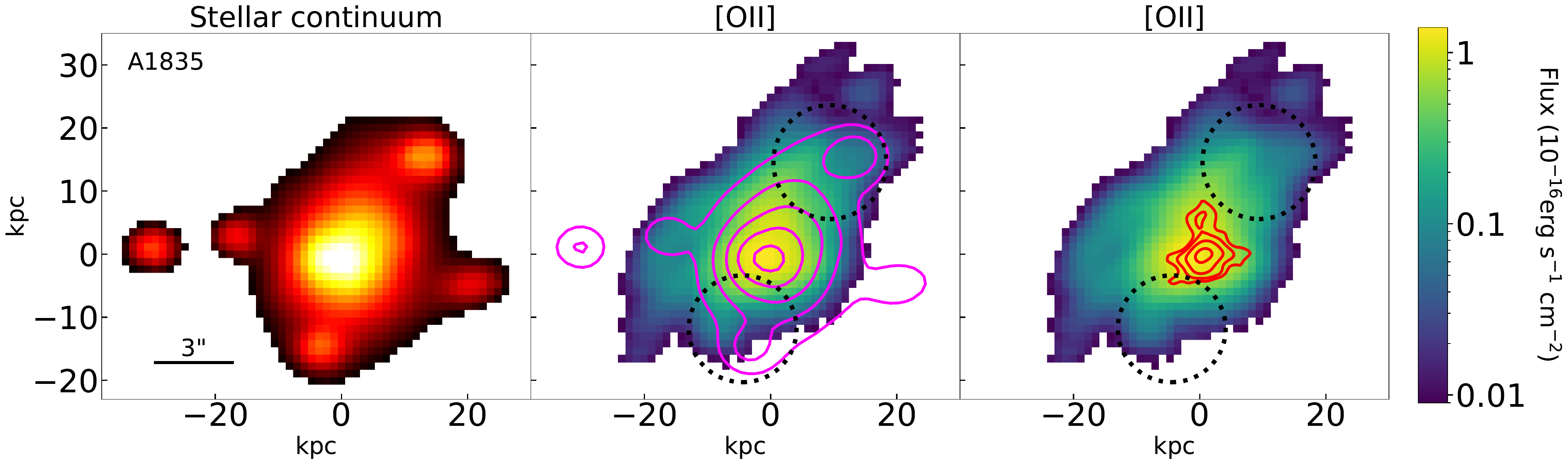}
    \caption{Left panel: Stellar continuum flux map for Abell 1835. The continuum flux colorbar ranges between $(2.5-80) \times 10^{-16}$ erg s$^{-1}$cm$^{-2}$. Middle and right panels: Continuum subtracted flux maps of [OII]3726,9$\lambda\lambda$ for Abell 1835, where $\text{S/N} \geq 5$. The magenta line shows stellar continuum flux contours for 5\%, 10\%, 20\%, 40\%, and 80\% of the maximum stellar continuum flux. The red contours indicate the flux of CO(3-2) emission, where the contours represent 10\%, 20\%, 40\%, and 80\% of the maximum CO(3-2) flux \citep{McNamara2014}. The black dotted ellipses show the position of the X-ray cavities \citep{McNamara2006,Olivares2019,Russell2019}. }
    \label{fig:flux:A1835}
\end{figure*}

The left panel of Fig.~\ref{fig:flux:A1835} shows the observed KCWI stellar continuum integrated over the rest-frame wavelength range of $3435-5035$~\AA. The central galaxy is flanked in projection by several fainter galaxies. Eight neighboring galaxies are observed within our full KCWI FOV. Apart from one galaxy, the projected galaxies' stellar velocities relative to the central galaxy differ by thousands of km s$^{-1}$. These high relative velocities are inconsistent with interactions with the central galaxy. Only one projected galaxy has a similar velocity as the central galaxy, although their projected separation of more than 60 kpc makes interactions between the two galaxies very unlikely. Therefore, we do not expect the projected galaxies to be interacting with the central galaxy or the surrounding gas. 

Above a flux of $2.5 \times 10^{-16}$ erg s$^{-1}$cm$^{-2}$ per spaxel, the stellar continuum is detected up to a radius of $\sim 20$ kpc from the nucleus of the central galaxy. Half of the total stellar flux is contained within a radius of $\sim 9$ kpc ($2.2 \arcsec$).

The middle and right panels of Fig.~\ref{fig:flux:A1835} show the detected [OII] flux in and around the central galaxy.
Half of the [OII] flux is emitted within a radius of 7.3 kpc (1.8\arcsec) from the nucleus of the central galaxy. The [OII] detected spans more than 45 kpc in RA and 50 kpc in Dec. The total [OII] luminosity is $4 \times 10^{42}$ erg s$^{-1}$; this is the most [OII] luminous object in our sample (see Table~\ref{tab:OII_lum}). We estimate more than $10^8$ M$_{\odot}$ of nebular gas in the inner region of Abell 1835 (see Table~\ref{tab:OII_lum}).

Fig.~\ref{fig:flux:A1835} shows that the peak of the stellar continuum, [OII] flux, and CO(3-2) flux all coincide. The stellar continuum and the [OII] emission overlap spatially. Approximately 90\% of the stellar continuum flux coincides with 96\% of the [OII] flux. Little nebular gas is detected outside of the region with detectable starlight.

Most of the observed molecular gas in Abell 1835 is concentrated around the nucleus, with the remaining molecular gas extending in roughly three filaments \citep{McNamara2014}, a prominent northern filament, and smaller eastern and western filaments.
The N and W filaments appear to be trailing the NW X-ray cavity, while the E filament seems to be trailing the SE cavity. The distribution of nebular gas generally follows the distribution of cold molecular gas but the nebular gas is far more extended, with a detected area 17 times greater than that of the detected cold gas (see Table~\ref{tab:OII_lum}). The [OII] morphology observed in the core of Abell 1835 provides compelling evidence for the impact of X-ray cavities on gas dynamics. The right panel of Fig.~\ref{fig:flux:A1835} shows that the nebular gas is extended along the direction of the two X-ray cavities. The detected [OII] emission spans 53 kpc along the axis of the bubbles, compared to 32 kpc in the perpendicular direction.

\subsubsection{Abell 1835: Kinematics}\label{sec:A1835:Kinematics}

\begin{figure*}
    \centering
    \includegraphics[width=\textwidth]{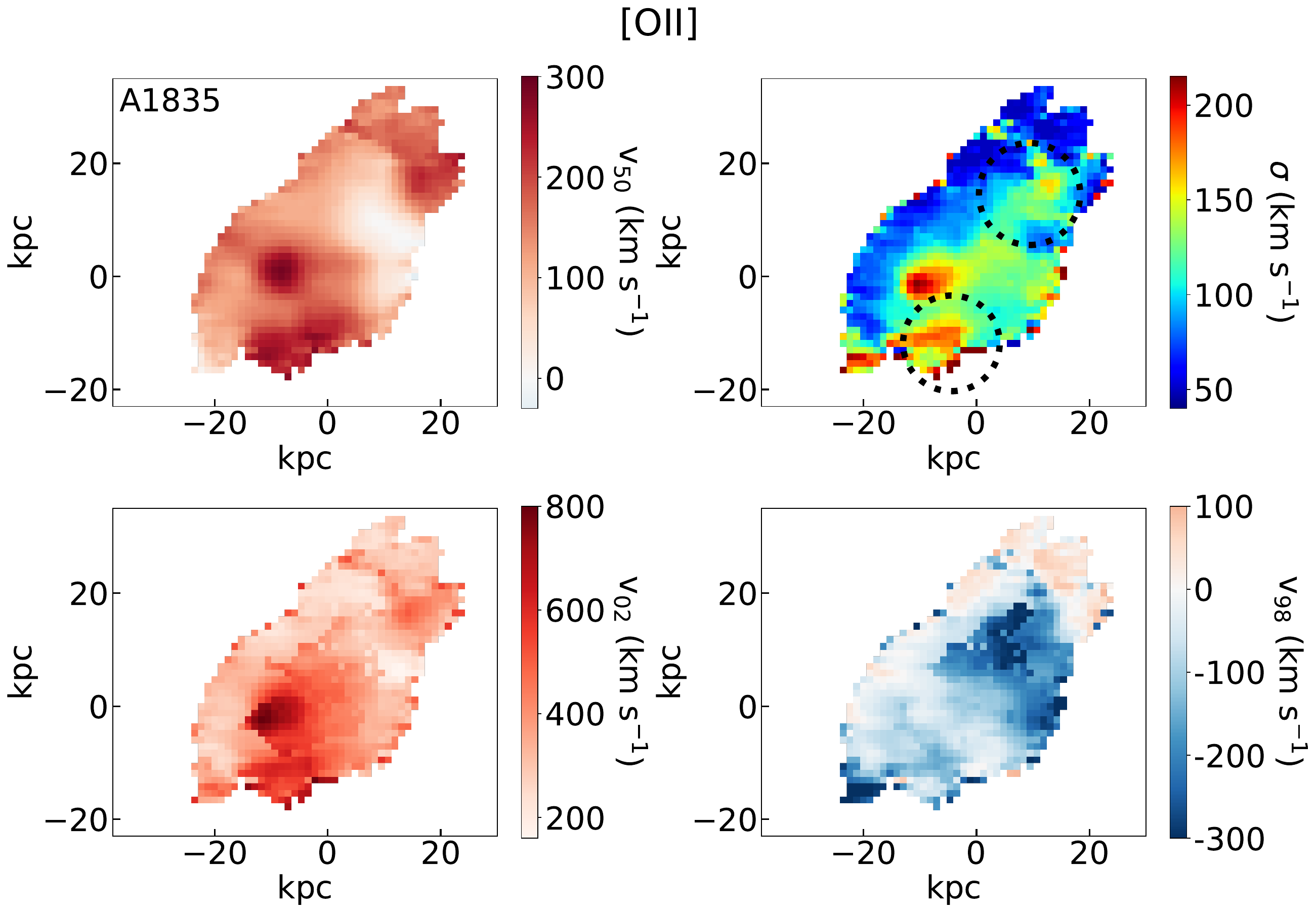}
    \caption{Kinematics of the [OII]3726,9$\lambda\lambda$ emission line doublet in Abell 1835. This figure is analogous to Fig.~\ref{fig:kinematics:A262}. In the upper right panel, the dotted ellipses show the position of the X-ray cavities.}
    \label{fig:kinematics:A1835}
\end{figure*}

\begin{figure*}
    \centering
    \includegraphics[width=\linewidth]{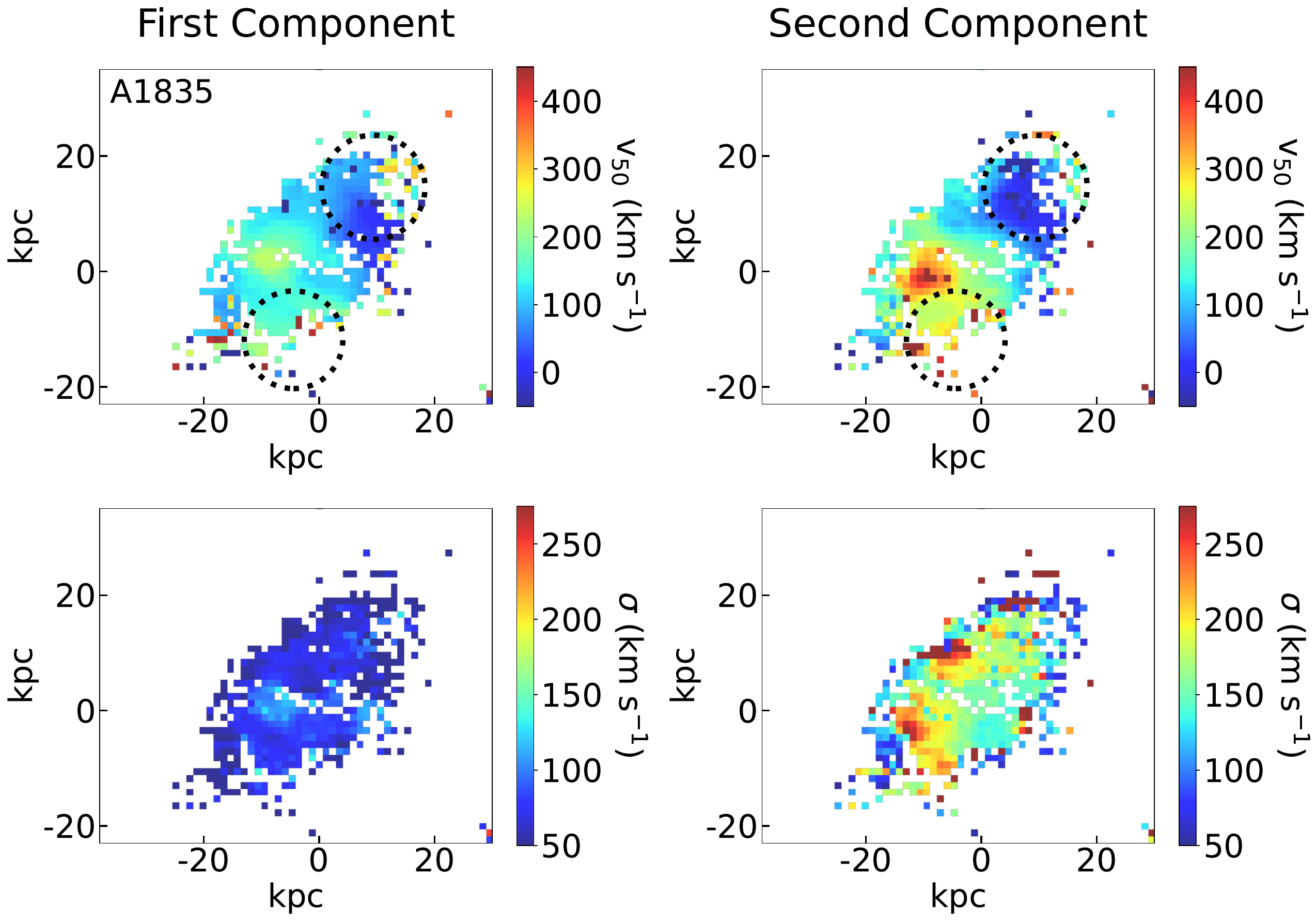}
    \caption{Median velocity maps (top) and velocity dispersion maps (bottom) for the first (left) and second (right) Gaussian components of the two kinematic component fits for Abell 1835. The dotted ellipses in the upper panels show the position of the X-ray cavities. Only spaxels where emission lines can be fitted by two kinematic components, each with $\text{S/N} \geq 3$, are shown.}
\label{fig:kinematics:A1835:Components}
\end{figure*}

Fig.~\ref{fig:kinematics:A1835} presents the LOS kinematics of the [OII] emission at each spaxel in Abell 1835. As detailed in Table~\ref{tab:target_properties}, the systemic redshift of this system is $z = 0.2514$. It is the most distant object in the sample. 

Fig.~\ref{fig:kinematics:A1835} reveals the complex gas dynamics surrounding the central galaxy. The median velocites of the nebular gas range between $-30$ and $+286$ km s$^{-1}$, with respect to the systemic velocity. It is worth noting that overall, the nebular gas is redshifted with respect to the systemic velocity, which is defined as the central stellar velocity. On average, the nebular gas is redshifted by an additional 150 km s$^{-1}$.

The gas with highest v$_{50}$ and $\sigma$ lies 8 kpc east of the nucleus. Analogously, on the other side of the nucleus, there is a low v$_{50}$ region. This could potentially indicate a bipolar outflow of gas caused by AGN activity. This system has two detected X-ray cavities, one on each side of the center along the SE-NW axis. The NW cavity partially overlaps with an extended low velocity filament while the SE cavity is projected onto substantially redshifted nebular gas with large $\sigma$. 

A sizable low $\sigma$ region is seen along the eastern and northern edges of the [OII] emission. This region which contributes 20\% of the total [OII] flux and spans a length of 75 kpc, has a velocity dispersion of less than 100 km s$^{-1}$. This gas is not being significantly disturbed. Finding undisturbed gas far from the nucleus implies that the main mechanism disrupting gas dynamics emanates from within the galaxy. 

Regions of interest in the kinematic maps are discussed in Appendix~\ref{appendix:A1835}. Properties of the regions, including flux-weighted mean velocities, are listed in Table~\ref{tab:A1835:regions}.

Abell 1835 is one of the two targets in our sample where two distinct kinematic components are detected, with most of the [OII] emission (82\%) coming from regions with two components. Fig.~\ref{fig:kinematics:A1835:Components} shows the v$_{50}$ and $\sigma$ separately for the first and second kinematic components of the regions requiring at least two components in the emission line fits.
For Fig.~\ref{fig:kinematics:A1835:Components}, the spaxels included in the velocity maps were chosen using softer criteria than for the total kinematic maps. To maximize the area where two kinematic components can be studied, we only require that each component of the resulting fit has a $\text{S/N} \geq 3$.
The second component is defined such that its velocity dispersion is always greater than that of the first component. 

Showing the kinematics of the two components separately clearly results in a different physical picture for each component. This is also shown in Fig.~\ref{fig:kinematics:A1835:regions:spectra} in Appendix~\ref{appendix:A1835} where significant differences between the first and second kinematic components are seen in the [OII] emission line fits. The top row of Fig.~\ref{fig:kinematics:A1835:Components} shows that substantial variations in gas velocities are mostly reflected in the second component of the fits. The main features of the v$_{50}$ map for the first component are a slight increase in velocities $\sim 8$ kpc east of the nucleus and lower velocities NW of the nucleus. The v$_{50}$ map of the second component (upper right panel) is very similar to the v$_{50}$ map of the total [OII] emission line. As seen in Fig.~\ref{fig:kinematics:A1835:Components}, the median velocity of the (broader) second component varies the most between different regions. The gradients in total median velocities, both the one along the E-W axis and the one along the axis of the cavities, are also observed in the v$_{50}$ map for the second component.

The bottom row of Fig.~\ref{fig:kinematics:A1835:Components} shows the velocity dispersion of the nebular gas associated with each kinematic component. By definition, the second component has larger $\sigma$. The velocity dispersion map of the first component (left panel) shows a slight increase in $\sigma$ in the highest v$_{50}$ region, with $\sigma \sim 100$ km s$^{-1}$, compared to the surrounding gas. We discussed above the NE region with low $\sigma$ on the outskirts of the detected [OII] emission. This low $\sigma$ region is due to the relatively quiescent gas in the first component, as a second component is not used when fitting this region. 

The bottom right panel, showing the velocity dispersion of the second component, varies more substantially. The region slightly east of the nucleus is particularly disturbed, with large v$_{50}$, $\sigma$ and v$_{02}$ values (see Fig.~\ref{fig:kinematics:A1835}). Another region with large velocity dispersion is located roughly 5 kpc east and 8 kpc north of the center. However, due to the low flux of the second component in this region, it is not seen as one of the prominent features on the total kinematics maps.
Actually, when applying the method described in Section~\ref{sec:Observations and data reduction: Data analysis: Fitting} to choose the optimal number of components, a one component fit is preferred for this region. Deeper observations are needed to thoroughly investigate the potentially complex gas dynamics obscured by the more quiescent gas.

The use of two kinematic component fits is required for this galaxy's structure with the most complex dynamics being fitted by the second component. The second component is much more disturbed that the first, as shown by the second component's large $\sigma$ and wide v$_{50}$ range. Both components contribute similarly to the total [OII] flux, with the second component emitting 55\% of the [OII] emission for spaxels using two kinematic components. The kinematics in Fig.~\ref{fig:kinematics:A1835:Components} are consistent with the galaxy having a quiescent gas component and a churned component which are modelled by the first and second kinematic components of our fits, respectively.
%The ionized gas seen in the second component emits a substantial fraction (45\%) of the total [OII] flux observed. 
%For spaxels which require the use of the two components fit, 55\% of the total [OII] emission comes from the churned component. 

\subsection{PKS 0745-191}\label{sec:Results:PKS0745}

PKS 0745-191 is a strong cool core cluster known for its powerful radio emission. First discovered as a radio source, this object has a radio luminosity (between 10 MHz and 5000 GHz) of $3.87 \times 10^{42}$ erg s$^{-1}$ \citep{Pulido2018}, more than ten times higher than the other objects in our sample (see Table~\ref{tab:target_properties}). The radio emission is diffuse and amorphous, showing no clear jets or radio lobes. Based on the lack of radio jets and other structures in the radio emission, jets, which inflated the detected X-ray cavities, may have been disrupted by the dense environment of the galaxy's core \citep{Russell2016}. \cite{Baum_ODea1991} suggested that a merger may have disrupted the jet by changing its orientation.

\subsubsection{PKS 0745-191: Morphology}

\begin{figure*}
    \centering
    \includegraphics[width=\textwidth]{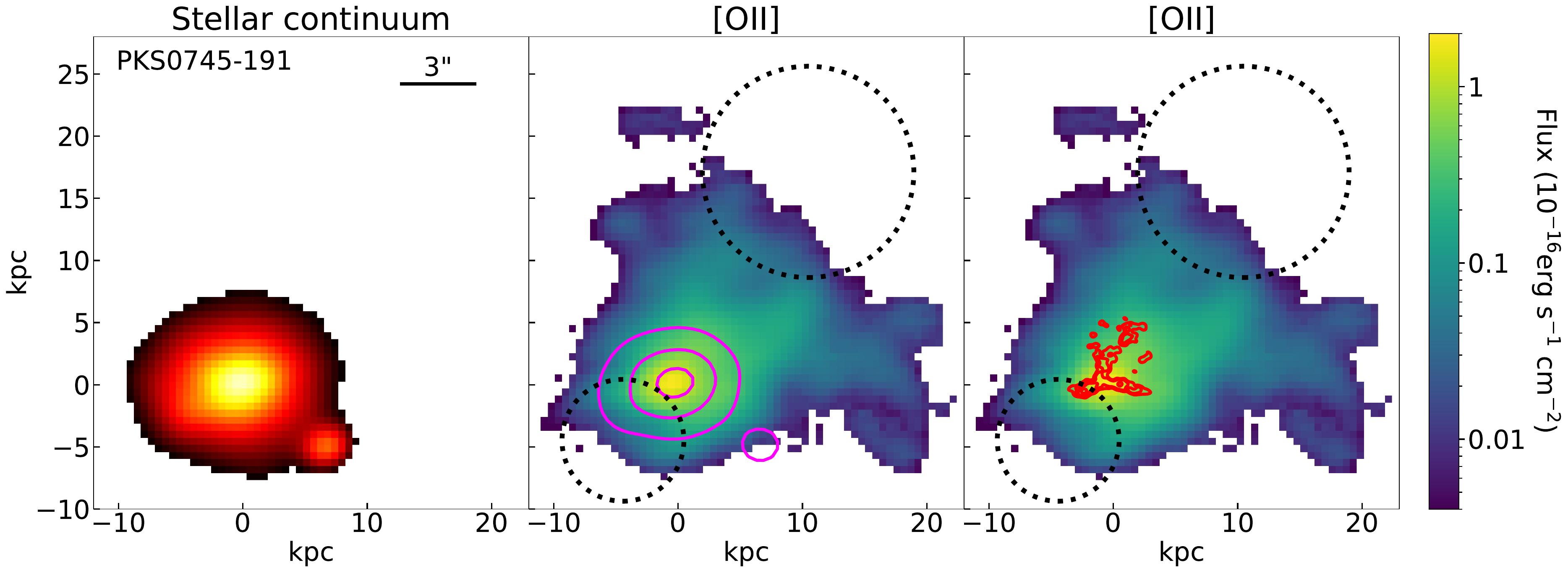}
    \caption{Left panel: Stellar continuum map for PKS 0745-191. The continuum flux colorbar ranges between $(2.5-40) \times 10^{-16}$ erg s$^{-1}$cm$^{-2}$. Middle and right panels: Continuum subtracted flux maps of [OII]3726,9$\lambda\lambda$, where $\text{S/N} \geq 5$. The magenta lines show stellar continuum contours for 20\%, 40\%, and 80\% of the maximum stellar continuum flux. The red contours map the flux of CO(3-2) emission, where the contours represent 10\%, 20\%, 40\%, and 80\% of the maximum CO(3-2) \citep{Russell2016}. The black dotted ellipses show the position of the two X-ray cavities \citep{Sanders2014,Russell2019}.}
    \label{fig:flux:PKS0745}
\end{figure*}

The left panel of Fig.~\ref{fig:flux:PKS0745} shows the stellar flux at rest-frame wavelengths between $3356-5017$ \AA. Above a stellar flux per spaxel of $2.5 \times 10^{-16}$ erg s$^{-1}$ cm$^{-2}$, the stellar continuum is detected in a roughly circular region with a radius of 8.3 kpc around the brightest stellar spaxel. This target has a total stellar luminosity of $1.78 \times 10^{43}$ erg s$^{-1}$, half of which is emitted within an inner region of 4.1 kpc in radius.

This target has one projected galaxy within the FOV shown in Fig.~\ref{fig:flux:PKS0745}, which is located 8.5 kpc SW of the nucleus of the central galaxy. The projected galaxy and the central galaxy may be interacting as they have similar stellar velocities, with a difference of only 34 km s$^{-1}$. The stellar flux map shown in the left panel of Fig.~\ref{fig:flux:PKS0745} does not show any obvious sign of interactions between the two galaxies. These dynamics will be discussed in depth in a future paper focused on the stellar properties of our targets.

The middle and right panels of Fig.~\ref{fig:flux:PKS0745} show the continuum-subtracted [OII] flux map.
The total [OII] luminosity is $4.7 \times 10^{41}$ erg s$^{-1}$, corresponding to $\sim 10^7$ M$_{\odot}$ of nebular gas (for $T_e \sim 10 000$ K and $n_e \sim 400$ cm$^{-3}$) as shown in Table~\ref{tab:OII_lum}. Half of the [OII] flux is emitted within a radius of 3.8 kpc (1.96\arcsec) from the nucleus of the central galaxy. The [OII] emission spans over 33 kpc in RA and 30 kpc in Dec, covering a total area of 542 kpc$^2$ (see Table~\ref{tab:OII_lum}). The [OII] morphology is related to the position of the X-ray cavities. As previously observed in many galaxies \citep{Russell2016,Olivares2019}, the [OII] emission extends from the nucleus towards and around the edges of the two X-ray cavities. The cavities, which are roughly aligned along the SE-NW axis, are not at the same distance from the nucleus. The NW cavity is more than three times as far from the center of the galaxy as the SE cavity. This asymmetry is reflected in the [OII] distribution. Along the SE-NW axis, the detected [OII] emission extends slightly past the inner edge of each cavity. Therefore, the different cavity heights are reflected in the asymmetric nebular gas distribution along the cavities' axis. This shows a clear relation between X-ray cavities and nebular gas morphology.

Fig.~\ref{fig:flux:PKS0745} shows that the peaks in stellar, [OII] and CO(3-2) emission roughly coincide spatially. The peak of CO(3-2) is 2.5 kpc east of our nucleus. This corresponds to a small offset at the scale studied. The central panel of Fig.~\ref{fig:flux:PKS0745} highlights the significant differences in the extent of the [OII] emission and the stellar continuum. The nebular emission is much more extended than the stellar emission towards the NW. Only 37\% of the area with [OII] emission also has detectable stellar emission. However, since most of the [OII] flux is emitted near the bright galaxy center, 82\% of the [OII] emission comes from regions with detected stellar emission. While we observe nebular emission to be more than twice as extended as the stellar continuum, most of the nebular gas coincides with starlight from the central galaxy.

The right panel of Fig.~\ref{fig:flux:PKS0745} shows contours of the CO(3-2) emission superimposed on the [OII] flux map. Most of the cold gas is located within a radius of 5 kpc from the nucleus. The molecular gas is mostly contained in three filaments \citep{Russell2016}. As we observed for the nebular gas, the cold gas filaments are extended in the direction of the cavities, one filament trailing the SE cavity and two filaments trailing the NW cavity. 

The detected CO(3-2) emission spans a much smaller area than the nebular gas. The detected [OII] emission covers a projected area 45 times larger than that of the CO(3-2) emission (Table~\ref{tab:OII_lum}). The distribution of molecular gas coincides with the brightest nebular gas, as shown by previous observations \citep{Russell2019,Olivares2019}. Deeper observations of the CO emission at the center of this galaxy may reveal a much larger cold gas reservoir.

\subsubsection{PKS 0745-191: Kinematics}

\begin{figure*}
    \centering
    \includegraphics[width=\textwidth]{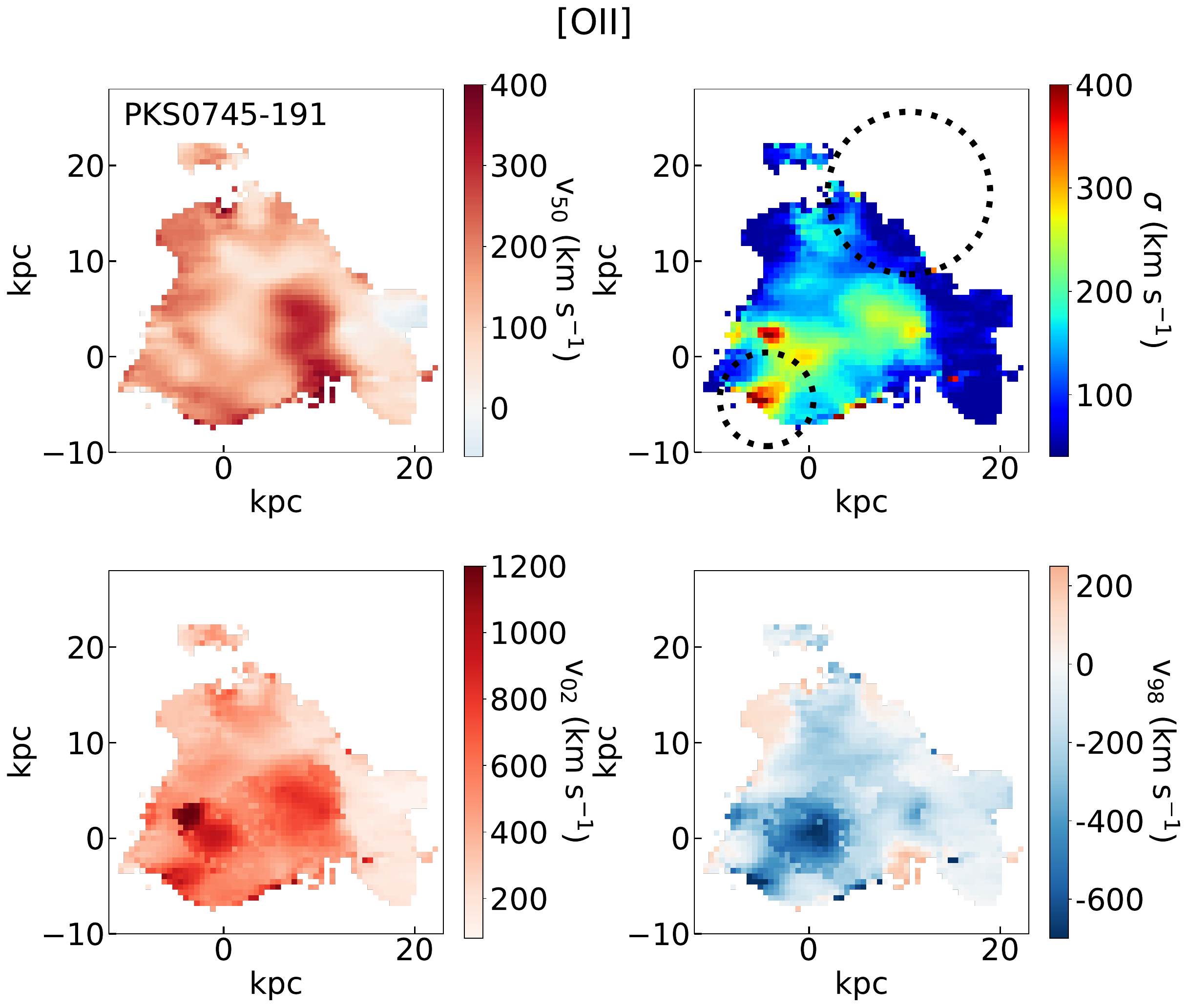}
    \caption{Kinematics of the [OII]3726,9$\lambda\lambda$ emission line doublet in PKS 0745-191. This figure is analogous to Fig.~\ref{fig:kinematics:A262}. In the upper right panel, the dotted ellipses show the position of the X-ray cavities.}
    \label{fig:kinematics:PKS0745}
\end{figure*}

\begin{figure*}
    \centering
    \includegraphics[width=\linewidth]{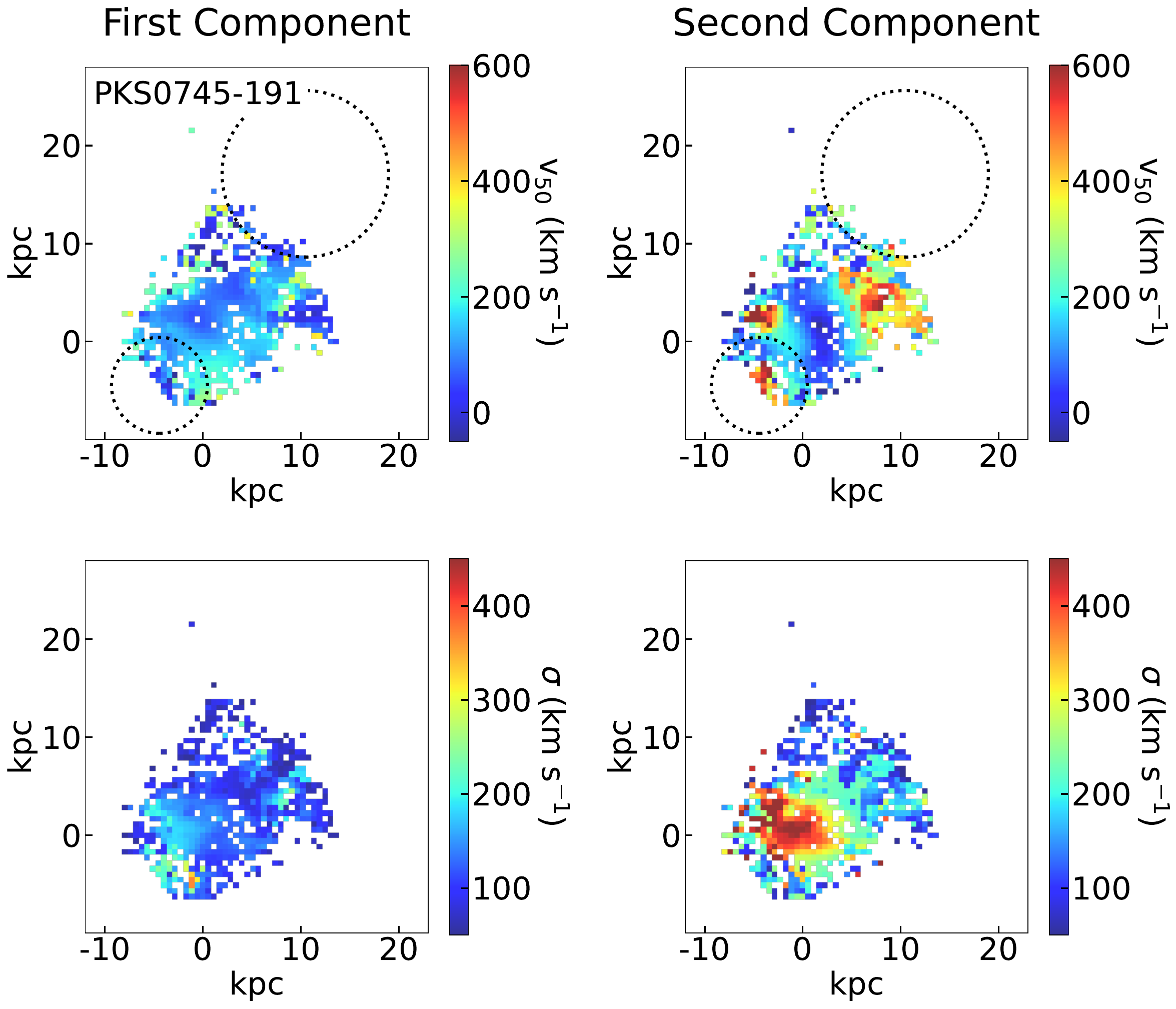}
    \caption{Similar to Fig.~\ref{fig:kinematics:PKS0745:Components}, v$_{50}$ (top) and $\sigma$ (bottom) maps for the first (left) and second (right) kinematic components for PKS 0745-191. Only spaxels where $\text{S/N} \geq 3$ for each component are included. The dotted ellipses in the top panels show the position of the cavities.}
\label{fig:kinematics:PKS0745:Components}
\end{figure*}

Fig.~\ref{fig:kinematics:PKS0745} shows the LOS kinematics of the [OII] emitting gas with maps of the median velocities (upper left panel), velocity dispersion (upper right), most-redshifted velocities (lower left) and most-blueshifted velocities (lower right). From the continuum absorption lines in the core of the galaxy, we measure a systemic redshift of 0.1024 (see Table~\ref{tab:target_properties}). As is the case for other targets in our sample, the nebular gas is redshifted with respect to the systemic velocity. On average, the nebular gas is redshifted by $\sim 135$ km s$^{-1}$. As shown in the top left panel of Fig.~\ref{fig:kinematics:PKS0745}, the total median velocities observed are mostly redshifted, with v$_{50}$ values between $-74$ km s$^{-1}$ and 402 km s$^{-1}$. 

The gas with the most redshifted total median velocities, with v$_{50}>225$ km s$^{-1}$, lies in a 20 kpc filament roughly 8.5 kpc from the center of the galaxy. The lowest v$_{50}$ gas is observed in the westernmost part of the detected emission.

Of the four central galaxies in our sample, PKS 0745-191 has the broadest emission lines observed. As seen in the top right panel of Fig.~\ref{fig:kinematics:PKS0745}, the nebular gas in the core of PKS 0745-191 has $\sigma$ values up to 400 km s$^{-1}$, which is almost twice that of our other targets. 

Two lines of high $\sigma$ nebular gas are seen in the right upper panel of Fig.~\ref{fig:kinematics:PKS0745}, forming a yellow cross roughly centered on the nucleus. The two regions with largest total velocity dispersion are located at the eastern end of the high $\sigma$ cross. They are located roughly 5 kpc NE and SE of the nucleus. The gas within these areas of the nebula is highly disturbed, having a total velocity dispersion greater than any other object in our sample. The lower left panel of Fig.~\ref{fig:kinematics:PKS0745} shows that some of the gas in the NE high $\sigma$ region is extremely redshifted. With v$_{02}$ greater than 1250 km s$^{-1}$, there may be an extremely powerful outflow in this region. 

This system has two known cavities within our FOV, roughly along the SE-NW axis, which are shown as dotted ellipses in the top right panel of Fig.~\ref{fig:kinematics:PKS0745}. The SE high $\sigma$ region is located within the SE cavity, while the upper part of the high v$_{50}$ filament partially overlaps with the NW cavity.

While the core of PKS 0745-191 has a large reservoir of nebular gas with $\sigma > 200$ km s$^{-1}$, the top right panel of Fig.~\ref{fig:kinematics:PKS0745} also shows low $\sigma$ gas in the outermost regions. While these low $\sigma$ regions are quite extended, with a total area of $\sim 200$ kpc$^2$, they are also quite faint such that gas with total $\sigma \leq 100$ km s$^{-1}$ contributes to only 5.5\% of the total [OII] flux. Other regions of interest are discussed more in depth in Appendix~\ref{appendix:PKS0745}.

Fig.~\ref{fig:kinematics:PKS0745:Components} shows v$_{50}$ and $\sigma$ for each of the kinematic components. For PKS 0745-191, we use a combination of the amplitude, position and width of each fitted gaussian component to differentiate between the first and second kinematic component. To be consistent with the analysis done for Abell 1835, we defined the more quiescent gas as the first component and the more churned-up and disturbed gas as the second component. The v$_{50}$ and $\sigma$ maps for the total emission lines (Fig.~\ref{fig:kinematics:PKS0745}) are different than the ones for the first and second components. While v$_{50}$ for the second kinematic component (top right panel Fig.~\ref{fig:kinematics:PKS0745:Components}) does not look like the total v$_{50}$ map (top left panel Fig.~\ref{fig:kinematics:PKS0745}), it does share similarities with the total $\sigma$ map (top right panel Fig.~\ref{fig:kinematics:PKS0745}). This indicates that the broadening of the [OII] emission line is predominantly caused by significant differences in the v$_{50}$ of the first and second components. This is clearly seen in the integrated spectra shown in Fig.~\ref{fig:kinematics:PKS0745:regions:spectra} in Appendix~\ref{appendix:PKS0745}. 

The significant differences in the v$_{50}$ and $\sigma$ maps for the total [OII] line (Fig.~\ref{fig:kinematics:PKS0745}) and the v$_{50}$ and $\sigma$ maps for each of the kinematic components (Fig.~\ref{fig:kinematics:PKS0745:Components}) highlights the importance of disentangling the different kinematic components.

The top right panel of Fig.~\ref{fig:kinematics:PKS0745:Components} shows the second components' v$_{50}$. It highlights the redshifted gas observed NE, NW and SE of the nucleus. In comparison, the median velocities of the first kinematic component, seen in the upper left panel, show little variation. 

The bottom right panel of Fig.~\ref{fig:kinematics:PKS0745:Components}, which maps $\sigma$ for to the second kinetic component, shows the distribution of the most disturbed gas in the central galaxy. The gas with the largest $\sigma$ lies around the nucleus and extends in the NE direction. This picture is consistent with a central mechanism, such as an AGN, disturbing gas around the nucleus. Comparing the v$_{50}$ and $\sigma$ second component maps, some of the central high $\sigma$ gas also has v$_{50} > 500$ km s$^{-1}$. The dynamics of the churned gas are consistent with an AGN driven outflow. 

The complex dynamics in the core of PKS 0745-191 requires the use of two component fits to properly model the [OII] emission lines. Fig.~\ref{fig:kinematics:PKS0745:Components} highlights the contrasting dynamics of the quiescent (first) component and the churned (second) component. A significant fraction of the nebular gas is churned, accounting for $\sim 39$ \% of the total [OII] emission and $\sim 49$ \% of the [OII] flux in spaxels with two kinematic components.

\section{Discussion}\label{sec:Discussion}

\subsection{Nebular Gas Structures: Peculiar Velocities With Respect to the Central Galaxy}\label{sec:Velocity_galaxies}

The v$_{50}$ maps, shown in the top left panel of Fig.~\ref{fig:kinematics:A262},~\ref{fig:kinematics:RXJ0820},~\ref{fig:kinematics:A1835}, and~\ref{fig:kinematics:PKS0745}, reveal little blueshifted gas in three of our four targets. Surprisingly, most of the nebular gas in Abell 1835, PKS 0745-191 and RXJ0820.9+0752 is redshifted by $\sim +140$ km s$^{-1}$ with respect to the central galaxy. Assuming that the nebular gas is bound to the central galaxy, we would expect its mean velocity to be close to that of the stars.
%We performed rigorous tests to ensure that our measurements are free of systematic bias. This is discussed in Section~\ref{sec:Observations and data reduction: Data analysis: Fitting} and in Appendix A. We are aware of no additional biases that would lead to an overall velocity shift.
%In three out of four targets, the nebular gas has a redshift of almost 150 km s$^{-1}$ with respect to the central galaxy. Although the nebular gas is redshifted, and not blueshifted, with respect to the systemic velocity in our three targets with significant velocity offsets, we must refrain from drawing any definitive conclusion regarding the direction of the velocity offset due to our small sample size. Nonetheless, our observations clearly show that the velocity of the central galaxy and that of the nebular gas can differ on the order of 100s of km s$^{-1}$
The relative velocities indicate that the nebular gas and the central galaxies are, to some degree, dynamically decoupled. 
%The velocity offset reflects overall motion of the nebular gas with respect to the central galaxy. 

Although initially surprising, velocity and positional offsets between nebular gas and central galaxies in the sky plane have been observed in many systems such as RXJ0820.9+0752, Abell 1991, Abell 3444 and Ophiuchus \citep{Hamer2012}.
Therefore, velocity offsets of this magnitude are likely common.

The nebular and molecular gas in RXJ0820.9+0752 \citep{Vantyghem2017} are spatially offset from the central galaxy indicating relative motion in the sky plane. Its LOS velocity offset of $\sim 150 ~\rm km~s^{-1}$ indicates redshifted motion similar to Abell 1835 and PKS 0745-191. 

A displacement of one kpc can be achieved in approximately $\sim 10$ Myr assuming a relative velocity of $100 ~\rm km~s^{-1}$. Therefore, the event that triggered the gas displacement would have occurred between a few tens to 100 Myr ago. Phenomena leading to galaxy oscillations include sloshing induced by infalling galaxies and halos \citep{Markevitch2007,Zuhone2016},
AGN outbursts, and larger scale dynamical oscillations \citep{vandenBosch2005, DePropris2021}.
 
%The feedback timescale may be estimated by the time since the last major AGN outburst in a system. X-ray cavity buoyancy ages lie between a few Myr to 120 Myr, which is consistent with our measurements.

%Central galaxies are expected to oscillate with respect to the barycenter by $\sim 100 ~\rm km~s^{-1}$ \citep{vandenBosch2005,DePropris2021}.
This motion could lead to several interesting effects. For example, cooling and star formation may be offset from the center of the galaxy. Motion of the central galaxy would distribute the energy released by the AGN more broadly throughout the atmosphere. This added heat circulation may help circumvent a potential problem with anisotropic heating models \citep{Heinz2006, Vernaleo2006, Babul2013}.
The relative motion may explain why radio bubbles are often misoriented with respect to the active radio source \citep{Fabian2006,Ubertosi2021} and are frequently amorphous rather than double-lobed.
%Ram pressure against radio jets and lobes may disrupt and redistribute the radio plasma into the amorphous radio emission seen in these and other systems. 

The peculiar motion of the galaxy with respect to the gas may have significant consequences for thermally unstable cooling and the observed gas kinematics. Assuming that the quiescent nebular gas recently cooled from the hot atmosphere and is precipitating onto the galaxy in the forms of atomic and molecular clouds, the gas and stars should acquire angular momentum with respect to the central galaxy. Moreover, the change in gravitational acceleration of cooling gas parcels will alter the conditions required for thermally unstable cooling \citep{Godon1994,Pizzolato2005,McCourt2012,Gaspari2013,Voit2015}, and will almost certainly enhance cooling. We expect peculiar motions with respect to the cooling X-ray atmospheres to be observed with X-Ray Imaging and Spectroscopy Mission (XRISM) calorimeter spectroscopy of hot atmospheres. Our understanding of thermally unstable cooling will require significant revision in light of these motions. 

%\subsection{Systems with Simple Gas Kinematics}\label{Nebular gas dynamics}

%The simplest gas motions are found in Abell 262 and RXJ0820.9+0752.  Much of the nebular gas and all of the molecular gas in Abell 262 is in rotation about the nucleus of the central galaxy along an E-W axis.
%Radio jets, oriented $\sim 25^{\circ}$ north of east, roughly align with the angular momentum axis of the disk. The velocity dispersion of the nebular gas sharply rises along the jets, as seen in Fig.~\ref{fig:sig_radio:A262}, indicating that the jets are interacting with and disturbing the nebula. A more complete analysis of the gas and stellar velocity fields in Abell 262 will be presented in a future paper.

%The nebular gas in RXJ0820.9+0752 is offset from the central galaxy in the sky plane and in velocity along the LOS.  The gas is largely quiescent.  Only a few regions are churning with $\sigma > 100$ km s$^{-1}$, and they are located on the edges of the [OII] nebula.  
%The higher $\sigma$ edge regions may reflect interactions with the secondary galaxy, with the star formation occurring there, or with the atmosphere itself due to their relative motion. A secondary galaxy lies at a projected distance of only $\sim 7.8$ kpc from the central galaxy, with relative one dimensional speeds  of $\sim 100$ km s$^{-1}$.
%These properties are consistent with the model proposed by \cite{Vantyghem2018} which suggests that sloshing motions caused by the secondary galaxy may have lifted gas from the core of the galaxy to altitudes where thermal instabilities may occur. 

\subsection{Powerful Feedback Systems with Complex Velocity Fields}\label{Powerful Feedback}

Abell 1835 and PKS 0745-191 were chosen to examine the impact of powerful radio AGN on their central galaxies. Both systems are home to some of the largest star formation rates known in central galaxies. 
Their nebular gas structures are similar. The total $\sigma$ maps reveal larger values in the inner regions indicating disturbed gas. The nebular gas at larger radii is more quiescent apart from regions near X-ray cavities or jets. The inner gas is apparently churned-up by the radio AGN. These disturbances are likely not the result of larger scale processes such as sloshing or mergers. 

Abell 1835 and PKS 0745-191 require two kinematic component fits. The second component is sensitive to high $\sigma$ and extreme LOS velocities. regions with two distinct kinematic components roughly coincide with the inner regions where churned gas is found. The total v$_{50}$ and $\sigma$ can be misleading when multiple kinematic components are present. Large total $\sigma$ values may indicate broad emission lines from disturbed gas or two narrow kinematic components with different median velocities. Where two components are required, v$_{02}$ maps, v$_{98}$ maps and individual component maps enable us to disentangle kinematic components. This in turn permits mapping of the distribution of disturbed gas which is sensitive to potential inflows and outflows. 

The v$_{50}$ of quiescent and churned gas along the same LOS can differ by hundreds of km s$^{-1}$. Furthermore, the median velocities of the churned and quiescent gases in the brightest spaxels, which contain the majority of nebular gas, also differ significantly. %Therefore, the flux-weighted median velocities for the churned and quiescent nebular gas are similar. 
For Abell 1835, $\bar{\textrm{v}}_{50}$ is $+121$ km s$^{-1}$ and $+175$ km s$^{-1}$ for the quiescent and the churned gas, respectively. For PKS 0745-191, $\bar{\textrm{v}}_{50}$ is $+126$ km s$^{-1}$ and $+161$ km s$^{-1}$ for the quiescent and the churned gas, respectively. 

\begin{figure}
    \centering
    \includegraphics[width=\columnwidth]{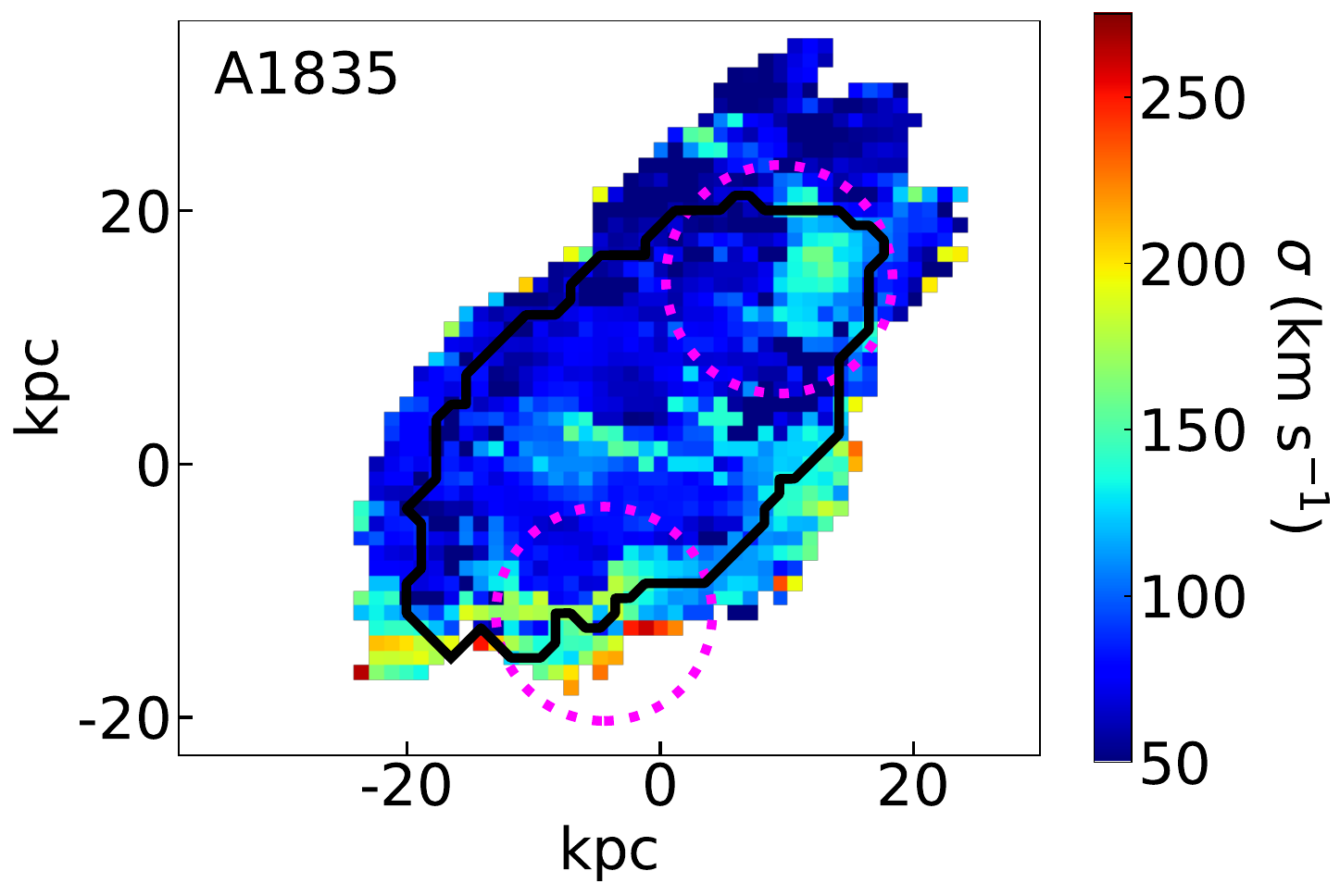}
    \caption{
    Velocity dispersion map of the quiescent nebular gas in the core of Abell 1835. For spaxels where two kinematic components were observed, the velocity dispersion of the first kinematic component is shown. Otherwise, the $\sigma$ from the one component fit is shown. The black contour roughly encompasses the region where two kinematic components can be detected. The dotted magenta circles show the position of the X-ray cavities.}
    \label{fig:A1835_v50_1comp}
\end{figure} 

Fig.~\ref{fig:kinematics:A1835:Components} and~\ref{fig:kinematics:PKS0745:Components} show the spaxels where two kinematic components can be disentangled along the LOS. Low velocity dispersion gas is associated with the first kinematic component while gas associated with the second component is significantly more disturbed and represents the churned part of the nebular gas.

%\color{red}
Generally, for spaxels where only one kinematic component is observed, the nebular gas has low $\sigma$ and has kinematics similar to the quiescent component of the two component fits as seen in Fig.~\ref{fig:A1835_v50_1comp}. The region enclosed by the solid black line in Fig.~\ref{fig:A1835_v50_1comp} shows the velocity dispersion of the quiescent gas for the inner region where two kinematic components are observed. Beyond this region, where only one kinematic component is observed, Fig.~\ref{fig:A1835_v50_1comp} shows the velocity dispersion from the one component fit. We see a smooth transition in the $\sigma$ map between the inner and outer region. This shows that when only one kinematic component is detected in the outer regions of the [OII] emission, the gas dynamics are consistent with those of the quiescent component.
%\color{black}

%Fig.~\ref{fig:A1835_v50_1comp} shows the median velocities of the quiescent nebular gas in the central galaxy of Abell 1835. The black contour roughly separates between the inner region, where two kinematic components are observed, and the outer region, where only one kinematic component is observed. We see a smooth transition in the v$_{50}$ map between the inner region, showing v$_{50}$ for the first kinematic component of the two components fit, and the outer region, showing the v$_{50}$ of the one component fit. 

\begin{figure}
    \centering
    \includegraphics[width=\columnwidth]{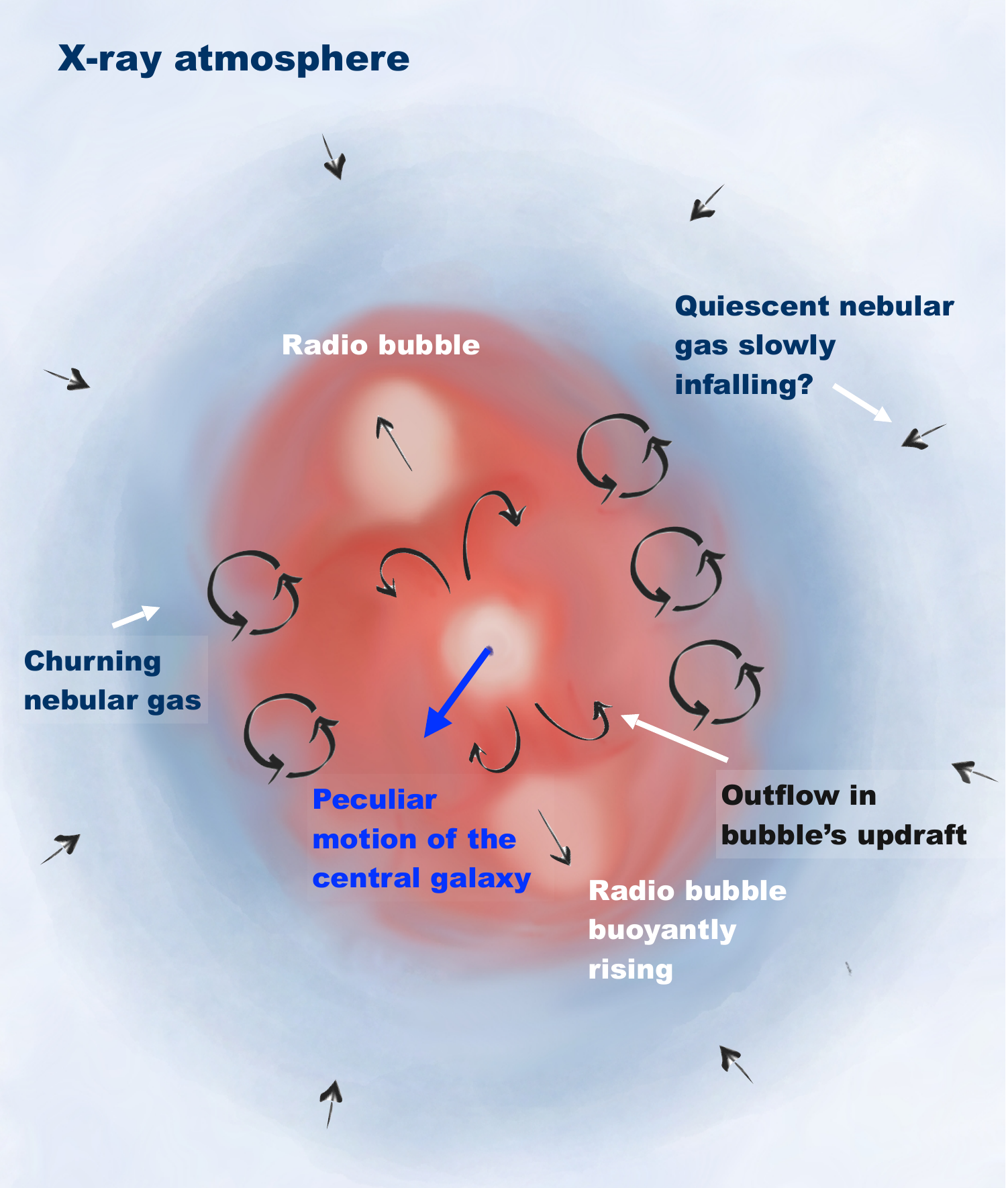}
    \caption{Schematic representation of the relationships between the dynamical states of the central galaxy (central spheroid) and its radio source (black lines), the large scale quiescent nebular emission (blue), and the churned inner nebular gas (red) in the vicinity of Abell 1835's central radio source.  PKS 0745-191 is likely similar but with the radio jets/lobes seen nearly pole-on.}
    \label{fig:diagram}
\end{figure}

Our observations are consistent with a model where extended quiescent nebular gas surrounds embedded churned-up gas which lies closer to the central AGN. A schematic representation of this configuration is shown in Fig.~\ref{fig:diagram}. Our model includes quiescent nebular gas (shown in blue) which extends to the outer parts of the nebula. This gas may have recently cooled from the X-ray atmosphere. If so, we expect it to have kinematics similar to the hot atmosphere.

%The inner gas is churned-up, associated with more extreme gas dynamics likely due largely but perhaps not exclusively, to interactions with the radio jets and lobes. 
As the outer gas moves into the inner regions, it may be lifted back out by the outward motion of the radio jets and bubbles. 
%This scenario is consistent with the dynamics of the different components of the nebular gas observed in Abell 1835 and PKS 0745-191. 
Similar velocity structures are observed in the Perseus cluster nebula, where churned-up gas is present in the inner 10 kpc near the radio bubbles, while quiescent low velocity dispersion gas extends out to 50 kpc \citep{Vigneron2024}.

This model will be further examined by XRISM high resolution spectroscopy which will reveal the velocity dispersion and LOS speed of the hot atmosphere to precision similar to the Keck data.

\subsection{Velocity Structures Near the Radio Lobes and X-ray Bubbles}\label{sec:AGN Energetics}

X-ray observations from the Chandra X-ray Observatory have played a pivotal role in our understanding of AGN mechanical feedback. Jets inflate massive X-ray cavities that rise through the atmosphere propelled primarily by buoyancy as they lift gas in their updrafts \citep{Birzan2004,McNamara2007}. 
However, the outward speeds of the bubbles and X-ray gas advected into their wakes have only been inferred \citep{Churazov2001,Pope2010} and could be measured directly using XRISM X-ray spectroscopy with great difficulty and only for nearby systems. Here, we use velocity profiles of the nebular emission to infer bubble speeds by measuring the motion of the nebular gas in the bubbles' wakes. 

\subsubsection{Abell 1835}

\begin{figure}
    \centering
    \includegraphics[width=\columnwidth]{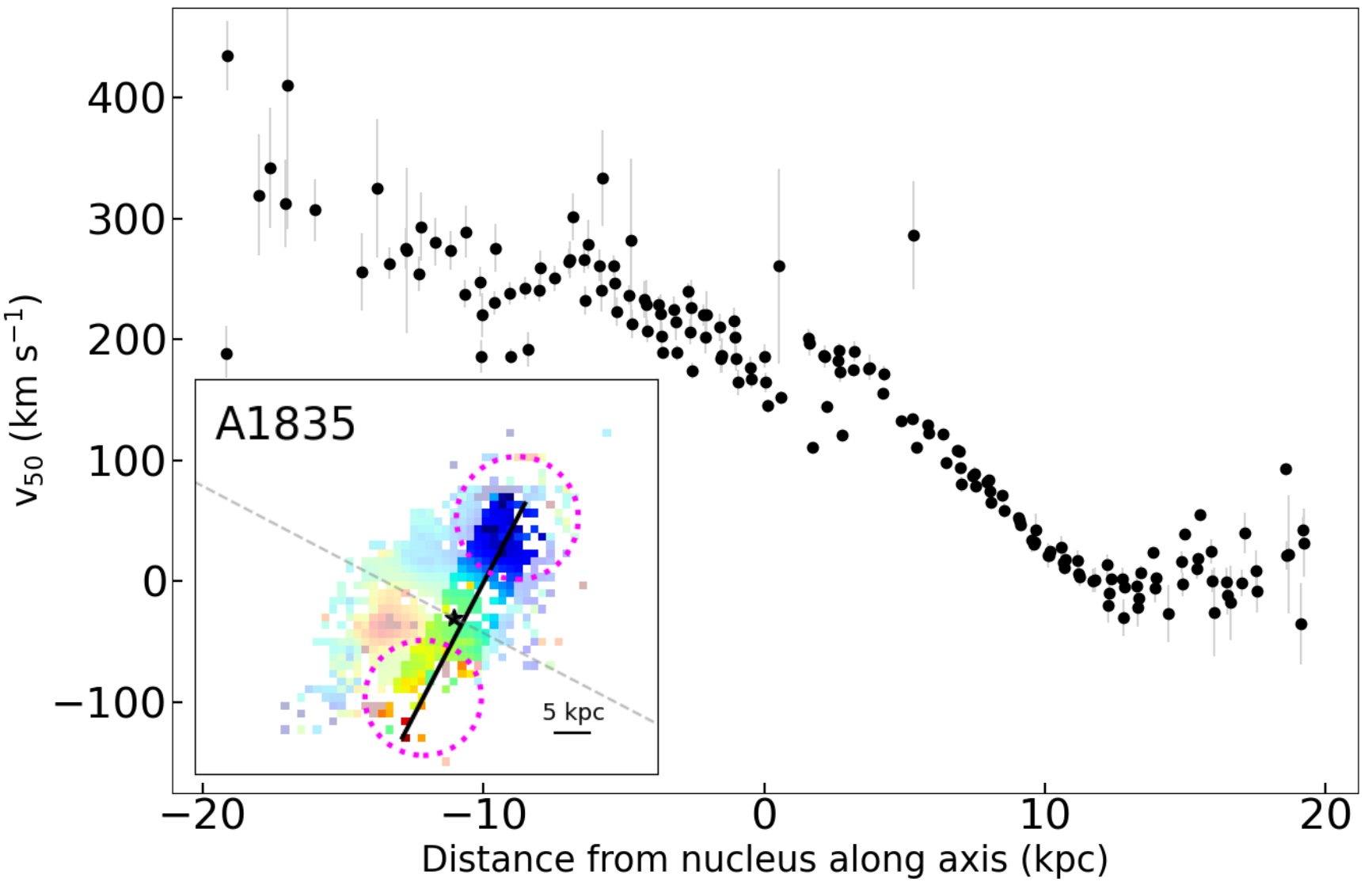}
    \caption{
    %\color{red}
    Median velocity of the second kinematic component of the [OII] emitting gas as a function of its projected distance along the cavities' axis for Abell 1835. The errors on v$_{50}$ are given by the grey error bars. The bottom left figure shows the v$_{50}$ map for the second component as in the top right panel of Fig.~\ref{fig:kinematics:A1835:Components}. The darker region shows the spaxels included in the v$_{50}$ vs distance plot and the black line shows the projection axis. The black star shows the position of the nucleus and the grey dashed line projects the position of the nucleus onto the cavities' axis.}
    
    \label{fig:A1835:dist_vs_v50_cavities}
\end{figure}

\begin{figure}
    \centering
    \includegraphics[width=\columnwidth]{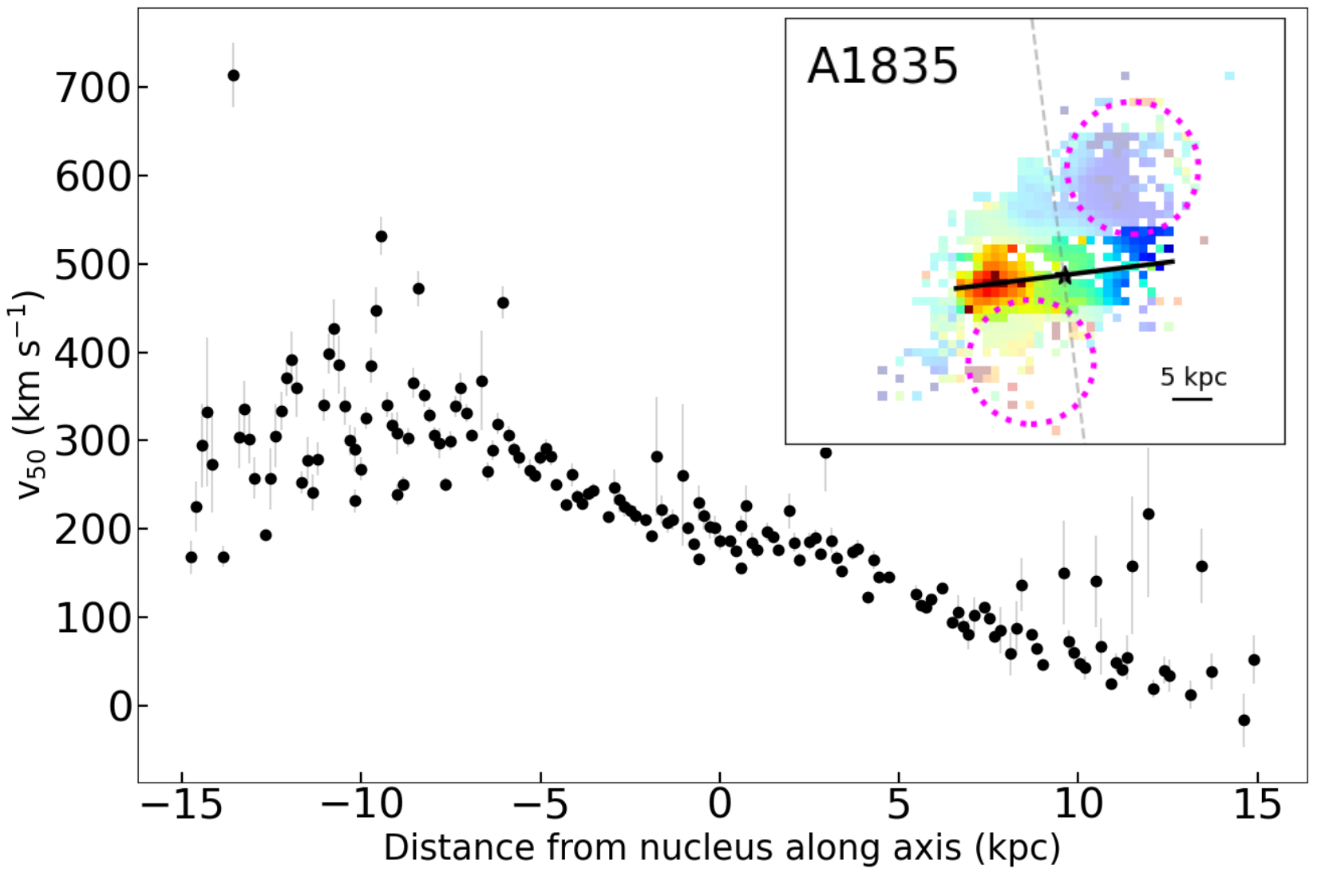}
    \caption{Median velocity of the second kinematic component of the [OII] emitting gas as a function of its projected distance along the axis shown in the top right figure. Symbol coding is the same as in Fig.~\ref{fig:A1835:dist_vs_v50_cavities}. }
    \label{fig:A1835:dist_vs_v50:x}
\end{figure}

%\begin{figure}
%    \centering
%    \includegraphics[width=\columnwidth]{PKS0745_dist_v50_NE_SW_errors.pdf}
 %   \caption{Median velocity of the second kinematic component of the [OII] emitting gas as a function of its projected distance along the axis shown in the embedded figure. The embedded figure shows the v$_{50}$ map for the second component of PKS 0745-191 as in the top right panel of Fig.~\ref{fig:kinematics:PKS0745:Components}. Symbol coding is the same as in Fig.~\ref{fig:A1835:dist_vs_v50_cavities}.}
    %\label{fig:PKS0745:dist_vs_v50:NE_SW}
%\end{figure}

\begin{figure}
   \centering
    \includegraphics[width=\columnwidth]{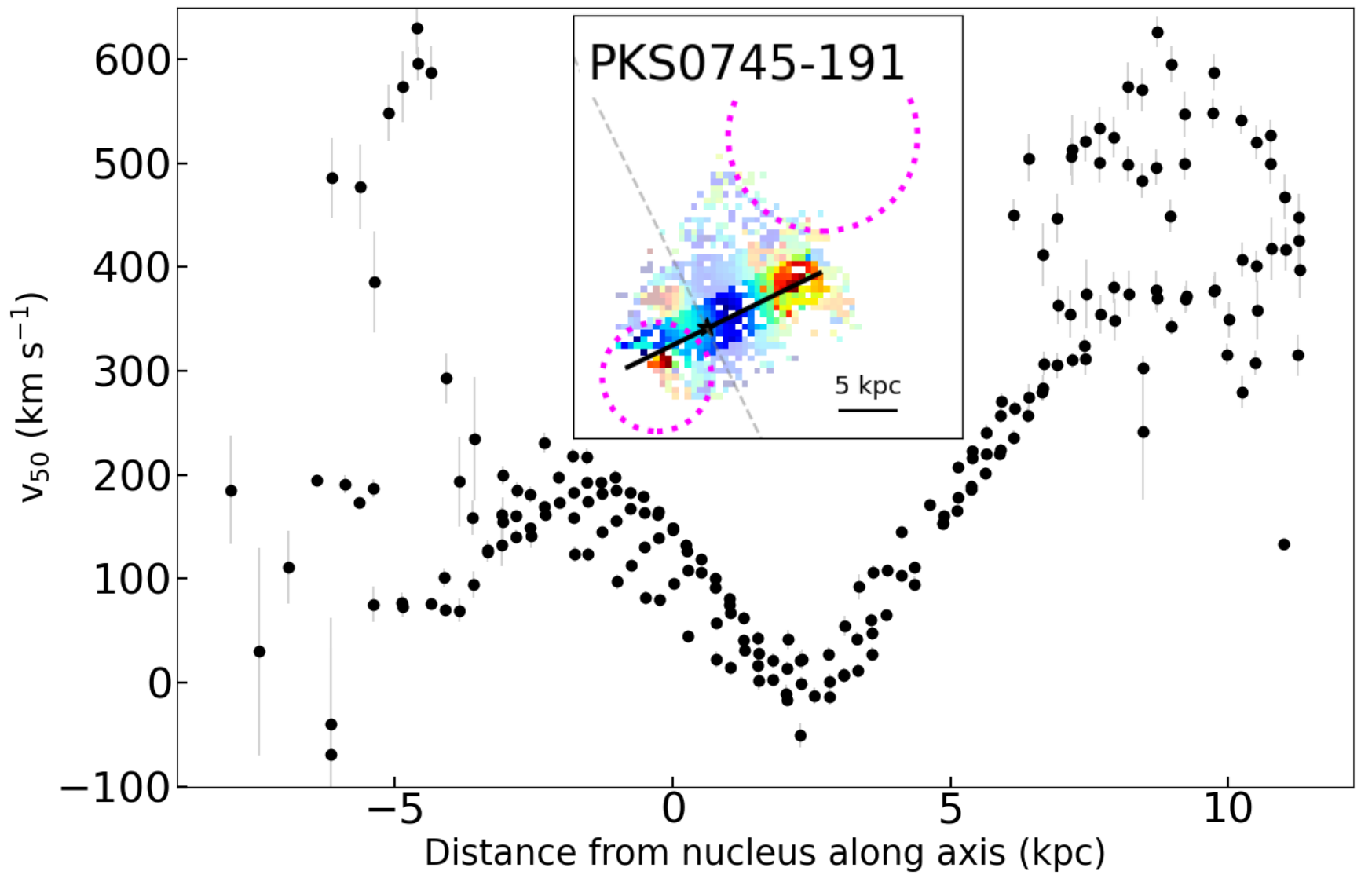}
   \caption{Median velocity of the second kinematic component of the [OII] emitting gas in PKS 0745-191 as a function of its projected distance along the axis of interest. The embedded figure shows the second kinematic component v$_{50}$ map for PKS 0745-191 as seen in the top right panel of Fig.~\ref{fig:kinematics:PKS0745:Components}. Other elements of the figure are analogous to those used in Fig.~\ref{fig:A1835:dist_vs_v50_cavities} and Fig.~\ref{fig:A1835:dist_vs_v50:x}.}
 \label{fig:PKS0745:dist_vs_v50:SE_NW}
\end{figure}

We focus on the second kinematic component of the emission line fits for Abell 1835. Using the second component v$_{50}$ maps (see the top right panel of Fig.~\ref{fig:kinematics:A1835:Components}), we searched for ordered motion and looked for signs of interactions between the [OII] gas and the AGN feedback process. We identified possible inflows and outflows by searching for velocity gradients in the churned v$_{50}$ maps. Once an axis is defined along some velocity gradient, we extracted the v$_{50}$ values along the projected axis and analyzed the evolution of nebular gas velocities along the chosen axis.

Different scenarios may be considered when testing possible gas inflows and outflows.
First, as nebular gas is observed in projection, clouds may lie on the near side or the far side of the central galaxy. Whether the gas associated with the observed velocity gradient, or filament, is an inflow or an outflow depends on whether the filament is in front or behind the core of the central galaxy. Similarly, the choice of rest-frame for the nebular gas, usually either the gas located at the nucleus or at the other end of the filament, impacts the inferred model. Therefore, we will generally consider four possible scenarios, with either 1) the velocity centered on the nucleus, or centered on the end of the filament and 2) the filament either in front of the central galaxy or behind it. Discarding unphysical models, we can identify possible inflows and outflows of nebular gas in our systems.

Fig.~\ref{fig:A1835:dist_vs_v50_cavities} shows the median velocities of the churned nebular gas in the central galaxy of Abell 1835 projected along the X-ray cavity axis, shown by the black solid line. The grey error bars show the uncertainty on the v$_{50}$ values obtained from the emission line fits. The v$_{50}$ map of the churned component is embedded in the bottom left of Fig.~\ref{fig:A1835:dist_vs_v50_cavities}. The median velocities are plotted as a function of their distance along the projected axis. The zero distance is defined as the projected position of the nucleus onto the cavities' axis, which is where the grey dashed line and the solid black line intersect. 

Fig.~\ref{fig:A1835:dist_vs_v50_cavities} shows a gradient in the velocity of the churned nebular gas along the cavities' axis. From the SE to the NW cavity, the gas velocity decreases from $+350$ km s$^{-1}$ to $-30$ km s$^{-1}$, with a central velocity of roughly $+165$ km s$^{-1}$. Centering the rest-frame velocity on the nebular gas velocity at the nucleus, the filament extending towards the SE cavity is redshifted, while the one extending towards the NW cavity is blueshifted with respect to the central galaxy.

For the negative distances in Fig.~\ref{fig:A1835:dist_vs_v50_cavities} that correspond to the gas SE of the nucleus, the projected v$_{50}$ increase linearly towards the SE cavity. This velocity gradient indicates that the gas is moving with a speed proportional to the distance from the nucleus. 
This velocity profile is inconsistent with infall where speeds should increase toward the nucleus. The profile may indicate gas accelerating away from the nucleus over time being pushed by a central force, perhaps due to a jetted outflow or gas being advected into the bubble wakes and being lifted out from the nucleus. While both Abell 1835 and PKS 0745 have powerful radio sources, jets are not observed. We therefore consider the latter model.
%In this model the gas must experience a nearly constant acceleration,  where we can calculate the average acceleration using:
%\begin{equation}\label{eq:acceleration}
%    a=\frac{v^2-v_0^2}{2 \Delta x},
%\end{equation}
%where $a$ is the acceleration, $v$ is the final %velocity, $v_0$ is the initial velocity and $\Delta x$ is the distance covered by the %nebular gas. 
%We calculate a mean acceleration of $1.28\times10^{-14}$ km s$^{-2}$ along the SE filament, for a velocity change ($\Delta$v) of 118 km s$^{-1}$ and a length of 17.6 kpc. If the SE filament is in front of the central galaxy, the gas, which is redshifted, must be moving towards the center of the galaxy and decelerating as it moves closer to the nucleus. This scenario can be discarded as it is unphysical and cannot be achieved through gravitational infall onto the central galaxy, jet activity or as a result of galaxy motion without extreme fine-tuning. On the other hand, if the SE filament is behind the central galaxy, the gas accelerates away from the central galaxy in an outflow. This is consistent with jet-driven outflows.

%The velocities of the cavities in Abell 1835 is

%We can also estimate cavity velocities in this system from X-ray observations \citep{Churazov2001,Birzan2004,Rafferty2006,Hogan2017,Pulido2018}. 
 
The velocity profiles shown in Fig.~\ref{fig:A1835:dist_vs_v50_cavities}
%and Fig.~\ref{fig:A1835:dist_vs_v50_cavities} 
are consistent with bubbles moving away from the galaxy nearly along the LOS, with the NW bubble located in front (blueshifted) and the SE bubble located behind (redshifted) the central galaxy. 
%shows another filament, NW of the nucleus, where gas velocities decrease away from the nucleus. With the same rest-frame velocity as before, the NW filament is blueshifted and accelerating away from the nucleus. Contrary to the SE filament, the curve in the velocity function indicates that gas acceleration is not constant, with slight changes in acceleration at both ends of the filament. With a $\Delta$v of 213 km s$^{-1}$ along a 19.2 kpc filament, we calculate a mean acceleration of $3.82\times10^{-14}$ km s$^{-2}$. If the NW filament is in front of the central galaxy, the gas is blueshifted, accelerating away from the nucleus, which is consistent with a jet-driven outflow. The NW filament cannot be behind the central galaxy as this would correspond to a decelerating inflow of gas, which is unphysical. 

%The velocity gradient observed along the cavities' axis in the central galaxy of Abell 1835 is consistent with a bipolar jet outflow. Both the NW and SE filaments are observed trailing known X-ray cavities, which supports the theory of bipolar jet driven outflows, including the entrainment and subsequent cooling of central gas by X-ray cavities. As expected for bipolar outflows, one cavity, NW, is on the near side of the central galaxy while the other one, SE, is on the far side. The dynamics of the churned disturbed gas along the axis connecting the SE and NW cavity is consistent with bipolar jet-driven outflows of gas and entrainment.

A similar analysis was applied to the axis perpendicular to the cavity axis shown in Fig.~\ref{fig:A1835:dist_vs_v50_cavities} to verify that the gas is indeed flowing along the bubble trajectory. As expected, the v$_{50}$ of the churned nebular gas show no correlation with distance along the axis orthogonal to the X-ray cavities.

%Fig.~\ref{fig:A1835:dist_vs_v50:x} is similar to Fig.~\ref{fig:A1835:dist_vs_v50_cavities} and shows the median velocities of the churned gas as a function of its projected distance from a different axis of interest, which is identified in the embedded figure. 

Fig.~\ref{fig:A1835:dist_vs_v50:x} shows a velocity gradient running from $\sim 13$ kpc east to $\sim 13 $ kpc west of the nucleus. The east-west v$_{50}$ decrease from $+380$ km s$^{-1}$ to $+15$ km s$^{-1}$. The scatter in the trend increases dramatically at 7 kpc to the east and to a lesser degree at a similar distance to the west.
The eastern increase in scatter roughly occurs at the region with a v$_{50}$ peak, identified as Region B in Appendix~\ref{appendix:A1835}. This region corresponds approximately to the location of known molecular clouds seen with ALMA \citep{McNamara2014}.
The amplitude of the velocity scatter exceeds $600 ~\rm km ~s^{-1}$ far from the nucleus, indicating a strong interaction or high impact event. The ALMA early science data are too crude to clearly interpret this apparent interaction, which must await deeper ALMA imaging.

The gas locations, velocity profiles, and speeds in Abell 1835 are consistent with gas lifted outward in the updraft of the rising X-ray bubbles at a low inclination angle with respect to the LOS. The highest speed gases nearest to the bubbles are presumably moving close to the bubble speeds, while the gas further from the bubbles was advected into the wake more recently and is moving outward with slower speeds. Adopting this model, we infer cavity velocities of about $185-195$ km s$^{-1}$. Depending on the inclination angle of the system, this corresponds to cavity speeds of less than 300 km s$^{-1}$.

Empty bubbles are expected to rise at their maximum terminal speeds
%$v_t=\left(\frac{2gV}{SC}\right)^{1/2}$ and the sound speed 
%$c_s=\left(\frac{\gamma kT}{\mu %m_H}\right)^{1/2}$ at the cavity's position 
\citep{Churazov2001,Birzan2004,Rafferty2006}. A detailed study of M87's bubble dynamics inferred speeds of roughly half the local sound speed \citep{Churazov2001}. Using atmospheric temperatures surrounding the bubbles from \cite{Hogan2017} and \cite{Pulido2018}, we find %terminal speeds of 1120 km s$^{-1}$ and 1240 km s$^{-1}$, and 
sound speeds of $840-910$ km s$^{-1}$. 
We therefore expect Abell 1835's bubbles to rise at roughly 450 km s$^{-1}$. After correcting for inclination, the implied bubble speeds lie below half the local sound speed, suggesting that mass loading by material dragged outward is slowing down the bubbles.

%This is within a few factors of the terminal velocities ($\sim 1100 - 1250$ km s$^{-1}$) and the sound speeds ($\sim 830 - 910$ km s$^{-1}$) at the cavities' positions, which can be estimated from X-ray observations \citep{Churazov2001,Birzan2004,Rafferty2006,Hogan2017,Pulido2018}. This is generally consistent with the results from \cite{Churazov2001} which states that for M87, the cavities rise at roughly half the speed of sound.
%\color{black}

\subsubsection{PKS 0745-191}

The second component v$_{50}$ map for PKS 0745-191,
%Nebular gas inflows and outflows are also seen in the second kinematic component velocity maps of PKS 0745-191. The second component v$_{50}$ map, 
shown in the top right panel of Fig.~\ref{fig:kinematics:PKS0745:Components}, appears to have two velocity gradients. We will focus on the velocity gradient which is closest to the cavities' axis which is shown in Fig.~\ref{fig:PKS0745:dist_vs_v50:SE_NW}. It shows the evolution of the median velocities of the churned nebular gas when roughly projected onto the SE-NW axis, shown by the solid black line in Fig.~\ref{fig:PKS0745:dist_vs_v50:SE_NW}. The embedded figure shows the v$_{50}$ map of the second component, as in the top right panel of Fig.~\ref{fig:kinematics:PKS0745:Components}. The symbols and lines overlaid on the velocity map are analogous to the ones in Fig.~\ref{fig:A1835:dist_vs_v50_cavities} and~\ref{fig:A1835:dist_vs_v50:x}.

Fig.~\ref{fig:PKS0745:dist_vs_v50:SE_NW} shows the median velocity of the nebular gas as a function of projected distance along the axis of the SE and NW X-ray cavities. 
Fig.~\ref{fig:PKS0745:dist_vs_v50:SE_NW} shows two (V-shaped) velocity gradients whose vertex lies $2.4$ kpc in projection NW of the nucleus. The gas velocity at the vertex is close to 0 km s$^{-1}$. The median velocities of the nebular gas along the SE cut increase smoothly away from the vertex from $-20$ km s$^{-1}$
to $+200 ~\rm km ~s^{-1}$ at $-3$ kpc.
%resulting in a mean acceleration of $1.2 \times 10^{-13}$ km s$^{-2}$. 
At larger negative distances, the scatter in v$_{50}$ dramatically increases, with values lying between $+625$ km s$^{-1}$ and $-100$ km s$^{-1}$. The explanation for this large increase in scatter near the bubble is unknown. %Projection effects may be responsible for this large scatter in v$_{50}$.

North-west of the vertex, the projected gas velocities increase from $-20$ km s$^{-1}$ to $+300$ km s$^{-1}$ at $+7$ kpc NW of the nucleus. 
%with a mean acceleration of $3.7 \times 10^{-13}$ km s$^{-2}$. 
Beyond $\sim 7$ kpc NW of the projected centroid, the scatter in median velocities again significantly increases, with v$_{50}$ between $+250$ km s$^{-1}$ and $+625$ km s$^{-1}$. The origin of this large scatter is also unknown.

Several scenarios were considered to explain the position of the X-ray cavities with respect to the central galaxy and the V-shaped velocity profile. They include
cavities launched at different epochs, and bubbles launched in different directions due to jet precession. None of these scenarios explain all aspects of the data.
%in other systems (Perseus \citep{Dunn2006}, Abell 2390 \citep{Augusto2006}, RBS797 \citep{Gitti2006}, RX J1532.9+3021 \citep{Hlavacek-Larrondo2013}). %However, while multiple outbursts and jet precession could explain the position of the cavities, we do not observe any indication of counter-jets associated with these outbursts, making this theory unlikely.
However, when combining the effects of the central galaxy's peculiar velocity along the LOS
with a lifting model similar to the one proposed for Abell 1835 but with the bubbles being launched nearly in the sky plane rather than nearly along the LOS, we naturally obtain a V-shaped velocity profile.
%However, the lifting model proposed for Abell 1835 appears to work well with a bubble launch orientation nearly in the sky plane rather than nearly along the line of sight. We incorporated the effects of the central galaxy's peculiar velocity along the sight line, which naturally leads a the V-shaped velocity profile.

\begin{figure}
    \centering
    \includegraphics[width=\columnwidth]{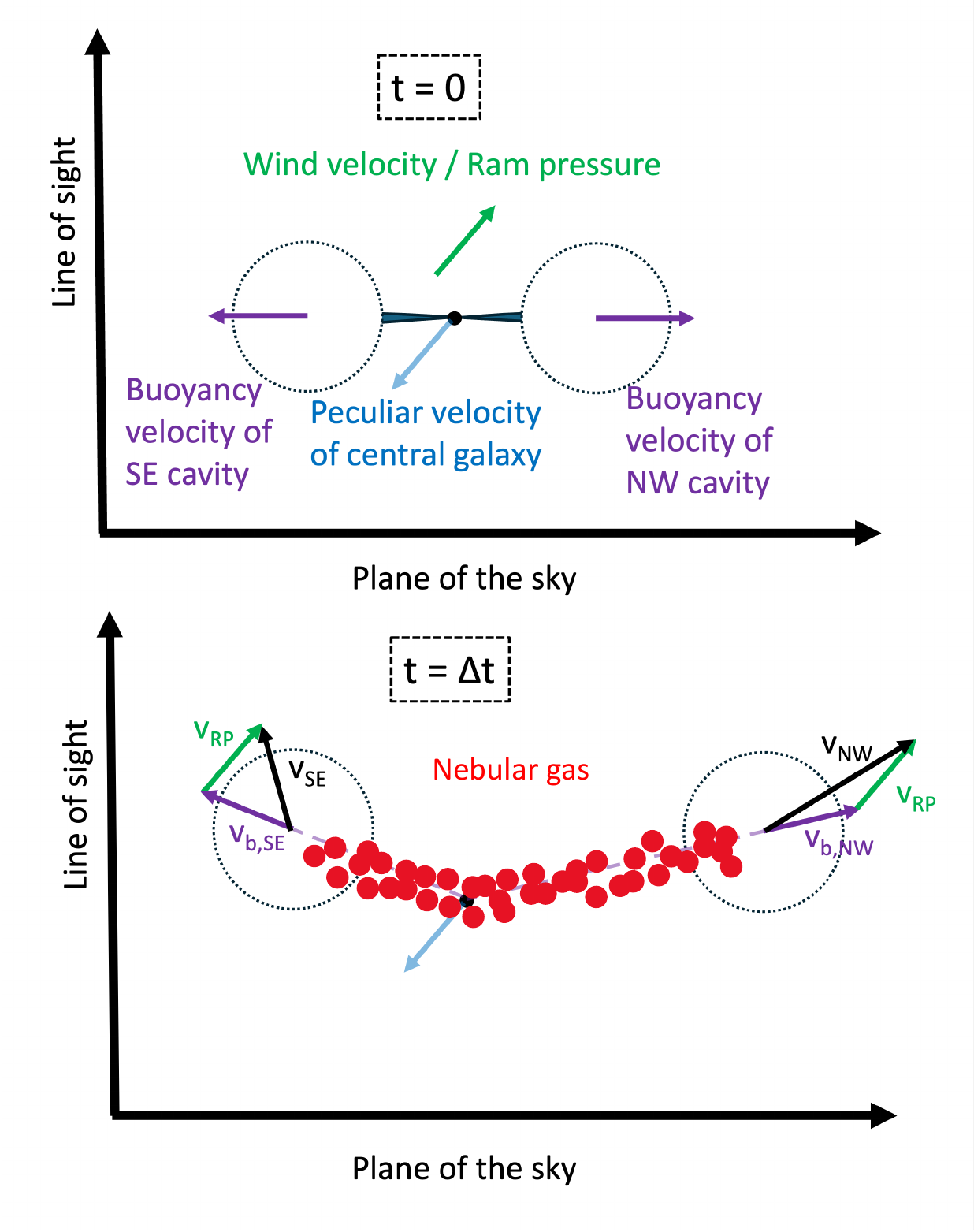}
    \caption{
    Diagram of the motion of the X-ray cavities and their surroundings in a system with significant motion of the central galaxy along our LOS. The top panel shows the system at the time when the AGN-driven jets inflate the X-ray cavities. The model assumes that the cavities are inflated along the plane of the sky. The top panel shows the position of the X-ray cavities at a time $\Delta$t after the cavities were inflated. The cavities are shown as dotted circles.
    The blue arrow shows the galaxy's peculiar velocity. The opposite wind or ram pressure resulting from the central galaxy's motion is shown in green. The purple arrows show the buoyancy velocity of the cavities, which always points away from the central galaxy. In the lower panel the black arrows show the resulting velocity on each cavity. The gas trailing the X-ray cavities is shown in red.}
    \label{fig:PKS0745:diagram:fig22}
\end{figure}

% We suggest that the V-shaped velocity gradient seen in Fig.~\ref{fig:PKS0745:dist_vs_v50:SE_NW} is caused by the peculiar motion of the central galaxy combined with AGN-driven jets and subsequent X-ray cavities inflated roughly in the plane of the sky.
Fig.~\ref{fig:PKS0745:diagram:fig22} shows a diagram of the motion of X-ray cavities, inflated in the plane of the sky, and their associated gas in a system with significant peculiar motion of the central galaxy.
%The cavities' (and their uplifted gas) motion is believed to be dictated by two main mechanisms: 

The top panel describes the dynamics in the core of PKS 0745-191 initially at time $t=0$ when the jet inflates the cavities. The lower panel shows the cavity trajectories governed by buoyancy and ram pressure vectors at a later time $\Delta$t. The red dots represent cool [OII] gas which, we assume, has been advected into the bubbles' paths and drawn outward beneath the bubbles creating the observed velocity gradients.
As seen in the lower panel of Fig.~\ref{fig:PKS0745:diagram:fig22}, both cavities are located behind the central galaxy, which is consistent with our observations. This model can explain how two cavities, which were inflated at the same time and by the same bipolar jet, can both lie behind the central galaxy. Fig.~\ref{fig:PKS0745:diagram:fig22} shows that the SE cavity appears closer to the nucleus in the sky plane than the NW cavity, which is consistent with observations \citep{Sanders2014}.

%\textbf{While this model is consistent with the morphology and the kinematics of the bipolar outflow seen in Fig.~\ref{fig:PKS0745:dist_vs_v50:SE_NW}, applying a similar scenario to Abell 1835 by combining the effects of the central galaxy's peculiar velocity to the buoyant model described in Section~\ref{sec:AGN Energetics} raises some issues. From the kinematics of the nebular gas with respect to the systemic velocity of the system, we infer that the central galaxy is blueshifted with respect to the nebular gas. This results in a redshifted wind acting on the buoyant cavities and their associated gas. From Fig.~\ref{fig:A1835:dist_vs_v50_cavities} in Section~\ref{sec:AGN Energetics}, we inferred that the SE cavity is behind the central while the NW cavity is in front. Combining the different forces acting of the cavities in Abell 1835 as we did for PKS 0745-191 in Fig.~\ref{fig:PKS0745:diagram:fig22}, }\textbf{I'm not sure. Wouldn't it simply result in the jump that we are seeing??}

%\color{red}
Using the LOS nebular gas speed closest to the cavities as an estimate of the cavities' speed, the NW and SE cavities have LOS velocities of $\sim 320$ km s$^{-1}$ and $\sim 220$ km s$^{-1}$, respectively.
The sound speed near the cavities in PKS 0745-191 based on temperatures from \citet{Hogan2017} and \citet{Pulido2018} %We calculate terminal velocities of 780 km s$^{-1}$ and 870 km s$^{-1}$, and
are $800-930$ km s$^{-1}$, depending on the cavities' position. The bubble velocities inferred in this paper for Abell 1835 and PKS 0745-191 lie well below half their sound speeds. Unreasonably large inclination angles would be required to explain this by projection.
%For an angle of 45$^{\circ}$, the velocities are 1.4 times greater than observed. 
A simple explanation consistent with observations is that the cavities are mass loaded and thus are moving more slowly.
%\color{black}

\subsection{Molecular Gas Masses}\label{molecules}

The targets presented here have been observed with ALMA for CO(1-0), CO(2-1), or CO(3-2), or a combination.
%\color{red}
Throughout the result section, we have highlighted similarities in the spatial distributions of nebular and molecular gases. The CO emission is significantly less extended than the [OII] emission. However, in regions where both CO and [OII] emission are detected, 
the regions with brightest CO emission also have the brightest [OII] emission.
%\color{black}
As [OII] is a more sensitive indicator of molecular and atomic gasses, we use [OII] as a proxy for CO and allow for the possibility that the molecular gas is more extended than observed but too faint to detect. 

To do so, we determine the ratio of the [OII] luminosity to M$_{\textrm{mol}}$ within the region where the CO integrated flux is at least 10\% of its peak value. This region corresponds to the areas enclosed by the CO contours in the right panel of Fig.~\ref{fig:flux:A262},~\ref{fig:flux:RXJ0820},~\ref{fig:flux:A1835}, and~\ref{fig:flux:PKS0745}. Using this ratio and the total [OII] luminosity within the FOV, we scale the observed molecular gas mass to estimate the total M$_{\textrm{mol}}$ that may be located in the extended nebula. Tables~\ref{tab:target_properties} and~\ref{tab:OII_lum} list the values used in our calculations. Under the assumption that the ratio of [OII] luminosity to molecular gas mass within an object is constant, we infer molecular gas masses of $1.3 \times 10^{11}$ M$_{\odot}$, $3.0 \times 10^{10}$ M$_{\odot}$, $1.3 \times 10^{9}$ M$_{\odot}$ and $1.1 \times 10^{11}$ M$_{\odot}$ for Abell 1835, PKS 0745-191, Abell 262, and RXJ0820.9+0752, respectively. These values indicate total gas reservoirs that may be several times more massive than inferred from direct CO observations.

\subsection{Molecular \& Nebular Gas Kinematics }\label{AGN Energetics}

Our attempt to compare the nebular and molecular gas velocities is complicated by differences in spatial resolution and velocity reference frame. This study uses the central galaxy's LOS speed as the reference velocity, which represents the best measure of the local rest-frame. However, lacking reliable stellar absorption line velocities, the ALMA studies of Abell 1835 \citep{McNamara2014}, PKS 0745-191 \citep{Russell2016}, and RXJ0820.9+0752 \citep{Vantyghem2018} used the central LOS velocity of either the molecular or nebular gas as the systemic velocity of the system.

%In comparing the kinematics of the nebular and molecular gas, it is worth noting that gas velocities depend on the choice of systemic velocity. Both the molecular and the nebular gas must be using the same systemic velocity. For our four targets, previous papers studying their molecular gas dynamics used emission lines, either from CO or from ionized gas, to determine the system's systemic velocity. We defined our systemic velocities based on stellar velocities in the core of the central galaxies. Our systemic redshifts are all lower than the ones used for the molecular gas dynamics \citep{McNamara2014,Russell2016,Olivares2019,Vantyghem2018}, with 
This resulted in rest-frame velocity differences of approximately 192 km s$^{-1}$, 111 km s$^{-1}$, 64 km s$^{-1}$ and 156 km s$^{-1}$ for Abell 1835, PKS 0745-191, Abell 262 and RXJ0820.9+0752.

\cite{McNamara2014} reported a possible bipolar flow of $10^{10} ~M_\sun $ of molecular gas to the north-west and east of Abell 1835's nucleus. Similar outflows to the ones depicted in Fig.~\ref{fig:A1835:dist_vs_v50_cavities} and~\ref{fig:A1835:dist_vs_v50:x} are observed in CO(1-0) (see Fig.5 in \cite{McNamara2014}). 
General velocity trends observed in the molecular gas maps are also seen in the nebular gas maps. The eastern high velocity gas clump and the NW lower velocity clump shown here were observed in the molecular gas by \cite{McNamara2014}. However, the increase in velocity in the eastern gas clump is more pronounced for the CO(1-0) map than for the [OII] map. Overall, similarities between the kinematics of molecular and nebular gas seem to indicate that the dynamics of the molecular gas are generally consistent with the nebular gas but not in detail. They do not move in lockstep.
PKS0745-191 contains almost $5\times10^9$ M$_{\odot}$ of molecular gas distributed in three filaments, N, SE and SW of the nucleus \citep{Russell2016}. The [OII] emission is much more extended than the CO emission, resulting in only a small area where nebular and molecular gas velocities can be compared. In these regions we find no clear clear relationship between the molecular and nebular gas velocities although they share similar magnitudes. This differentiation would indicate that the nebular gas is not solely composed of the ionized skins of molecular clouds in the churning region.

%For RXJ0820.9+0752 and Abell 262, the median velocities of the nebular gas show similar trends as for the molecular gas \citep{Vantyghem2018,Olivares2019}. 

The molecular gas systemic velocity of RXJ0820.9+0752 \citep{Vantyghem2018} in CO(1-0) and CO(3-2) align closely to the nebular velocity maps. 
%\cite{Vantyghem2018} detected two kinematic components, in both CO(1-0) and CO(3-2), while we detected only one kinematic component in our observations. However, deeper observations of the warm ionized gas may reveal a second kinematic component, allowing for the comparison of molecular and nebular gas velocities for each kinematic component.
The rotating disk of nebular and molecular gas in Abell 262 were easily registered about the zero velocity and their speeds closely align.
%support the idea that the circumnuclear disk around the central AGN contains multiphase gas, with both molecular and nebular gas coexisting. 
However, the nebular gas disk is more extended than the molecular disk and appears dynamically disturbed in its outer regions. %han the molecular gas, with $\sigma_{\textrm{[OII]}}$ up to 193 km s$^{-1}$, while the molecular gas velocity dispersion plateaus around 130 km s$^{-1}$.

Estimates of the gas kinetic energy indicate the fraction of the jet's power dissipated into motion of the surrounding gas. As the gas mass is dominated by the molecular gas, molecular gas dominates the energy budget. Typically only a few percent or less of the jet power is dissipated in lifting the molecular gas \citep{Tamhane2022}. The sizes, approximate locations, and speeds of the molecular gas flows in PKS 0745-191 and Abell 1835 \citep{Tamhane2022} are similar to the outwardly-rising velocity profiles of the nebular gas between the X-ray bubbles found here.  However, the nebular gas with velocities exceeding $300 ~\rm km$ s$^{-1}$ and upward of $600~ \rm km$ s$^{-1}$ are not detected in CO. This very high speed gas involves a small fraction of the mass and thus would have little impact on the kinetic energy budget.

\color{black}
\section{Summary} \label{sec:Summary}

We investigated the dynamics of the warm nebular gas in and around the central galaxies of four clusters, Abell 1835, PKS 0745-191, Abell 262 and RXJ0820.9+0752, using KCWI observations. We explore both the morphology and the dynamics of these four galaxies, focusing on the [OII] nebular emission. We combine our observations with previous multiwavelength studies, including radio and X-ray observations, delving into the dynamics of the multiphase gas in clusters cores.

\begin{itemize}
    \item We detect significant line-of-sight velocity differences, roughly $+150$ km s$^{-1}$, between the nebular gas and the stars in the core of the central galaxy in three of our four targets: Abell 1835, PKS 0745-191 and RXJ0820.9+0752. Comparing the morphology and the kinematics of the central galaxy and its nebular gas, we conclude that the central galaxy has a peculiar velocity with respect to its surroundings. Peculiar motion of central cluster galaxies in the plane of the sky have been inferred to explain spatial offsets between central galaxies and their associated nebular gas as is seen here in RXJ0820.9+0752. The LOS velocity offsets measured here are consistent with inferred velocities in the sky plane required to explain spatial offsets and asymmetries in galaxy morphologies. 
    
    \item Two distinct velocity components are found in the nebular gas surrounding the Abell 1835 and PKS 0745-191 central galaxies.  Higher velocity dispersion churned gas is found towards and behind X-ray cavities surrounded by lower velocity dispersion quiescent gas extending to higher altitudes. Median velocities of the quiescent and churned-up nebulae differ by $\sim 45$ km s$^{-1}$. Energy input from the AGN is likely churning up the gas.

    \item Molecular and nebular gas in central galaxies are spatially correlated  such that the brightest CO and [OII] emission are coincident. However, the detected [OII] gas is much more extended by tens of kpc. Three to six times the molecular gas mass of the bright centrally located molecular clouds may lie in a diffuse state. The velocity structures of the molecular and nebular gas are broadly similar but do not match in detail.
    \item Coherent motions are observed beneath X-ray cavities in Abell 1835 and PKS 0745-191. The velocites are consistent with gas being lifted behind the X-ray bubbles whose kinematics are governed by buoyancy and ram pressure produced by the peculiar motion of the central galaxy with respect to the surrounding atmosphere. The speeds of the advected gas nearest to the X-ray cavities imply cavity speeds of several hundred kilometers per second, but below the buoyancy speeds of empty bubbles. The slower speeds imply mass loading of gas lifted by the cavities. 
    %These speeds are lower than those expected from X-ray observations, indicating that factors, such as non-empty cavities, must be slowing down the cavities.   
    \item  High velocity dispersion nebular gas (Fig.~\ref{fig:sig_radio:A262}) is observed along the radio jets in the central galaxy of Abell 262, indicating a strong interaction between the jets and the surrounding gas. The jets are emerging along the angular momentum axis of a disturbed, possibly nascent circumnuclear disk.  
    
\end{itemize}

\begin{acknowledgments}
The data presented herein were obtained at the W. M. Keck Observatory, which is operated as a scientific partnership among the California Institute of Technology, the University of California, and the National Aeronautics and Space Administration. The observatory was made possible by the financial support of the W. M. Keck Foundation. We wish to recognize and acknowledge the very significant cultural role and reverence that the summit of Maunakea has always had within the indigenous Hawaiian community. We are most fortunate to have the opportunity to conduct observations from this mountain. We thank Prathamesh Tamhane and Adrian Vantyghem for generously providing us with CO detection maps. We thank Tom Rose for the helpful discussions. B.R.M acknowledges support from the Natural Sciences and Engineering Council of Canada and the Canadian Space Agency Space Science Enhancement Program. A.~L.~C. acknowledges support from the Ingrid and Joseph W. Hibben endowed chair at UC San Diego. Finally, we thank the anonymous referee who graciously took the time to review our paper and give us constructive comments which improved our paper.
\end{acknowledgments}

\vspace{5mm}

\software{Astropy \citep{Astropy1,Astropy2,Astropy3},  Python \citep{Python3}, Numpy \citep{Numpy1,Numpy2}, Matplotlib \citep{Matplotlib}, Scipy \citep{SciPy}, 
          }
          
\appendix

\section{Systemic Velocity Measurements}

\begin{figure}
    \centering
    \includegraphics[width=\columnwidth]{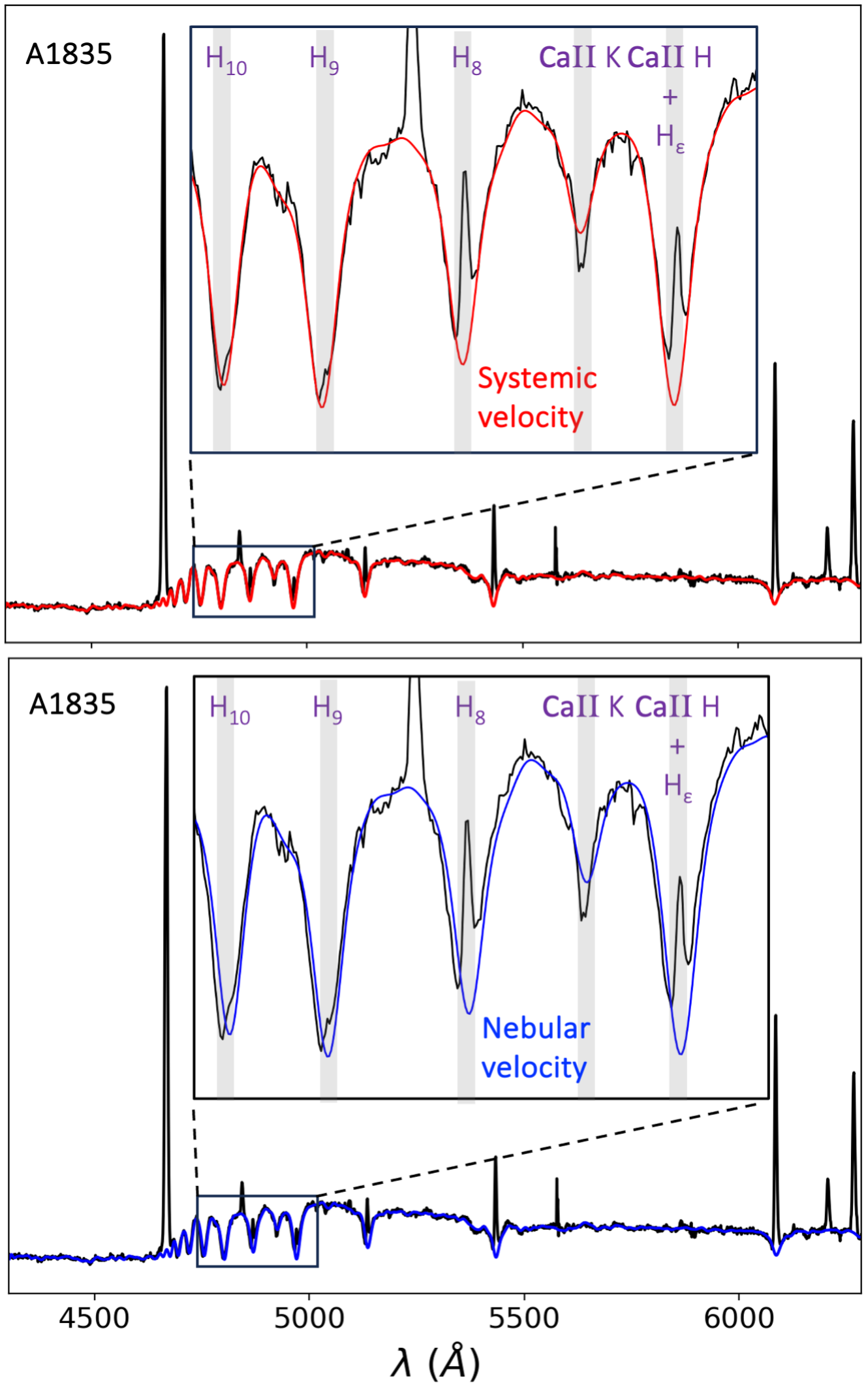}
    \caption{Fit of the spatially-integrated stellar continuum at the core of the central galaxy of Abell 1835. The top panel shows the spectrum, in black, and the stellar continuum fit, in red, from which the systemic velocity of the system was obtained. The bottom panel shows the same spectrum and, in blue, the stellar fit shifted to the flux-weighted median v$_{50}$. The fitting method used is described in Section~\ref{sec:Observations and data reduction: Data analysis: Fitting}.}
    \label{fig:appendix:A1835:zsys}
\end{figure}

\begin{figure}
    \centering
    \includegraphics[width=\columnwidth]{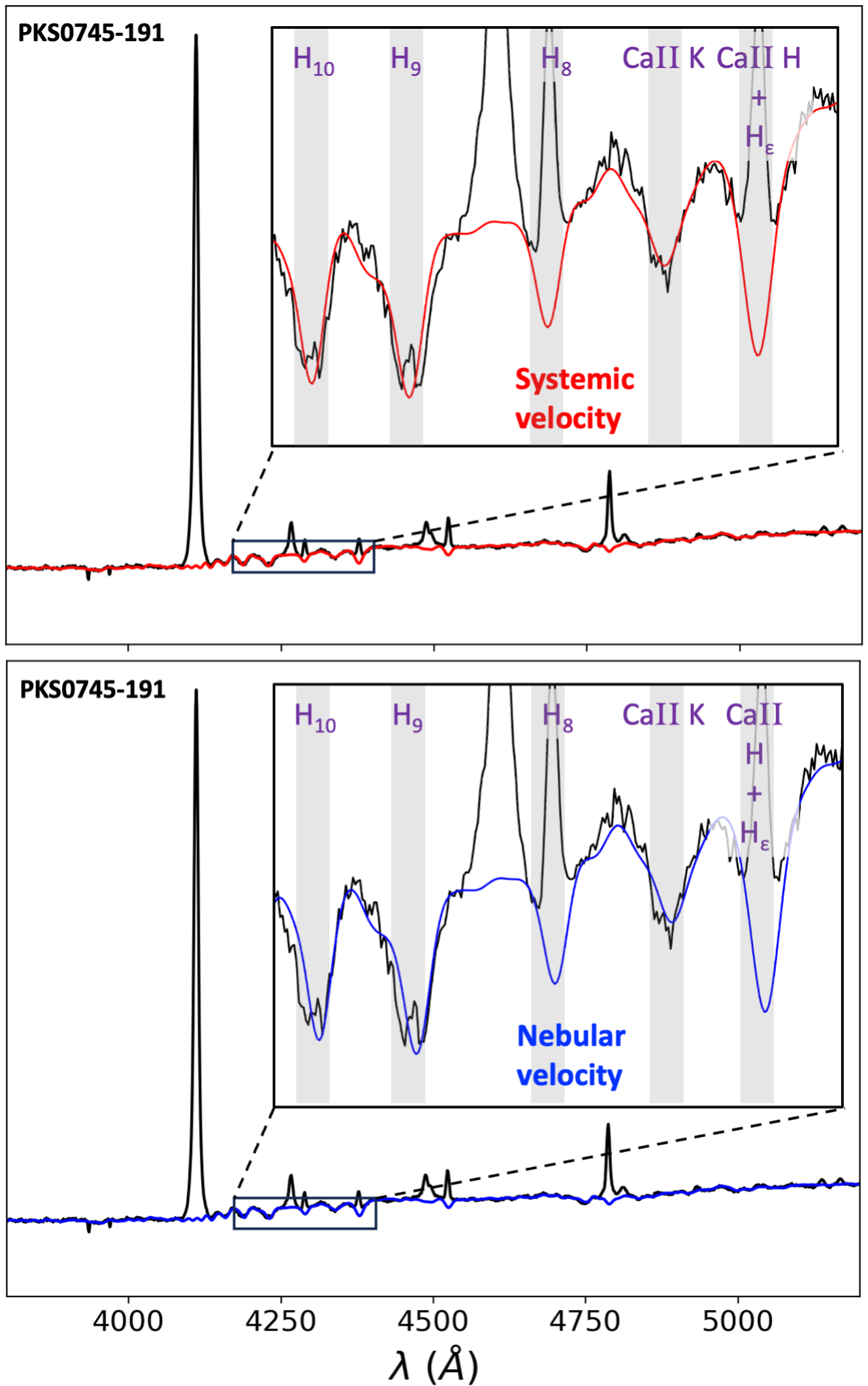}
    \caption{Fit of the spatially-integrated stellar continuum in the core of the central galaxy of PKS 0745-191. As in Fig.~\ref{fig:appendix:A1835:zsys}, the top panel shows the stellar fit at the systemic velocity and the bottom panel shows the same fit shifted to the flux-weighted median v$_{50}$.}
    \label{fig:appendix:PKS0745:zsys}
\end{figure}

\begin{figure}
    \centering
    \includegraphics[width=\columnwidth]{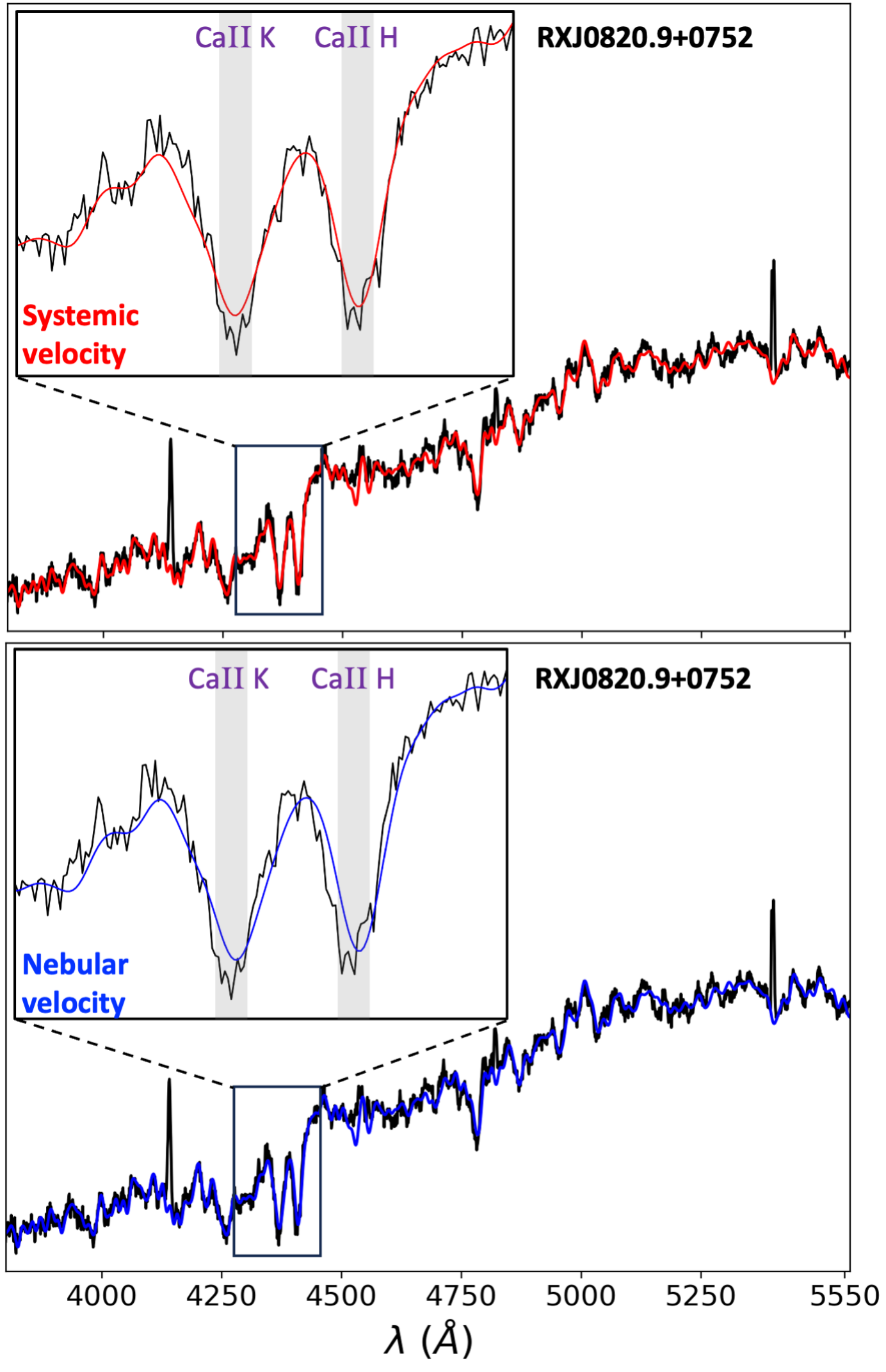}
    \caption{Fit of the spatially-integrated stellar continuum in the core of the central galaxy of RXJ0820.9+0752. As in Fig.~\ref{fig:appendix:A1835:zsys}, the top panel shows the stellar fit at the systemic velocity and the bottom panel shows the same fit shifted to the flux-weighted median v$_{50}$.}
    \label{fig:appendix:RXJ0820:zsys}
\end{figure}

The measurement of systemic velocities followed the procedure outlined in Section~\ref{sec:Observations and data reduction: Data analysis: Fitting}. 
In our approach, we employ stellar absorption lines and stellar population models to derive the stellar velocity of the systems, operating under the assumption that these stars are gravitationally bound to the central galaxy, with the central stellar velocity reflecting the velocity of the central galaxy. Spaxels surrounding the peak of the stellar emission are spatially-integrated, yielding a high quality central spectrum. Then, masking the emission lines, the stellar continuum is fitted using a combination of stellar templates. The mean stellar velocity in the core of the galaxy is obtained from this fit and used as the systemic velocity. Heliocentric corrections are then applied to these velocities to obtain the corrected redshifted listed in column 4 of Table~\ref{tab:target_properties}. The uncorrected systemic redshifts are 0.25131, 0.10245, 0.01606, and 0.11033 for Abell 1835, PKS 0745-191, Abell 262 and RXJ0820.9+0752, respectively.

The top panels of Fig.~\ref{fig:appendix:A1835:zsys},~\ref{fig:appendix:PKS0745:zsys} and~\ref{fig:appendix:RXJ0820:zsys} show the central stellar fits obtained for Abell 1835, PKS 0745-191 and RXJ0820.9+0752, respectively. For Abell 1835 and PKS 0745-191, the stellar continuum shows strong Balmer absorption lines which are caused by a strong intermediate age stellar population. This raised concern that this population, if kinematically distinct from the older stellar population, may bias our systemic velocity and not properly represent the velocity of the central galaxy. To verify that the older stellar population, which is expected to be strongly bounded to the core of the central galaxy, is correctly fitted by our stellar model, we looked at the goodness of the fit for the Ca H \& K lines, which originate primarily from old stars. The figures embedded in the upper panels of Fig.~\ref{fig:appendix:A1835:zsys},~\ref{fig:appendix:PKS0745:zsys} and~\ref{fig:appendix:RXJ0820:zsys} show that the Ca H \& K lines are correctly fitted by the model, showing that the intermediate age stellar population is not biasing our systemic velocities.

Fig.~\ref{fig:appendix:A1835:zsys},~\ref{fig:appendix:PKS0745:zsys} and~\ref{fig:appendix:RXJ0820:zsys} show velocity differences of $\sim 150$ km s$^{-1}$ between the central nebular gas and the systemic velocities  in three out of our four target systems, as discussed in Section~\ref{sec:Velocity_galaxies}. The lower panels of 
Fig.~\ref{fig:appendix:A1835:zsys},~\ref{fig:appendix:PKS0745:zsys} and~\ref{fig:appendix:RXJ0820:zsys} show the stellar continuum models shifted to the central nebular gas velocity. In all instances, the shifted stellar models do not fit the observed spectrum as well as the non-shifted one, thus confirming the reality of the peculiar nebular velocities.

\section{Notes on uncertainties}\label{appendix:Uncertainties}

Throughout this paper, for the emission line maps, S/N cuts for the [OII] doublet are used as the main criterion to determine which spaxels are included in the flux and kinematic maps as described in Section~\ref{sec:Observations and data reduction: Data analysis: Fitting}. S/N cuts are also used when deciding between the one or two kinematic component fits and as a threshold for the spaxels to include in the two component kinematic maps (Fig.~\ref{fig:kinematics:A1835:Components} and Fig.~\ref{fig:kinematics:PKS0745:Components}). 

The precision of the fits obtained can also be assessed from the errors on v$_{50}$ and $\sigma$, both measured for the one and two kinematic component fits. Generally, the regions with higher flux have smaller uncertainties, with errors of only a few km s$^{-1}$. The uncertainties in v$_{50}$ and $\sigma$ increase away from the brightest spaxels, with the largest uncertainties being around the edges of the detected [OII] emission. This increase in uncertainties can be seen on the edges of the v$_{50}$ and $\sigma$ maps for Fig.~\ref{fig:kinematics:A1835:Components} and~\ref{fig:kinematics:PKS0745:Components}, where the velocity variations between adjacent spaxels is larger. While spaxels on the edges of the detected [OII] emission can have larger uncertainties, reaching values of $20-50$ km s$^{-1}$, consistent motion seen over multiple neighbouring spaxels can be trusted and is very unlikely to be caused by uncertainties in the velocities. For example, the higher $\sigma$ region around the origin in the $\sigma$ map of RXJ0820.9+0752 (top right panel of Fig.~\ref{fig:kinematics:RXJ0820}) is believed to be physical and not caused by an increase in uncertainties on the edges of the nebula. However, the individual $\sigma$ values for the dark red spaxels in this region, $\sigma > 170$ km s$^{-1}$, are more uncertain, with errors between $16-47$ km s$^{-1}$.

%However, the individual $\sigma > 175$ km s$^{-1}$ spaxels have uncertainties in $\sigma$

\section{Additional comments on kinematics}

\subsection{v$_{02}$ and v$_{98}$}\label{appendix:v02_v98}

\begin{figure}
    \centering
    \includegraphics[width=\columnwidth]{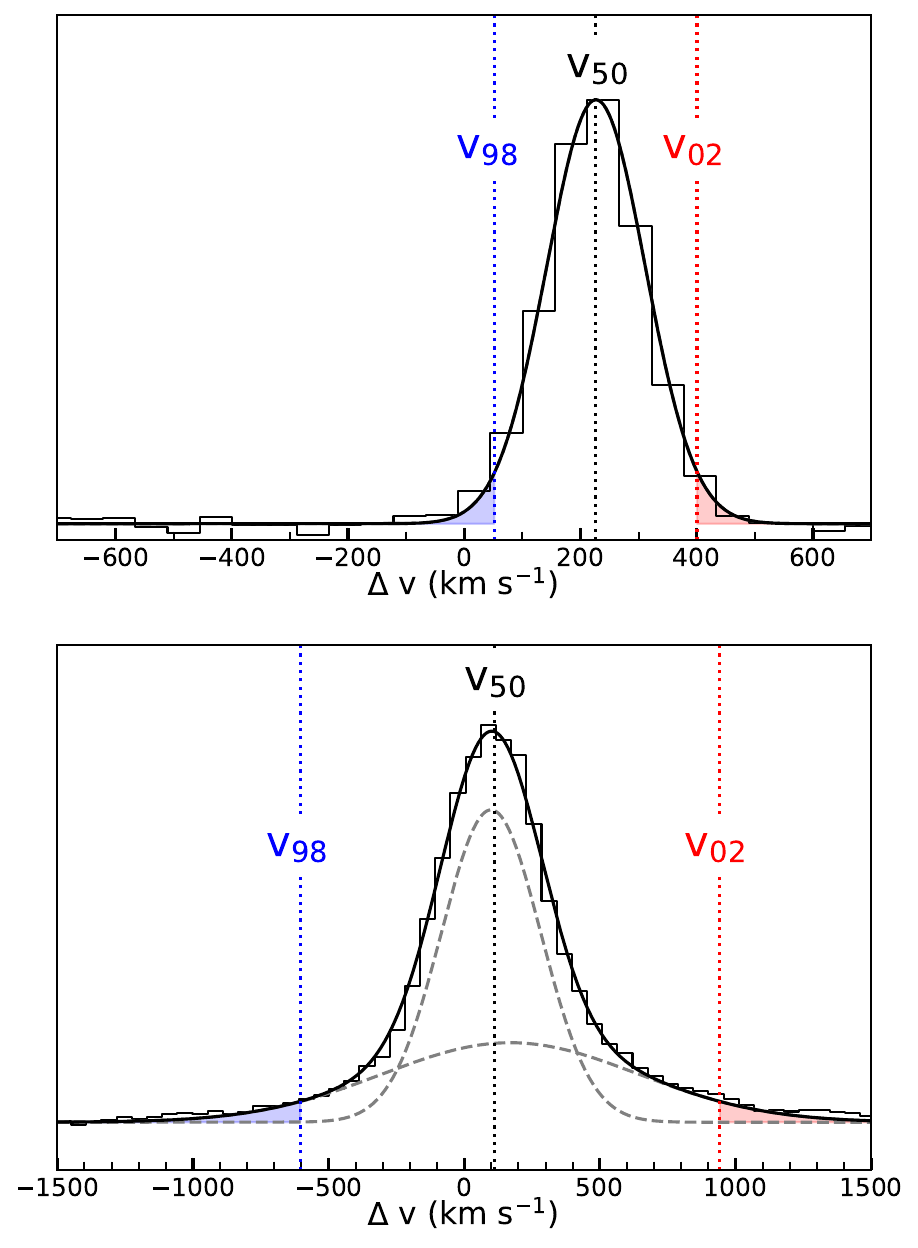}
    \caption{Emission line fits with one kinematic component (top panel) and with two kinematic components (lower panel) to visualize the significance of the v$_{02}$ and v$_{98}$ parameters from some of our kinematic maps (e.g. Fig.~\ref{fig:kinematics:RXJ0820} and~\ref{fig:kinematics:A1835}). The histograms show the observed emission lines, the black solid lines show the total emission line fit and the dashed lines in the lower panel show each of the fitted kinematic components. The vertical dashed lines show the position of v$_{02}$, v$_{50}$ and v$_{98}$ and the shaded red and blue regions show the regions under the fit which are in the top 2\% redshifted or blueshifted part of the cumulative velocity functions. The fluxes are plotted as a function of their offset velocities ($\Delta$v) where the zero velocity corresponds to the systemic velocity of the system. The example emission lines are for a single emission line, as opposed to the [OII] doublet which is comprised of two lines.}
    \label{fig:v02_v98}
\end{figure}

When visualizing the kinematics for our four targets as in Fig.~\ref{fig:kinematics:A262},~\ref{fig:kinematics:RXJ0820},~\ref{fig:kinematics:A1835} and~\ref{fig:kinematics:PKS0745}, we map four velocities for the total [OII] doublet emission line: the median velocity v$_{50}$, the velocity dispersion $\sigma$, the 2\% most-redshifted velocity v$_{02}$ and the 2\% most-blueshifted velocity v$_{98}$. Fig.~\ref{fig:v02_v98} shows examples of emission line fits for a one kinematic component fit (on the top panel) and for a two kinematic component fit (on the bottom panel) to visualize the meaning of the v$_{02}$ and v$_{98}$ parameters. The shaded blue (or red) area under the curve corresponds to 2\% of the total area under the fitted curve. Only 2\% of the fitted emission line's flux has a velocity greater than v$_{02}$. Similarly, only 2\% of the fitted emission line's flux has a velocity lower than v$_{98}$. As seen in the lower panel of Fig.~\ref{fig:v02_v98}, v$_{02}$ and v$_{98}$ are especially useful for systems with two kinematic components, where they highlight asymmetries in emission lines. v$_{02}$ and v$_{98}$ maps are critical for identifying highly blueshifted and redshifted tails in nebular emission.

\subsection{Abell 1835}\label{appendix:A1835}

\begin{figure*}
    \centering
    \includegraphics[width=\textwidth]{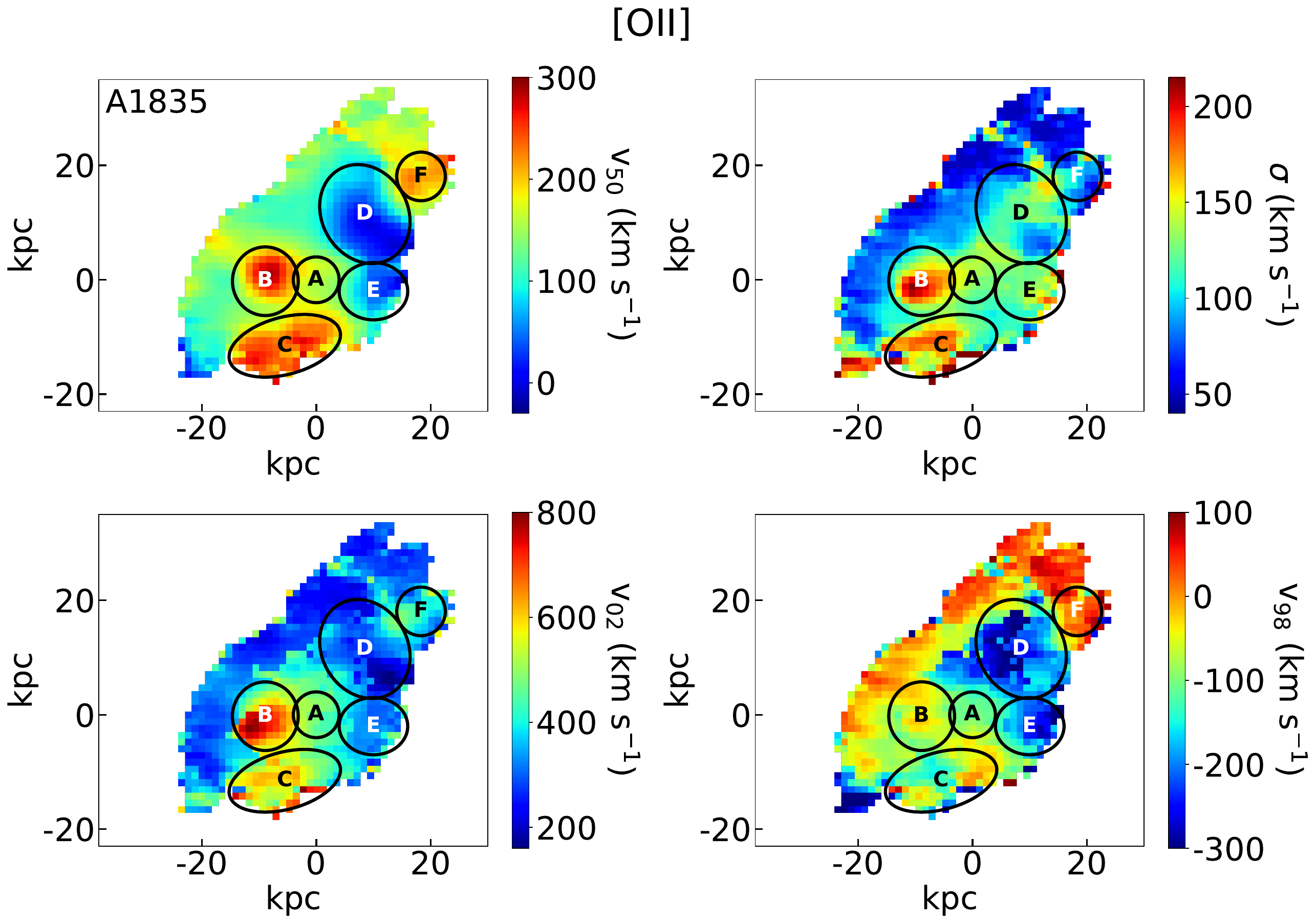}
    \caption{Kinematics of the [OII]3726,9$\lambda\lambda$ emission line doublet in Abell A1835 as in Fig.~\ref{fig:kinematics:A1835} with regions of interest listed in Table~\ref{tab:A1835:regions} identified by solid ellipses.}
    \label{fig:kinematics:A1835:regions}
\end{figure*}

\begin{deluxetable*}{ccccccccc}
        \tablecaption{Abell 1835 regions}
    \tabletypesize{\scriptsize}
    \tablewidth{0pt} 
    \tablehead{ \colhead{Region}&\colhead{v$_{50}$}&\colhead{$\sigma$}&\colhead{v$_{02}$}&\colhead{v$_{98}$}&\colhead{Fraction L$_{\rm[OII]\rm}$}&\colhead{Dist}&\colhead{$\theta$}&\colhead{Area}\\ \colhead{} & \colhead{(km s$^{-1}$)} & \colhead{(km s$^{-1}$)} & \colhead{(km s$^{-1}$)} & \colhead{(km s$^{-1}$)} & \colhead{(\%)} & \colhead{(kpc)} & \colhead{($^{\circ}$)} & \colhead{(kpc$^2$)} }
\label{tab:A1835:regions}
\colnumbers
\startdata 
A & 165 & 136 & 482 & $-102$ & 21.08 & 0.0 & 0 & 52 \\
B & 227 & 164 & 633 & $-61$ & 12.61 & 8.3 & 0 & 83 \\
C & 224 & 150 & 536 & $-83$ & 2.82 & 13.2 & 297 & 122 \\
D & 51 & 111 & 312 & $-236$ & 15.73 & 13.5 & 52 & 201 \\
E & 87 & 125 & 352 & $-175$ & 5.51 & 10.7 & 264 & 88 \\
F & 211 & 100 & 416 & 5 & 0.42 & 25.1 & 45 & 39 \\
\enddata
\tablecomments{\\(1) Region identified in Fig.~\ref{fig:kinematics:A1835:regions}.\\
(2) [OII] flux weighted mean of the median velocity within the region.\\
(3) [OII] flux weighted mean of the velocity dispersion within the region.\\
(4) [OII] flux weighted mean of the v$_{02}$ within the region.\\
(5) [OII] flux weighted mean of the v$_{98}$ within the region.\\
(6) Fraction of the total [OII] luminosity observed from the spaxels included in the region.\\
(7) Distance between the central point of the region and the nucleus.\\
(8) Average angle between the center of the galaxy and the region. The angle is 0$^{\circ}$ when pointing east and increases clockwise.\\
(9) Area of the region.\\
}
\end{deluxetable*}

\begin{figure}
    \centering
    \includegraphics[width=\columnwidth]{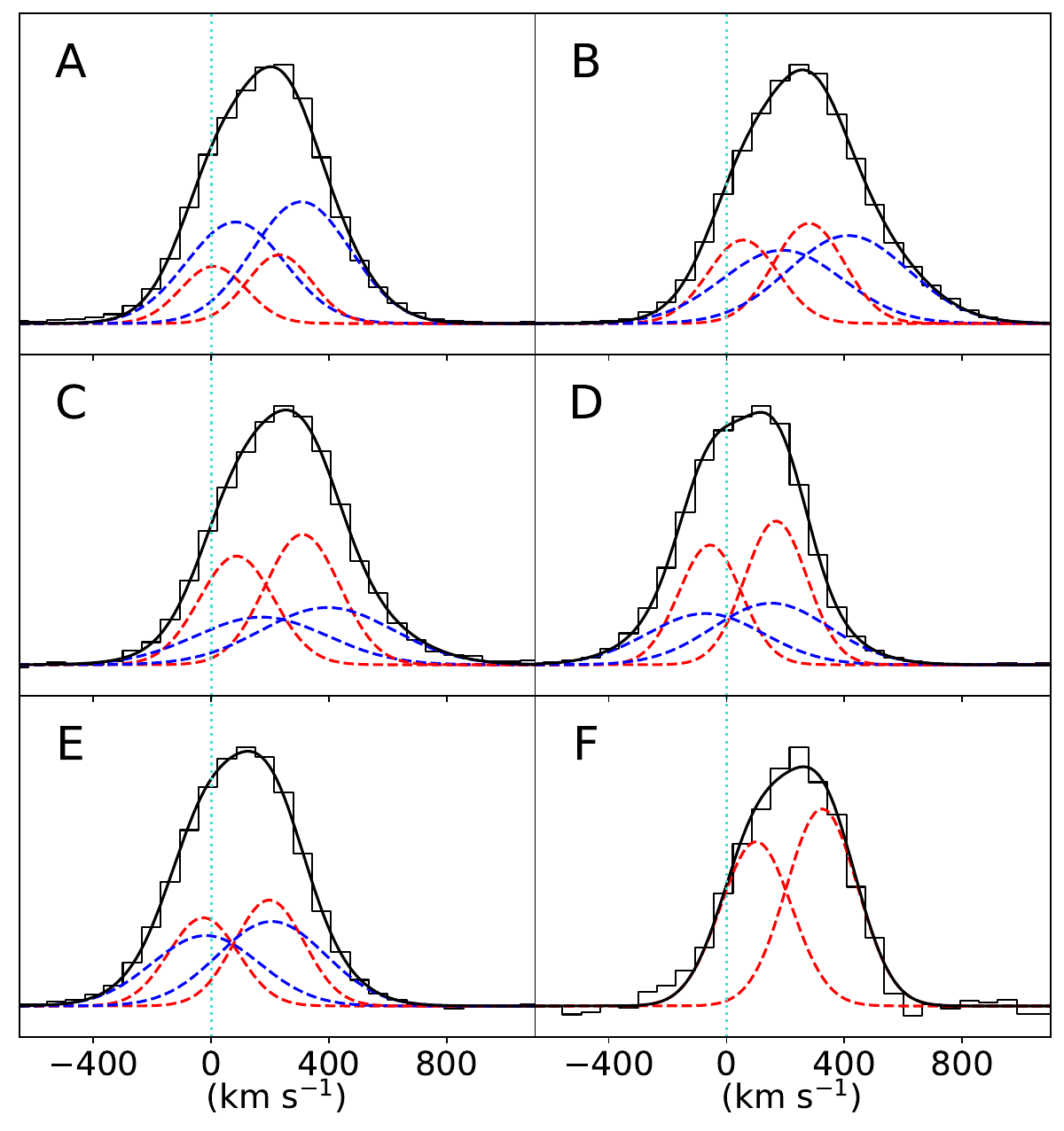}
    \caption{Fits of the [OII] emission line from the spatially-integrated spectrum of each of the six regions identified in Fig.~\ref{fig:kinematics:A1835:regions} and Table~\ref{tab:A1835:regions}. The black histogram shows the observed spectrum, the red and blue dashed lines show the first and second Gaussian kinematic components of the fit, respectively. The black solid line shows the resulting total fit of the emission line. The turquoise dotted line identifies the systemic velocity used for this target.}
    \label{fig:kinematics:A1835:regions:spectra}
\end{figure}

Fig.~\ref{fig:kinematics:A1835:regions} shows the same panels as in Fig.~\ref{fig:kinematics:A1835} using different color scale bars to emphasize differences. Six regions, A-F, are identified on the velocity maps as regions of interest. Various properties of these regions are listed in Table~\ref{tab:A1835:regions}, including position, size, flux-weighted mean velocities (see equation~\ref{eq:flux-weighted_mean}), and relative contribution to [OII] luminosity. 

To get a clear picture of the kinematics in each region, Fig.~\ref{fig:kinematics:A1835:regions:spectra} shows the [OII] line doublet and the best-fit model for each of the regions identified in Fig.~\ref{fig:kinematics:A1835:regions}. It is worth noting that since [OII]3726,9 is a doublet, each component is comprised of two Gaussians, one for each doublet line. The higher signal-to-noise in the spatially-integrated spectra can reveal a second kinematic component within a region which is not always robustly detected in the individual spaxels.

Region A is a circle with a radius of $1 \arcsec$ centered on the nucleus. The ionized gas at the very center of the galaxy does not share the same velocity as its projected stellar population. The central gas is overall redshifted by 165 km s$^{-1}$ with respect to the systemic velocity.  

The nebular gas in region B has the largest v$_{50}$, $\sigma$ and v$_{02}$ values observed in this system. Region B is located just north of the edge of the SE cavity, although it is not located along the axis of the cavities. The high velocity dispersion, $\geq 150$ km s$^{-1}$, indicates that the gas is disturbed. Fig.~\ref{fig:kinematics:A1835:regions:spectra} shows the two component fit for region B. The second component (shown in blue) is significantly more redshifted ($\Delta \textrm{v}_{50} > +130$ km s$^{-1}$) and broader ($\Delta \sigma \sim 95$ km s$^{-1}$) than the first component (shown in red). 

Region E is located on the opposite side of the nucleus, roughly at the same distance from the center as region B. While the gas in region B is redshifted with respect to the gas in the nucleus, the gas in region E is blueshifted. The spectral fits of regions A, B and E in Fig.~\ref{fig:kinematics:A1835:regions:spectra} show that both kinematic components are redshifted from region A to region B, and conversely are blueshifted from region A to region E. The second, broader kinematic component has the most significant velocity differences, of order of hundreds of km s$^{-1}$. The dynamics of the nebular gas in regions A, B and E seem to indicate motion of the nebular gas along the E-W axis passing through the center.

Regions C and D are another interesting pair of regions which spatially overlap with the SE and NW X-ray cavities, respectively. Both regions are roughly 13 kpc from the nucleus. The nebular gas in region C is redshifted with respect to the central nebular gas, while the nebular gas in region D is blueshifted. The spectral fits of the [OII] doublet for regions A, C and D (see Fig.~\ref{fig:kinematics:A1835:regions:spectra}) show significant shifts in velocities, especially for the second component with v$_{50}$ differences of more than 150 km s$^{-1}$. Additionally, there is substantial broadening of the second component in region C compared to region A, and slight broadening of the second component in region D compared to region A. The spatial overlap with X-ray cavities and the substantial velocity shifts and line broadenings are consistent with the cavities playing an important role in the nebular gas dynamics of regions C and D. 

The nebular gas in region F has higher v$_{50}$ and slightly higher $\sigma$ than the surrounding gas.  While this region  overlaps partially with one of the projected galaxies, it is not expected to significantly affect the gas dynamics in this region, as the velocity difference of almost 1000 km s$^{-1}$ between the projected galaxy and the nebular gas argues against any interaction between them. Region F also overlaps with the northern edge of the NW cavity. However, as the edges of X-ray cavities are particularly challenging to precisely define, we cannot confidently ascertain a relation between the NW cavity and this region.

\subsection{PKS 0745-191}\label{appendix:PKS0745}

\begin{figure*}
    \centering
    \includegraphics[width=\textwidth]{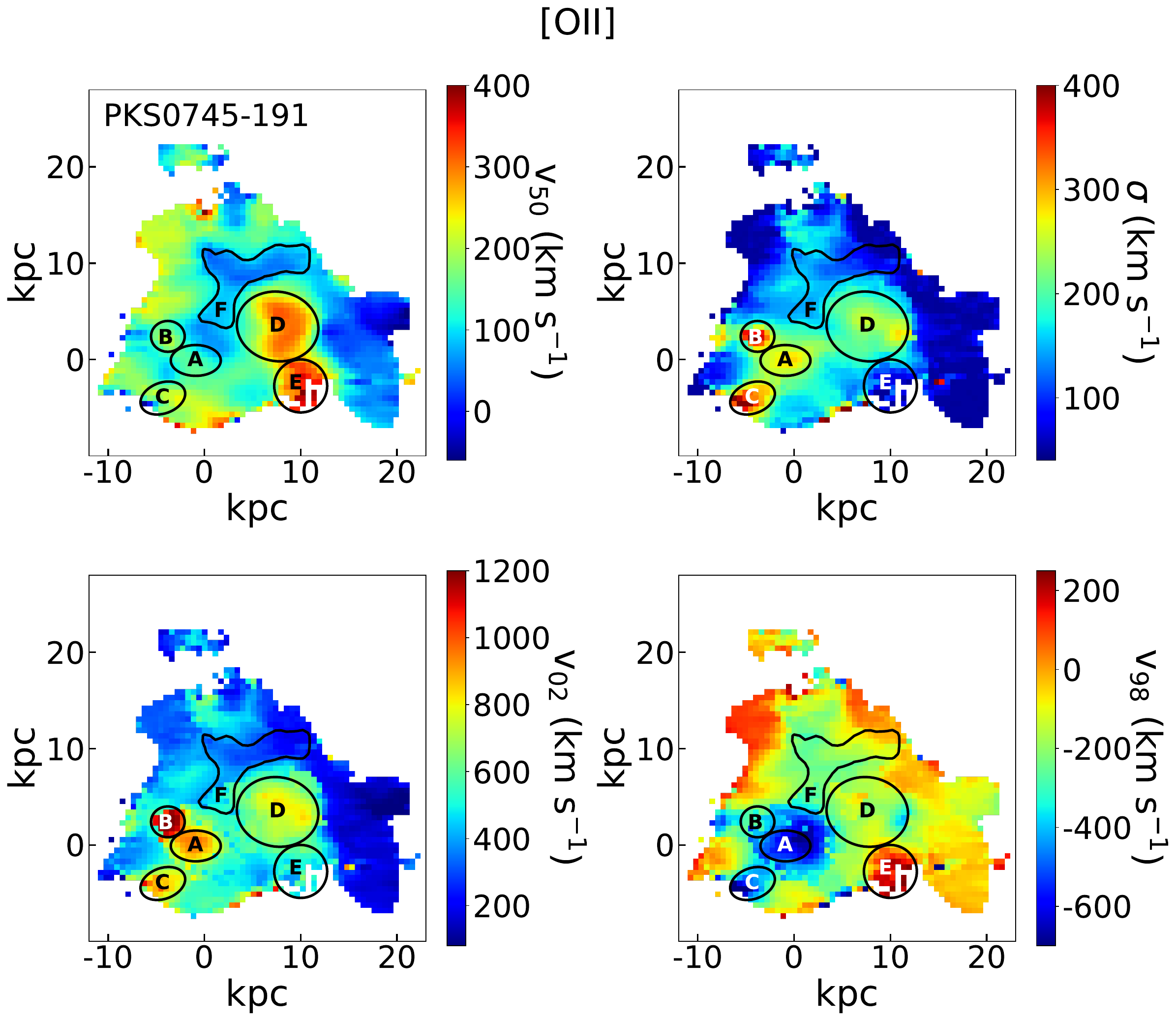}
    \caption{Kinematics of the [OII]3726,9$\lambda\lambda$ emission line doublet in PKS 0745-191 as in Fig.~\ref{fig:kinematics:PKS0745} with regions of interest mentioned in Table~\ref{tab:PKS0745:regions} identified by solid ellipses.}
    \label{fig:kinematics:PKS0745:regions}
\end{figure*}

\begin{deluxetable*}{ccccccccc}
        \tablecaption{PKS 0745-191 regions}
    \tabletypesize{\scriptsize}
    \tablewidth{0pt} 
    \tablehead{ \colhead{Region}&\colhead{v$_{50}$}&\colhead{$\sigma$}&\colhead{v$_{02}$}&\colhead{v$_{98}$}&\colhead{Fraction L$_{\rm[OII]\rm}$}&\colhead{Dist}&\colhead{$\theta$}&\colhead{Area}\\ \colhead{} & \colhead{(km s$^{-1}$)} & \colhead{(km s$^{-1}$)} & \colhead{(km s$^{-1}$)} & \colhead{(km s$^{-1}$)} & \colhead{(\%)} & \colhead{(kpc)} & \colhead{($^{\circ}$)} & \colhead{(kpc$^2$)} }
\label{tab:PKS0745:regions}
\colnumbers
\startdata 
A & 130 & 267 & 889 & $-597$ & 22.81 & 1.1 & 0 & 11 \\
B & 166 & 309 & 1145 & $-350$ & 1.78 & 4.3 & 148 & 6\\
C & 189 & 304 & 814 & $-439$ & 0.63 & 5.6 & 315 & 9 \\
D & 246 & 222 & 718 & $-187$ & 9.69 & 8.7 & 23 & 39 \\
E & 304 & 99 & 509 & 99 & 0.32 & 9.9 & 257 & 16 \\
F & 81 & 146 & 429 & $-260$ & 9.56 & 9.7 & 69 & 53 \\
\enddata
\tablecomments{\\(1) Region identified in Fig.~\ref{fig:kinematics:PKS0745:regions}.\\
(2) [OII] flux weighted mean of the median velocity within the region.\\
(3) [OII] flux weighted mean of the velocity dispersion within the region.\\
(4) [OII] flux weighted mean of the v$_{02}$ within the region.\\
(5) [OII] flux weighted mean of the v$_{98}$ within the region.\\
(6) Fraction of the total [OII] luminosity observed from the spaxels included in the region.\\
(7) Distance between the central point of the region and the nucleus.\\
(8) Average angle between the center of the galaxy and the region.\\
(9) Area of the region.\\
}
\end{deluxetable*}

\begin{figure}
    \centering
    \includegraphics[width=\linewidth]{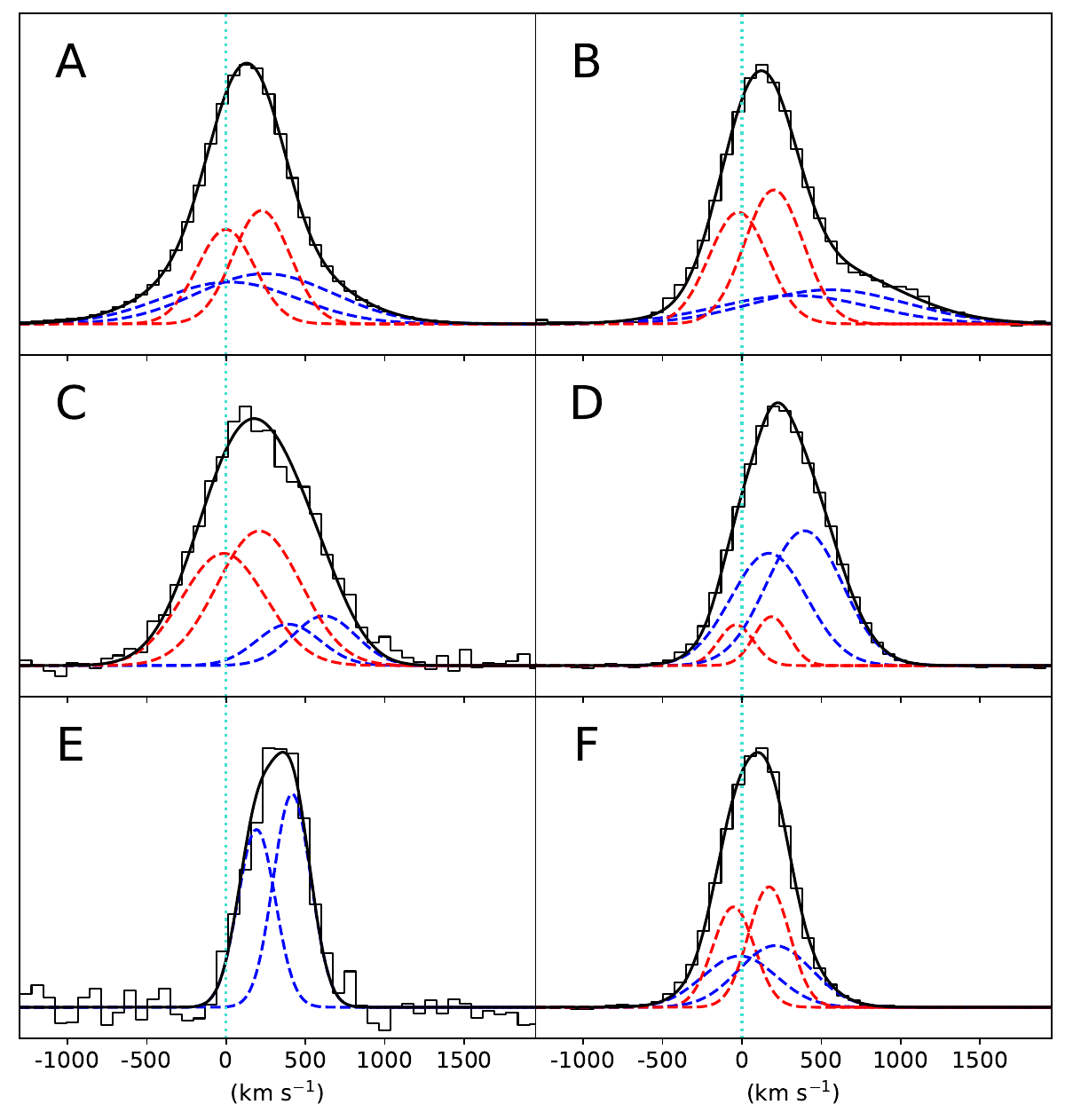}
    \caption{Spectra for the summed regions identified in Fig.~\ref{fig:kinematics:PKS0745:regions} and Table~\ref{tab:PKS0745:regions}. The turquoise dotted line identifies the systemic velocity used for this target.}
    \label{fig:kinematics:PKS0745:regions:spectra}
\end{figure}

Fig.~\ref{fig:kinematics:PKS0745:regions} defines six regions of interest in the LOS kinematics map, shown as black solid ellipses. Otherwise, Fig.~\ref{fig:kinematics:PKS0745:regions} is identical to Fig.~\ref{fig:kinematics:PKS0745}. Following the same method as in Appendix~\ref{appendix:A1835}, Table~\ref{tab:PKS0745:regions} lists some properties of the highlighted regions and Fig.~\ref{fig:kinematics:PKS0745:regions:spectra} shows the best-fit model of the spatially-integrated spectrum around the [OII] doublet for each region. 

Region A is the central region with high velocity dispersion gas and large v$_{02}$ and v$_{98}$ values. This region includes the center of the galaxy, defined using the stellar flux. The center of region A is offset $\sim 1$ kpc east of the nucleus. The nebular gas in this central region is redshifted by +130 km s$^{-1}$. Fig.~\ref{fig:kinematics:PKS0745:regions:spectra} shows that the gas associated with the second component of the fit is very disturbed, with a velocity dispersion $\sim 420$ km s$^{-1}$. Even the narrow component of the fit for region A is quite broad, with $\sigma = 162$ km s$^{-1}$. The most-redshifted gas velocities of almost $+900$ km s$^{-1}$ and the most-blueshifted velocities of almost $-600$ km s$^{-1}$ result from the extreme velocities attained by the very disturbed central gas.

Region B has the nebular gas with the largest total velocity dispersion and the most-redshifted velocities observed in this galaxy. Fig.~\ref{fig:kinematics:PKS0745:regions:spectra} shows that the large $\sigma$ and v$_{02}$ values are caused by the broad and redshifted second kinematic component. The nebular gas modelled by the second kinematic component, which emits 40\% of the total flux in region B, has v$_{50}$ and $\sigma$ values greater than 450 km s$^{-1}$. These dynamics are consistent with a very powerful outflow.

The gas in region C has $\sigma$ values similar to the gas in region B. However, the most-redshifted gas in region C is not as redshifted as in region B, with flux-weighted mean v$_{02}$ of $+814$ km s$^{-1}$ and $+1145$ km s$^{-1}$, respectively. Fig.~\ref{fig:kinematics:PKS0745:regions:spectra} shows similarities in the integrated spectra of regions B and C. The gas associated with the first component have similar kinematics. In both regions, the second kinematic component is significantly redshifted, with v$_{50} \sim +518$ km s$^{-1}$ for region C. The main difference in the gas kinematics of regions B and C is the velocity dispersion of the second component. The gas associated with the second component in region C is disturbed, with $\sigma = 183$ km s$^{-1}$, albeit significantly less than the high velocity gas in region B where the velocity dispersion is more that 2.5 times larger. The large gas velocities of the second kinematic component in region C are consistent with an inflow or outflow of gas. The significant difference in the velocity dispersion of the high velocity gas in regions B and C hint to different mechanisms responsible for the gas dynamics in each region, with a more gentle process occurring in region C. Region C lies within the SE cavity, which may be disturbing surrounding gas and causing the several hundred km s$^{-1}$ gas motions.

Regions D and E encompass the large v$_{50}$ filament as seen in the top left panel of Fig.~\ref{fig:kinematics:PKS0745:regions}. The $\sigma$, v$_{02}$ and v$_{98}$ maps in Fig.~\ref{fig:kinematics:PKS0745:regions} show clear differences between the upper and lower regions of the $\sim +280$ km s$^{-1}$ filament. Region D corresponds to the northern, more disturbed, part of the filament while region E represents the southern more quiescent part.

Region D encompasses the northernmost section of the redshifted filament. The gas in region D is quite disturbed, with $\sigma > 200$ km s$^{-1}$. Most of the gas in region D is modelled by the second component of the fit as shown in Fig.~\ref{fig:kinematics:PKS0745:regions:spectra}. The majority of the nebular gas in region D is redshifted by $\sim +300$ km s$^{-1}$ and significantly disturbed. Part of region D lies within and on the edge of the NW cavity. The large v$_{50}$ filament lies south of the NW cavity. It does not align with the nucleus-cavity axis where we would expect gas entrained by X-ray cavities to lie.

Region E contains the southern, more quiescent, part of the high v$_{50}$ filament. Fig.~\ref{fig:kinematics:PKS0745:regions:spectra} shows that region E is the only region whose spatially-integrated spectrum is fitted using a single kinematic component. In region D, the first kinematic component (red) is quite faint. It becomes too faint to be detected in region E. Region E has the most redshifted total median velocities, although the gas associated with the second kinematic component in regions B and C is more redshifted. Interestingly, the gas in region E has large v$_{50}$ but low $\sigma$, such that while the nebular gas is moving at a speed of $\sim 300$ km s$^{-1}$, it is not being significantly disturbed.

Region F is the region NW of the nucleus which has low v$_{50}$ (see Fig.~\ref{fig:kinematics:PKS0745:regions}, upper left panel) and low $\sigma$ compared to surrounding gas (see Fig.~\ref{fig:kinematics:PKS0745:regions}, upper right panel). Region F partially overlaps with the NW cavity. Although the edges of the cavities are quite uncertain, the edge of the NW cavity roughly coincides with the place where the low v$_{50}$ filament separates into two filaments. The gas in region F has low velocity dispersion compared to the gas in regions A, B, C and D. However, as seen in Fig.~\ref{fig:kinematics:PKS0745:regions:spectra}, the broad component of the [OII] emission line fit in region F has $\sigma > 200$ km s$^{-1}$. A significant fraction of the nebular gas in region F is disturbed, albeit not as much as some of the other regions studied. The NW cavity may be playing an important role in the kinematics of region F, although we cannot confidently establish a link between the two structures.

\subsection{RXJ0820.9+0752}\label{appendix:RXJ0820}

\begin{figure*}
    \centering
    \includegraphics[width=\textwidth]{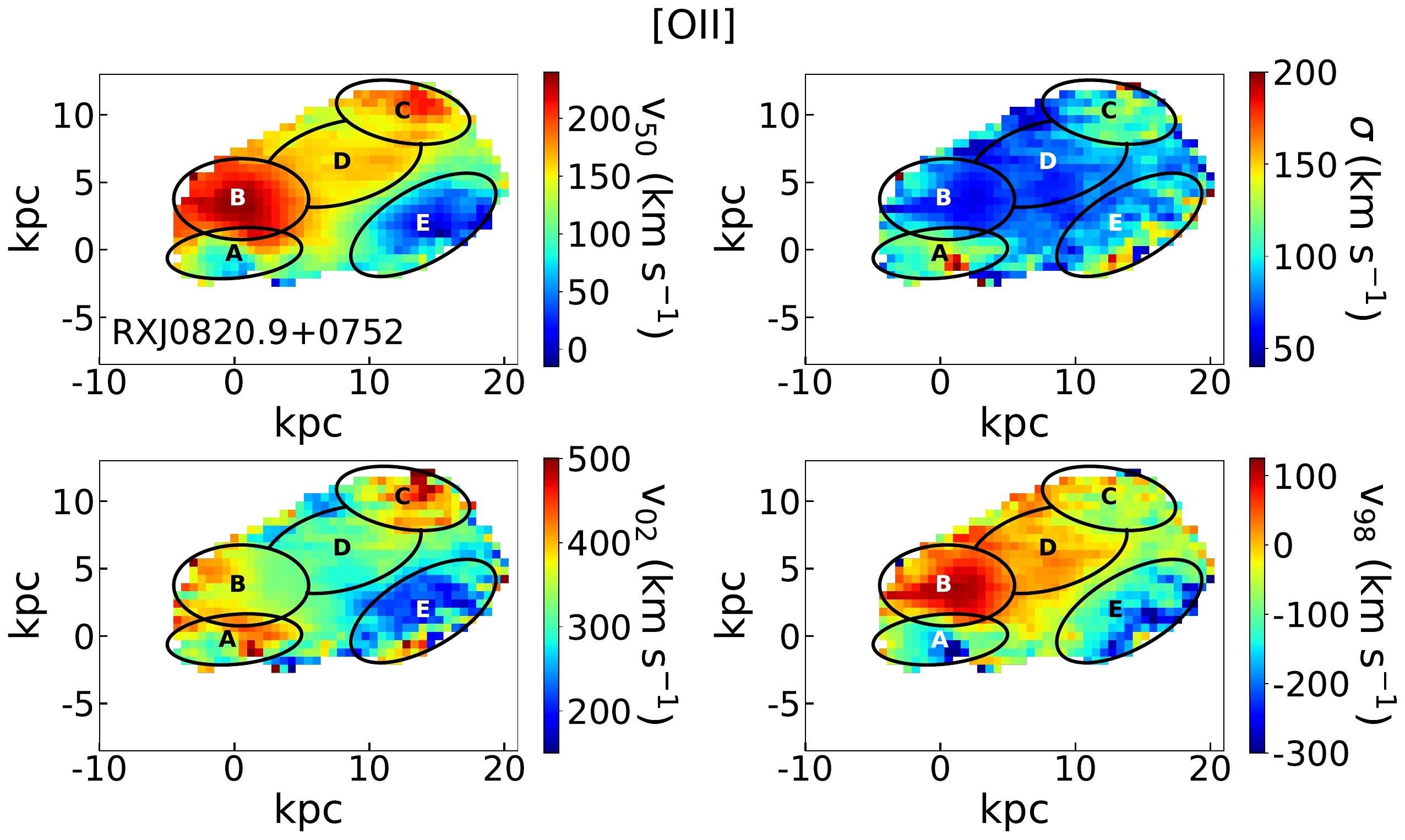}
    \caption{Kinematics of the [OII]3726,9$\lambda\lambda$ emission line doublet in RXJ0820 as in Fig.~\ref{fig:kinematics:RXJ0820} with regions of interest mentioned in Table~\ref{tab:RXJ0820:regions} identified by solid black ellipses.}
    \label{fig:kinematics:RXJ0820:regions}
\end{figure*}

\begin{deluxetable*}{ccccccccc}
        \tablecaption{RXJ0820.9+0752 regions}
    \tabletypesize{\scriptsize}
    \tablewidth{0pt} 
    \tablehead{ \colhead{Region}&\colhead{v$_{50}$}&\colhead{$\sigma$}&\colhead{v$_{02}$}&\colhead{v$_{98}$}&\colhead{Fraction L$_{\rm[OII]\rm}$}&\colhead{Dist}&\colhead{$\theta$}&\colhead{Area}\\ \colhead{} & \colhead{(km s$^{-1}$)} & \colhead{(km s$^{-1}$)} & \colhead{(km s$^{-1}$)} & \colhead{(km s$^{-1}$)} & \colhead{(\%)} & \colhead{(kpc)} & \colhead{($^{\circ}$)} & \colhead{(kpc$^2$)} }
\label{tab:RXJ0820:regions}
\colnumbers
\startdata 
A & 139 & 122 & 391 & $-114$ & 9.65 & 0.6 & 0 & 28 \\
B & 210 & 74 & 362 & 57 & 24.47 & 3.7 & 81 & 50 \\
C & 170 & 112 & 400 & $-61$ & 7.81 & 16.9 & 38 & 28 \\
D & 160 & 79 & 324 & $-4$ & 22.30 & 10.8 & 38 & 41 \\
E & 47 & 104 & 262 & $-168$ & 12.28 & 14.1 & 7 & 45 \\
\enddata
\tablecomments{\\(1) Region identified in Fig.~\ref{fig:kinematics:RXJ0820:regions}.\\
(2) [OII] flux weighted mean of the median velocity within the region.\\
(3) [OII] flux weighted mean of the velocity dispersion within the region.\\
(4) [OII] flux weighted mean of the v$_{02}$ within the region.\\
(5) [OII] flux weighted mean of the v$_{98}$ within the region.\\
(6) Fraction of the total [OII] luminosity observed from the spaxels included in the region.\\
(7) Distance between the central point of the region and the nucleus.\\
(8) Average angle between the center of the galaxy and the region. The angle is 0$^{\circ}$ pointing east and increases clockwise.\\
(9) Area of the region.\\
}
\end{deluxetable*}

Fig.~\ref{fig:kinematics:RXJ0820:regions} depicts the same kinematic maps as in Fig.~\ref{fig:kinematics:RXJ0820}, using a different colormap to highlight differences in velocity. The ellipses identify five regions of interest whose properties are detailed in Table~\ref{tab:RXJ0820:regions}. 

Region A is one of two regions where the stellar continuum is detected. This region encompasses the nebular gas closest to the nucleus. With a [OII] flux-weighted mean v$_{50}$ of $+139$ km s$^{-1}$, region A reiterates the significant differences in kinematics between the nebular gas and the stellar continuum. This region has the largest velocity dispersion. The few spaxels with $\sigma > 150$ km s$^{-1}$ are only $\sim 1.9$ kpc SW of the nucleus.

Region B is the second region where stellar continuum is observed. Region B has the gas with the largest median velocities. Although the gas in region B has, on average, velocities greater than $+200$ km s$^{-1}$, the gas is quiescent with the lowest mean velocity dispersion ($\sigma = 74$ km s$^{-1}$) of the five regions examined. Therefore, there is either a process displacing almost 25\% of the observed nebular gas at $\sim 200$ km s$^{-1}$ while barely disrupting it, and/or the stellar population in the core of the system, from which we measure the systemic velocity, is moving at $\sim 200$ km s$^{-1}$ with respect to the nebular gas in region B. 

Region C is another region with large median velocities, located roughly 15 kpc away from region B. Contrary to region B, the gas in region C is somewhat disturbed, with $\sigma = 112$ km s$^{-1}$. Region B and C, the two highest v$_{50}$ regions, appear to be connected through a $\sim 160$ km s$^{-1}$ filament, which is identified in Fig.~\ref{fig:kinematics:RXJ0820:regions} as region D. The velocity dispersion in region D is below 80 km s$^{-1}$, indicating that the nebular gas in region D is fairly quiescent. This ``filament" contains $\sim 22\%$ of the observed nebular gas.

Region E contains the nebular gas with the smallest velocities with respect to the systemic velocity. The gas is slightly disturbed, with $\sigma \sim 104$ km s$^{-1}$. The $\sim 100$ km s$^{-1}$ change in velocities between the gas in region E and the surrounding gas may explain the increase in velocity dispersion.

\bibliography{sample631}{}
\bibliographystyle{aasjournal}



\end{document}